\documentclass[12pt,epsf]{article}
\pdfoutput=1
\usepackage{amsmath}
\usepackage{graphicx}
\usepackage{epstopdf}

\begin{document}

\newcommand{\bra}[1]{\left\langle #1 \right|}
\newcommand{\ket}[1]{\left| #1 \right\rangle}
\def \cJ{{ \cal J}}
\newcommand{\boldk}{\bm{k}}
\newcommand{\boldn}{\bm{n}}
\newcommand{\boldp}{\bm{p}}
\newcommand{\boldx}{\bm{x}}
\newcommand{\boldI}{\bm{I}}
\newcommand{\boldJ}{\bm{J}}
\newcommand{\boldr}{\bm{r}}
\newcommand{\boldR}{\bm{R}}
\newcommand{\boldtheta}{\bm{\theta}}
\newcommand{\boldomega}{\bm{\omega}}
\newcommand{\boldalpha}{\bm{\alpha}}
\newcommand{\boldpi}{\bm{\pi}}
\def\<{\langle}
\def\>{\rangle}
\def \beq{\begin{equation}}
\def \eeq{\end{equation}}
\def \ben{\begin{eqnarray}}
\def \een{\end{eqnarray}}
\def \bea{\begin{array}}
\def \eea{\end{array}}
\def \bem{\begin{displaymath}}  
\def \eem{\end{displaymath}}
\def \frac#1#2{ { #1 \over #2} }
\def \braketa#1{\langle #1 \rangle}
\def \braketb#1#2{\langle #1 | #2 \rangle}
\def \braketc#1#2#3{\langle #1 | #2 | #3 \rangle}
\def \gtsim{\mathrel{\hbox{\raise0.2ex
     \hbox{$>$}\kern-0.75em\raise-0.9ex\hbox{$\sim$}}}}
\def \ltsim{\mathrel{\hbox{\raise0.2ex
     \hbox{$<$}\kern-0.75em\raise-0.9ex\hbox{$\sim$}}}}
\def \dE{{ \mit\Delta E}}
\def \dI{{ \mit\Delta I}}
\def \cJ{{ \cal J}}
\def \cH{{ \cal H}}
\def \cQ{{ \cal Q}}
\def \cB{{ \cal B}}
\def \cF{{ \cal F}}
\def \cG{{ \cal G}}
\def \cW{{ \cal W}}
\def \cU{{ \cal U}}
\def \data{{ \rm data}}
\def \Nils{{ \rm Nils}}
\def \self{{ \rm self}}
\def \rexp{{ \rm exp}}
\def \rcal{{ \rm cal}}
\def \rms{{ \rm rms}}
\def \osc{{ \rm osc}}
\def \rot{{ \rm rot}}
\def \RPA{{ \rm RPA}}
\def \max{{ \rm max}}
\def \intr{{ \rm intr}}
\def \evod{{ {\rm e}\hbox{\raise-0.15ex\hbox{\rm -}}{\rm o} }}
\def \oneqp{{ 1\hbox{\raise-0.15ex\hbox{\rm -}}{\rm q.p.} }}
\def \nmax{{ n_{\rm max}}}
\def \colon{{ \mbox{:} }}
\def \maru#1{ {\mathop{#1}^{\circ}} }
\def \T{{ \rm T}}
\newcommand{\idp}{\int \frac{{\rm d}^3\bp}{(2\pi)^3}}
\newcommand{\idk}{\int \frac{{\rm d}^3\bk}{(2\pi)^3}}
\newcommand{\ip}{\int \frac{{\rm d}^4p}{(2\pi)^4}}
\newcommand{\ik}{\int \frac{{\rm d}^4k}{(2\pi)^4}}
\newcommand\bx{{\bf x}}
\newcommand\nn{\nonumber \\}
\renewcommand{\d}{\partial}
\newcommand{\dbeta}{\frac{\partial}{\partial\beta}}
\newcommand{\dgamma}{\frac{\partial}{\partial\gamma}}
\newcommand{\ddbeta}{\frac{{\partial}^2}{\partial{\beta}^2}}
\newcommand{\ddgamma}{\frac{{\partial}^2}{\partial{\gamma}^2}}
\newcommand{\dss}[1]{#1 \hspace{-0.45em}/} 
\newcommand{\ds}[1]{#1 \hspace{-0.5em}/}  
\newcommand{\Ds}[1]{#1 \hspace{-0.55em}/} 
\newcommand{\Dds}[1]{#1 \hspace{-0.6em}/} 
\newcommand\bzeta{\mbox{\boldmath$\zeta$}}
\newcommand\bgamma{\mbox{\boldmath$\gamma$}}
\newcommand\bsigma{\mbox{\boldmath$\sigma$}}
\newcommand\bSigma{\mbox{\boldmath$\Sigma$}}
\newcommand\btau{\mbox{\boldmath$\kappa$}}
\newcommand\btheta{\mbox{\boldmath$\theta$}}
\newcommand\bk{{\bf k}}
\newcommand\bq{{\bf q}}
\newcommand\E{\epsilon}
\newcommand\brho{\mbox{\boldmath$\rho$}}
\newcommand\bp{{\bf p}}
\def\rvec{\vec{r}}
\def\xvec{\vec{x}}
\def\half{{1\over 2}}
\def\sla{\slash{\!\!\!} }
\newcommand{\vp}{\mbox{\boldmath $p$}}
\newcommand{\vq}{\mbox{\boldmath $q$}}
\newcommand{\vr}{\mbox{\boldmath $r$}}
\newcommand{\vk}{\mbox{\boldmath $k$}}
\newcommand{\vP}{\mbox{\boldmath $P$}}
\newcommand{\vR}{\mbox{\boldmath $R$}}
\newcommand{\tpsp}{\hspace{1.5em}}
\newcommand{\del}{\partial}
\newcommand{\Hhat}{\hat{H}}
\newcommand{\Nhat}{\hat{N}}
\newcommand{\Qhat}{\hat{Q}}
\newcommand{\Phat}{\hat{P}}
\newcommand{\Ghat}{\hat{G}}
\newcommand{\That}{\hat{\Theta}}
\newcommand{\Nt}{\tilde{N}}
\newcommand{\Hc}{{\cal H}}
\newcommand{\bg}{\beta,\gamma}
\newcommand{\Dhatp}{\hat{D}^{(+)}}
\newcommand{\Dp}{D^{(+)}}
\newcommand{\Ddotp}{\dot{D}^{(+)}}
\newcommand{\adag}{a^{\dagger}}
\newcommand{\brared}[1]{\langle #1 ||}
\newcommand{\ketred}[1]{|| #1 \rangle}
\renewcommand{\thefootnote}{\fnsymbol{footnote}}

\begin{center}
{\bf\Large 
Microscopic derivation of the Bohr-Mottelson collective Hamiltonian and
its application to quadrupole shape dynamics
}\\
\end{center}

\vspace{5mm}

\begin{center}
Kenichi Matsuyanagi,$^{1,2}$
Masayuki Matsuo,$^{3}$ 
Takashi Nakatsukasa,$^{1,4}$
Kenichi Yoshida,$^{5}$
Nobuo Hinohara,$^{4,6}$ 
and Koichi Sato$^{1}$
\footnote{Present address: Department of Physics, Osaka City University, Osaka 558-8585, Japan}     
\end{center}

\begin{center}
$^{1}$ RIKEN Nishina Center, Wako 351-0198, Japan\\
$^{2}$ Yukawa Institute for Theoretical Physics, Kyoto University, Kyoto 606-8502, Japan \\
$^{3}$ Department of Physics, Faculty of Science, Niigata University, Niigata 950-2181, Japan \\
$^{4}$ Center for Computational Sciences, University of Tsukuba, Tsukuba 305-8571, Japan \\
$^{5}$ Graduate School of Science and Technology, Niigata University, Niigata 950-2181, Japan \\
$^{6}$ FRIB Laboratory, Michigan State University, East Lansing, Michigan 48824, USA 
\end{center}

\abstract{
We discuss the nature of the low-frequency quadrupole vibrations 
from small-amplitude to large-amplitude regimes. 
We consider full five-dimensional quadrupole dynamics 
including three-dimensional rotations restoring the broken symmetries 
as well as axially symmetric and asymmetric shape fluctuations. 
Assuming that the time-evolution of the self-consistent mean field 
is determined by five pairs of collective coordinates and collective momenta, 
we microscopically derive the collective Hamiltonian of Bohr and Mottelson,  
which describes low-frequency quadrupole dynamics. 
We show that the five-dimensional collective Schr\"odinger equation 
is capable of describing large-amplitude quadrupole shape dynamics 
seen as shape coexistence/mixing phenomena. 
We summarize the modern concepts of microscopic theory of large-amplitude collective motion,  
which is underlying the microscopic derivation of the Bohr-Mottelson collective Hamiltonian.  
}
\vspace{7mm}

\section{Introduction} 

The subject of this review has a long history more than 50 years. 
Instead of describing the whole history, we mainly discuss the recent progress 
in the microscopic derivation of the Bohr-Mottelson collective Hamiltonian 
from a viewpoint of microscopic theory of large-amplitude collective motion. 
Special emphasis will be put on the development of fundamental concepts 
underlying the collective model. It is intended to motivate future studies  
by younger generations on the open problems suggested in this review.
\\

\newpage   
\noindent      
{\bf Progress in fundamental concepts of the collective model}\\

Vibrational and rotational motions of a nucleus can be described as 
time-evolutions of a self-consistent mean field. 
This is the key idea of the collective model of Bohr and Mottelson, 
which opened up a new field of contemporary physics,
{\it quantum many-body theory of nuclear collective dynamics.}  
The central theme in this field is to describe 
the single-particle and collective motions 
in finite quantum systems in a unified manner. 
After the first paper in 1952 \cite{boh52}, the basic concepts underlying the unified model 
of Bohr and Mottelson has been greatly developed.  
The progress achieved until 1975 is summarized 
in their text book \cite{boh69, boh75} and Nobel lectures in 1975 \cite{boh76, mot76}.

The unified description of complementary concepts such as 
the collective and single-particle motions in nuclei possess 
a great conceptual significance in theoretical physics in general.    
Needless to say, understanding the coexistence of complementary concepts 
(such as particle-wave duality) 
constitutes a central theme in theoretical physics.  
The physics underlying the Bohr-Mottelson unified model is deep and wide.  
Among rich subjects pertinent to this model, we select and focus on the subject of  
microscopic derivation of the Bohr-Mottelson collective Hamiltonian: 
that is,  we concentrate on the collective Hamiltonian $H_{\rm coll}$ 
in the unified model Hamiltonian  
\begin{equation}
H_{\rm uni} = H_{\rm part} + H_{\rm coll} + H_{\rm coupl}, 
\label{eq:H_unified}
\end{equation}
where $H_{\rm part}$ describes the single-particle motions in a self-consistent mean field 
and $H_{\rm coupl}$ is the coupling Hamiltonian generating the interplay between 
the single-particle motions and collective motions. 
Specifically, we focus on the low-frequency quadrupole collective motions 
and call the quadrupole collective Hamiltonian 
{\it the Bohr-Mottelson collective Hamiltonian}. 
We discuss its generalized form as described in their textbook, 
where the mass parameters (collective inertial masses) 
appearing in the collective Hamiltonian 
are not constant but functions of deformation variables. 
In our point of view, it is desirable to adopt  
this general definition of the Bohr-Mottelson collective Hamiltonian 
in order to respect the conceptual progress achieved 
by collaborative efforts of many researchers worldwide during 1952 to 1975.   
In this connection, we would like to quote a sentence from their Nobel lectures: 
``The viewpoints that I shall try to summarize gradually emerged 
in this prolonged period." \cite{boh76}. 
\\

\noindent
{\bf Brief remarks about the history} \\

Soon after the introduction of the collective model by Bohr and Mottelson in 1952-1953,
attempts to formulate a microscopic theory of the collective model started.  
This became one of the major subjects of theoretical physics in 1950's 
and greatly stimulated to open up a new fertile field, the nuclear many-body theory,  
to derive the collective phenomena starting from the 
nucleon degrees of freedom constituting the nucleus.  
 
The major approach at that time was to introduce collective coordinates explicitly 
as functions of coordinates of individual nucleons 
and separate collective shape degrees of freedom from the rest. 
From among numerous papers, we refer Tomonaga theory \cite{tom55,miy56} 
and a similar work by Marumori {\it et al.} \cite{mar55} as representative examples. 
In spite of their conceptual significance, however,  
it turned out that these approaches fail for description
of low-energy modes of shape fluctuations.   
The main reasons of this failure are 1) the assumption that the collective coordinates 
are given by local one-body operators (such as mass-quadrupole operators) leads to 
the inertial masses of irrotational fluids \cite{boh75}, in contradiction to experimental data 
which suggest that the inertial masses of the first excited quadrupole vibrational states   
are much larger than the irrotational masses, 
and, as we shall discuss in this review, 
2) the quantum shell structure of the single-particle motion 
in the self-consistent mean field  and  
the pairing correlations among nucleons near the Fermi surface  
play essential roles in the emergence of the 
low-frequency quadrupole modes of excitation in nuclei.       
Interestingly, it became clear much later that the Tomonaga theory is applicable 
to high-frequency giant resonances, rather than 
the originally intended low-frequency quadrupole vibrations \cite{row85,row15,suz80,kur96}. 
One of the important lessons we learned from these early attempts is
that, it is not trivial at all to define microscopic structure of collective coordinates
appropriate to low-energy shape vibrations. 

In 1960, the quasiparticle random-phase approximation (QRPA) 
based on the Bardeen-Cooper-Schrieffer (BCS) theory 
of superconductivity was introduced in nuclear structure theory \cite{bar60,mar60}.  
This was a starting point of the modern approach to determine, 
on the basis of the time-dependent mean field picture,  
the microscopic structures of the collective coordinates and their conjugate momenta  
without postulating them by physical intuition.    

After the initial success of the BCS+QRPA approach for small amplitude oscillations 
in 1960's and its extensions by boson expansion methods \cite{kle91} in succeeding years, 
attempts to construct a microscopic theory of large-amplitude collective motion (LACM) 
started in mid 1970's \cite{abe83}.  
At that time, time-dependent Hartree-Fock (TDHF) calculations 
for heavy-ion collisions also started \cite{dav85}. 
These attempts introduced collective coordinates 
as parameters specifying the time evolution of the self-consistent mean field, 
instead of explicitly defining them as functions of coordinates of individual nucleons. 
{\it This was a historical turning point in basic concept
of collective coordinate theory}: In these new approaches,
it is unnecessary to define {\it global} collective operators as
functions of coordinates of individual nucleons. 
In this paper, we shall discuss the basic ideas of such modern approaches and 
describe how to derive, in a microscopic way, 
the quadrupole collective Hamiltonian of Bohr and Mottelson
on the basis of the moving self-consistent mean-field picture.  
\\

\noindent
{\bf Contents of this review}\\

Our major aim is to review the progress in the fundamental concept of  
``{\it collective motion, collective coordinates, and collective momenta}" 
which have been acquired during the long-term efforts of many researchers 
to give a {\it microscopic foundation} of the Bohr-Mottelson collective model.  
Special emphasis will be put on the developments 
during the 40 years after the Nobel Prize of 1975 to Bohr, Mottelson, and Rainwater 
\cite{boh76,mot76,rai76}.  
Although we focus on the quadrupole collective motions, 
the techniques and underlying concepts are general and applicable to 
other collective motions at zero temperature as well,  
including octupole collective motions \cite{but96} 
and various kinds of many-body tunneling phenomena 
of finite quantum systems, 
such as spontaneous fissions \cite{bra72} and subbarrier fusion reactions \cite{hag12}. 

\noindent
In Sec.~\ref{sec:QuadrupoleCollectivePhenomena}, 
we summarize the basic properties of the low-frequency quadrupole collective excitations. \\
In Sec.~\ref{sec:BMcollH}, 
the Bohr-Mottelson collective Hamiltonian is recapitulated. \\
In Sec.~\ref{sec:QRPAandBeyond}, 
we discuss the basics of the microscopic theory of nuclear collective motion. 
We start from the QRPA as a small amplitude approximation of 
the time-dependent Hartree-Fock-Bogoliubov (TDHFB) theory, 
which is an extension of the TDHF to superfluid (superconducting) systems 
taking into account the pairing correlations. 
(We use the terms, superfluidity and superconductivity, in the same meaning.)  
We then discuss how to extend the basic ideas of the QRPA to treat LACM
as seen in shape coexistence/mixing phenomena widely observed in nuclear chart. \\
In Sec.~\ref{sec:microBM}, 
we introduce the {\it local} QRPA method and describe  
how to derive the Bohr-Mottelson collective Hamiltonian 
in a microscopic way on the basis of the TDHFB theory. \\
In Sec.~\ref{sec:Illustration}, 
we present an illustrative example of numerical calculation
for a shape coexistence/mixing phenomenon.\\
In Sec.~\ref{sec:OtherApproaches}, 
we briefly remark on other microscopic approaches to derive 
the Bohr-Mottelson collective Hamiltonian. \\
In Sec.~\ref{sec:SCC}, 
we review fundamentals of the microscopic theory of LACM. 
The basic concepts underlying the adiabatic self-consistent coordinate (ASCC) method 
and the local QRPA method will be summarized. \\
In Sec.~\ref{sec:OpenProblems}, 
we discuss future outlook. Possible extensions to new regions of 
nuclear structure dynamics will be suggested. \\
In Sec.~\ref{sec:Conclusion}, 
we conclude this review emphasizing future subjects awaiting 
applications and further extensions of the collective Hamiltonian approach.       
  
This review is an extended version of the short article published quite recently \cite{mat16}
\footnote{
Accordingly, there is some similarity between sections 2.3, 4.3, 7.2, 7.3 and 8.1, and reference~\cite{mat16}.}
and provides more detailed discussions on the basic ideas and fundamental 
concepts in the microscopic derivation of the Bohr-Motteson collective Hamiltonian. 
As we develop the basic concepts of collective motion 
on the basis of the time-dependent self-consistent mean-field theory, 
we refer the review  by Pr{\'o}chniak and Rohozi{\'n}ski \cite{pro09}
for other microscopic approaches and a detailed account of the techniques 
of treating the Bohr-Motteson collective Hamiltonian. 
For analyses of a wide variety of quadrupole collective phenomena 
in terms of various models related to the Bohr-Motteson collective Hamiltonian, 
we refer the recent review by Frauendorf \cite{fra15}.  
Because our major aim is the microscopic derivation of the collective inertial masses, 
we leave out discussions on phenomenological models 
reviewed by Cejnar, Jolie, and Casten \cite{cej10}, 
where the inertial masses are treated as parameters. 

\section{Low-frequency quadrupole collective motions}
\label{sec:QuadrupoleCollectivePhenomena}

In this section, we first summarize the basic properties of the low-frequency quadrupole 
collective excitations and then introduce the Bohr-Mottelson collective Hamiltonian.  

\subsection{Nature of the first excited $2^+$ modes}

Except for doubly magic nuclei 
in the spherical $j$-$j$ coupling shell-model picture,
the first excited states in almost all even-even nuclei 
(consisting of even numbers of neutrons and protons)
have angular momentum two with positive parity ($I^\pi=2^+$).
Systematics of experimental data for these first $2^+$ states shows 
that their excitation energies are very low 
in comparison to the energy gap 2$\Delta$ that characterizes 
nuclei with superfluidity (superconductivity), 
and that their electric quadrupole ($E2$) transition probabilities to the $0^+$ ground states 
are very large compared to those associated with single-particle transitions. 
For nuclei whose mean fields are spherical,  the first excited $2^+$ states
can be characterized as collective vibrations of finite quantum systems with superfluidity 
\cite{bri05}.
They are genuine quantum vibrations that are essentially different from 
surface oscillations of a classical liquid drop. 
In other words, the superfluidity and shell structure play indispensable roles in their emergence.

In axially deformed nuclei, whose mean fields break the rotational symmetry 
but conserve the axial symmetry,
the first excited $2^+$ state can be interpreted as quantum rotational states  
whose mean fields are uniformly rotating about an axis perpendicular to the symmetry axis.  
Regular rotational spectra appear when the amplitudes of quantum shape fluctuation 
are smaller than the magnitude of equilibrium deformation.  
Nuclei that have very small ratios, $E(2^+)/2\Delta$, of the $2^+$ excitation energies to the energy gaps, 
 (less than, as a rule of thumb, 0.1) belong to this category. 
The rotational moments of inertia evaluated from $E(2^+)$
are found to be about half of the rigid-body value.
This deviation of the moment of inertia from the rigid-body value is one of the most clear evidences 
that the ground states of nuclei are in a superfluid phase.
Large portion of nuclei exhibiting regular rotational spectra have the prolate (elongated spheroidal) shape.
Origin of prolate shape dominance over oblate (flattened spheroidal) shape is 
an interesting fundamental problem (prolate-oblate asymmetry)
\cite{ari12,ari15}.

When the mean field of the ground state conserves the rotational symmetry, 
the first excited $2^+$ state have been regarded as
quadrupole vibrational excitation of a spherical shape with frequencies lower than the energy gap.
They are more lowered as the numbers of neutrons and/or protons 
deviate from the spherical magic numbers.  
Eventually the vibrational $2^+$ states turn into the rotational $2^+$ states discussed above. 
Thus, one may regard low-lying quadrupole vibrations as soft modes of the quantum phase transition 
that breaks the spherical symmetry of the mean field.
In finite quantum systems such as nuclei, however, this phase transition takes place rather gradually  
for a change of nucleon number, 
creating a wide transitional region in the nuclear chart.  
Low-energy excitation spectra of these nuclei exhibit
intermediate characters between the vibrational and rotational ones.
Softer the mean field toward the quadrupole deformation, larger the amplitude 
and stronger the nonlinearity of the vibration.

\subsection{Quantum shape transitions in nuclei}
\vspace{5mm}
\noindent
{\bf Why are nuclei deformed ?} \\

As is well known, the equilibrium shape of the classical liquid drop is spherical. 
When it rotates, it becomes oblate due to the effect of the centrifugal force. 
In contrast, most nuclei favor the prolate shape, 
except for nuclei situated near the closed shells of the $j$-$j$ coupling shell model 
(whose proton number $Z$ and/or neutron number $N$ are near 
the spherical magic numbers). As we shall discuss later, even such nuclei 
whose ground state are spherical, deformed states appear in their excited states. 
The appearance of the deformed shapes in nuclei is due to quantum-mechanical shell 
effects associated with the single-particle motions in the mean field. 
Let us first start with what this means.
\\
 
\noindent
{\bf Deformable mean field and deformed shell structure}\\

It is well known from the success of the $j$-$j$ coupling shell model \cite{may55} 
that the concept of {\it single-particle motion in a mean field} holds in nuclear structure. 
Differently from electrons in an atom, the shell-model potential is  
collectively generated by all nucleons constituting the nucleus. 
In other words, the single-particle picture of the shell model emerges 
as a result of collective effects of all nucleons generating the  self-consistent mean field.  
It implies that the single-particle potential of the nucleus is a deformable quantum object 
\cite{boh76, mot76, rai76}.  
In fact, as we shall discuss below, the self-consistent mean field possesses 
{\it collective predisposition} to generate a variety of 
vibrational and rotational modes of excitation.

Because the nucleus is a finite quantum system, the single-particle states 
form a shell structure. 
The spherical shell structure in the $j$-$j$ coupling shell model 
gradually changes with the growth of deformation in the mean field 
and generates {\it deformed shell structures} and {\it deformed magic numbers} 
at certain deformed shapes \cite{rag78,nil95}.  
The gain of binding energies associated with deformed magic numbers 
appearing at various deformed shapes for certain combinations of $(Z, N)$   
stabilizes the deformed shape. 
For instance, for a nucleus whose $(Z, N)$ are far from the spherical magic numbers 
but near the deformed magic numbers associated with a certain prolate shape, 
it is energetically favorable for this nucleus to take the prolate shape. 
This is the major origin of the appearance of a rich variety of deformed shapes 
in nuclei. The deformed shell-structure effects are clearly seen 
in the appearance of superdeformed nuclei having prolate shapes 
with the axis ratio about 2:1 \cite{nol88, jan91}.

\begin{figure}[tbp]
\begin{center}
\includegraphics[width=0.5\textwidth,bb=0 0 455 566,clip]{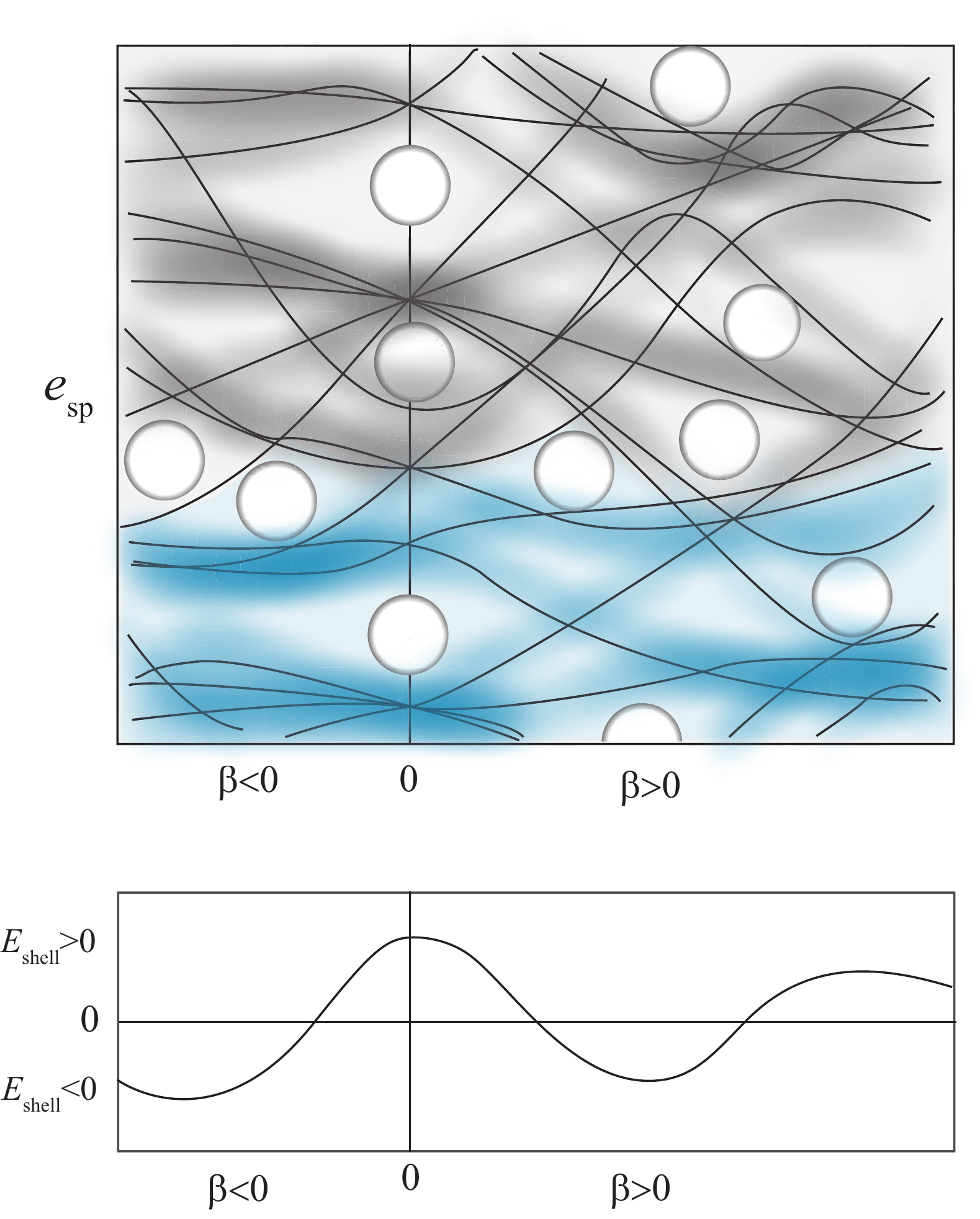}
\caption{
Illustration of shell structure and its change with increasing deformation. 
The level densities are drawn as functions of the single-particle energy $e_{sp}$ 
and the deformation parameter $\beta$. 
Positive and negative $\beta$ corresponds to the prolate and oblate shapes, respectively.  
Regions with high- and low-level densities are shown by shading  
(shade for high and light for low). 
Occupied and unoccupied regions are indicated by blue and gray shades, respectively.
The circles indicate appearance of 
(spherical and deformed) magic numbers. 
The contribution of  the shell structure to the binding energy (shell structure energy) 
for a fixed neutron (or proton) number changes as a function of $\beta$.  
Generally, it exhibits an oscillating pattern. 
This conceptual figure is drawn on the basis of the realistic calculation \cite{naz85} 
by using the deformed Wood-Saxon potential for $^{80}$Sr.  
}
\label{fig:shell_structure}
\end{center}
\end{figure}

The shell structures can be defined, in a general concept, 
as regularly oscillating gross structures in the distribution of 
single-particle-energy eigenvalues \cite{boh75,bra72,bra97}. 
It is very important to notice that those structures are quite sensitive 
to the shape of the mean-field potential. 
The oscillation pattern changes following the variation of the deformation parameter. 
Figure \ref{fig:shell_structure} illustrates this concept. 

Although one can easily calculate single-particle-energy eigenvalues 
for a given shape of the mean field,  such a quantum-mechanical calculation 
does not tell the origins of appearance of such gross structures. 
For a deeper understanding of the origins, one can make use of 
the semi-classical theory of (deformed) shell structure. 
For further discussions on this subject, we refer 
the textbook by Brack and Bhaduri \cite{bra97} 
and the review by Arita \cite{ari15}.  
\\

\noindent
{\bf Emergence of collective rotational motions restoring broken symmetries}\\

The central concept of the BCS theory of superconductivity is 
spontaneous gauge-symmetry breaking and emergence of associated collective modes. 
The massless collective modes that restore the broken symmetry   
are called Anderson-Nambu-Goldstone (ANG) modes   
\cite{bri05,and58,nam60}. 
Nuclear rotations are manifestations of this dynamics in finite quantum systems,  
as pointed out by Bohr and Mottelson \cite{boh75,boh76}; they are ANG modes 
restoring the spherical symmetry broken by the self-consistently generated mean field. 

The spontaneous breaking of the spherical symmetry (deformation) in the self-consistent 
mean field enables us to define the orientation degrees of freedom that specify 
the orientation of the body-fixed (intrinsic) frame relative to the laboratory frame. 
The body-fixed (intrinsic) frame can be defined as a principal-axis frame of the deformed
self-consistent mean field generated by all nucleons constituting the nucleus. 
It is important to keep in mind that {\it the spontaneous breaking of symmetry can be hidden  
in finite quantum systems such as nuclei};  that is, the experimental measurements probe 
the states in the laboratory frame, which preserves the symmetries of the original Hamiltonian.
Thus, nuclear rotations may be viewed as rotational motions of the self-consistent mean field 
relative to the laboratory frame. 

``The spontaneous breaking of the rotational symmetry in the self-consistent mean field" 
is the key concept to a unified description of the single-particle motion and 
the collective rotational motion.  With this concept we can generalize the notion of 
the single-particle motion in a spherical mean-field to that in a deformed mean field. 
At the same time the deformed mean field is rotating to restore the broken symmetry. 
Thus, extension of the concept of single-particle excitation with spontaneous breaking 
of the symmetry and appearance of new collective excitation restoring the broken symmetry 
are {\it dual concepts} that underline the quantum many-body theory of nuclear structure. 
We shall discuss in Sec.~\ref{sec:SCC} how to generalize 
this concept of {\it particle-collective duality} to slowly vibrating mean fields 
where the time scales of the single-particle and vibrational motions are separated 
in a good approximation. 
\\   

\noindent
{\bf Excitation spectra in the transitional region} \\ 

The low-frequency quadrupole vibrations can be regarded as 
soft modes of the quantum phase transition generating 
equilibrium deformations in the mean field. 
In nuclei situated in the transitional region from spherical to deformed,   
the amplitudes of quantum shape fluctuation about the equilibrium shape increase significantly.  
The large shape fluctuations occur also in weakly deformed nuclei 
where the binding-energy gains due to the symmetry breaking are comparable 
in magnitude to the vibrational zero-point energies.  
Such transitional situations are abundant in nuclear chart and 
those transitional nuclei show quite rich excitation spectra
(see {\it e.g.,} \cite{sak91}).   
Existence of wide transitional regions is a characteristic feature 
of finite quantum systems and provides an invaluable opportunity 
to investigate the process of the quantum phase transition through 
the change of quantum spectra with nucleon number.  
A detailed account of instability phenomena and strong anharmonicity effects   
in the transitional region is given in Chap. 6 of \cite{boh75}.  

\subsection{Quadrupole collective dynamics}

Before introducing the Bohr-Mottelson collective Hamiltonian, 
we add some remarks on quadrupole collective phenomena that await its application. 
\\

\noindent
{\bf Interplay of low-frequency shape-fluctuations and rotational motions}\\

In finite quantum systems such as nuclei, the rotational ANG modes may couple 
with quantum shape-fluctuation modes rather strongly. 
For example, even when the self-consistent mean field acquires  
a deep local minimum at a finite value of $\beta$ in this direction,   
the deformation energy surface may be flat in the $\gamma$ direction.  
In this case,  the nucleus may exhibit a large-amplitude shape fluctuation 
in the $\gamma$ degree of freedom.  
(Here, $\beta$ and $\gamma$ represent the magnitudes of 
axial and triaxial quadrupole deformations.)   
Actually, such a situation, called $\gamma$ soft, is widely observed in experiments. 
In nuclei which preserve the axial symmetry, 
the quantum-mechanical collective rotation  about the symmetry axis is forbidden. 
Once the axial symmetry is dynamically broken by quantum shape fluctuations,  
however, the rotational degrees of freedom about three principal axes are all activated. 
As a consequence, the rotational spectra in such $\gamma$-soft nuclei 
do not exhibit a simple $I(I+1)$ pattern of an axial rotor.  
Such an interplay of the shape-fluctuation and rotational modes may be regarded as 
a characteristic feature of finite quantum systems and provides an invaluable opportunity 
to investigate the process of the quantum phase transition through analysis of quantum spectra. 

Thus, we need to treat the two kinds of collective modes  
(symmetry-restoring ANG modes and quantum shape fluctuation modes)  
in a unified manner to describe low-energy excitation spectra of nuclei. 
\\

\noindent
{\bf Quantum shape fluctuations and shape coexistence}\\

When different kinds of quantum eigenstates associated with different shapes coexist 
in the same energy region, 
we call them `{\it shape coexistence phenomena.}'
This situation is realized when shape mixing due to tunneling motion is weak and 
collective wave functions retain their localizations about different equilibrium shapes. 
In contrast, when the shape mixing is strong, 
large-amplitude shape fluctuations ({\it delocalization} of the collective wave functions) 
extending to different local minima may occur. 

When a few local minima of the mean field with different shapes appear in the same energy region, 
LACM tunneling through potential barriers and extending between local minima may take place. 
These phenomena may be regarded as a kind of macroscopic quantum tunneling. 
Note that the barriers are not given by external fields but are self-consistently generated as 
a consequence of quantum dynamics of the many-body system.  
Quantum spectra of low-energy excitation that needs such concepts have been observed 
in almost all regions of the nuclear chart \cite{abe90,woo92,hey11}. 
\\

\section{Bohr-Mottelson collective Hamiltonian}
\label{sec:BMcollH}

Bohr and Mottelson introduced the five-dimensional (5D) quadrupole collective Hamiltonian    
describing the quadrupole vibrations and rotations in a unified manner 
\cite{boh75}.   
It is written as 
\begin{equation}
H_{\rm coll}=T_{\rm vib}+T_{\rm rot}+V(\bg), 
\label{eq: BMcoll}
\end{equation}
\begin{equation}
T_{\rm vib}=\frac{1}{2}D_{\beta\beta}(\bg)\dot \beta^2 
 + D_{\beta\gamma}(\bg)\dot \beta \dot \gamma
 + \frac{1}{2}D_{\gamma\gamma}(\bg)\dot \gamma^2, 
\label{eq:classicalTvib}
\end{equation}
\begin{equation}
T_{\rm rot}=\frac{1}{2}\sum_{k=1}^3\cJ_k(\bg)\dot{\varphi}_k^2. 
\label{eq:classicalTrot}
\end{equation}
Here, $\varphi_k$ are components of the rotational angle on the three intrinsic axes.   
The quadrupole deformations $(\bg)$ and the rotational angles $\varphi_k$ are treated 
as dynamical variables, and 
$(\dot \beta, \dot \gamma)$ and $\dot{\varphi}_k$ represent their time-derivatives. 
The $\dot{\varphi}_k$ are called angular velocities. 
We shall define in Sec.~\ref{sec:microBM} the $(\bg)$ deformations  
through the expectation values of the quadrupole operators 
with respect to the time-dependent mean-field states.  
The quantities ($D_{\beta\beta}, D_{\beta\gamma}$, and $D_{\gamma\gamma}$) 
appearing in the kinetic energies of vibrational motion, $T_{\rm vib}$, 
represent inertial masses of the vibrational motion. 
They are functions of $\beta$ and $\gamma$.  
The quantities $\cJ_k(\bg)$ in the rotational energy $T_{\rm rot}$
represent the moments of inertia with respect to the intrinsic (body-fixed) axes.
The intrinsic axes may be defined by the principal axes of the 
body-fixed frame that is attached to the instantaneous shape of 
the time-dependent mean field. 
The term, $V(\bg)$, represents the potential energy as a function of $\beta$ and $\gamma$.

The Bohr-Mottelson collective Hamiltonian (\ref{eq: BMcoll}) is often referred to 
in relation to the liquid drop model.  It should be emphasized, however, that, 
the analogy with the {\it classical} liquid drop is irrelevant 
to low-frequency quadrupole collective motions. 
Already in 1950's, it was recognized that 
the nucleus is ``an unusual idealized {\it quantum} fluid" and 
``one is dealing with a most interesting new form of matter" \cite{hil53}. 
Indeed, as discussed in Sec.~\ref{sec:QuadrupoleCollectivePhenomena}, 
most of nuclei may be regarded as a superfluid of extremely small size 
(with a radius of a few femtometer), 
and the nature of nuclear deformation is essentially different from 
that of surface shape oscillations of the classical liquid drop; 
that is, the nuclear deformation is associated with quantum shell structure 
and spontaneous breaking of the spherical symmetry in the self-consistent mean field.    

The form of the collective Hamiltonian (\ref{eq: BMcoll}) is quite general and applicable 
to various finite many-body systems, 
but the specific dynamical properties of the system of interest are revealed by 
the values and the ($\bg$)-dependence of 
the collective inertia masses  ($D_{\beta\beta}, D_{\beta\gamma}$, $D_{\gamma\gamma}$, 
$\cJ_k$) as well as the potential energy $V(\bg)$. 
For understanding the dynamical properties of the nucleus, therefore,   
it is imperative to derive these quantities in a microscopic way 
and compare with what experimental data indicate.  
We shall show in this review that the collective Hamiltonian (\ref{eq: BMcoll}) 
with the collective inertial masses and the potential energy 
microscopically evaluated on the basis of the moving superfluid mean-field picture 
describes very well the low-frequency quadrupole collective dynamics of the nucleus. 
Furthermore, quantum correlations beyond the mean field are nicely described  
by quantizing the collective variables that govern the time-evolution of 
the self-consistent mean field. 

The classical Hamiltonian (\ref{eq: BMcoll}) is given in terms of 
the five curvilinear coordinates ($\beta, \gamma$ and 
the three Euler angles which are connected with $\varphi_k$ by a linear transformation)    
and their time derivatives.   
For quantization in curvilinear coordinates, 
we can adopt the so-called Pauli prescription~\cite{eis87}. 
(For convenience of readers, we recapitulate this prescription in Appendix A.)  
We shall discuss on its foundation in Sec.~\ref{sec:microBM} describing 
the microscopic derivation of the Bohr-Mottelson collective Hamiltonian.   
The quantized 5D quadrupole collective Hamiltonian takes the following form: 
\begin{equation}
{\hat H}_{\rm coll}={\hat T}_{\rm vib}+{\hat T}_{\rm rot}+V(\beta,\gamma).  
\label{eq: quantized H_BMcoll}
\end{equation}
Here, 
the vibrational kinetic energy term ${\hat T}_{\rm vib}$ is given by  
\begin{eqnarray}
{\hat T}_{\rm vib}=-\frac{1}{2\sqrt{WR}}\left\{ \frac{1}{\beta^4} 
\left[\dbeta \left(\beta^2\sqrt{\frac{R}{W}}D_{\gamma\gamma}
\dbeta\right)-\dbeta \left(\beta^2\sqrt{\frac{R}{W}}D_{\beta\gamma}\dgamma\right)\right] 
\right. 
\nonumber 
\\ \left.
+\frac{1}{\beta^2\sin 3\gamma}\left[-\dgamma 
\left(\sqrt{\frac{R}{W}}\sin 3\gamma D_{\beta\gamma}\dbeta\right)
+\dgamma \left(\sqrt{\frac{R}{W}}\sin 3\gamma D_{\beta\beta}\dgamma\right)
\right] 
\right\}, 
\label{eq:quantumTvib}
\end{eqnarray}
and the rotational energy term ${\hat T}_{\rm rot}$ is given by 
\begin{equation}
\hat{T}_{\rm rot}=\sum_{k=1,2,3}\frac{\hat{I}_k^2}{2\cJ_k(\bg)} 
\label{eq:quantumTrot}
\end{equation}
with $\hat{I}_k$ denoting three components of the angular-momentum operator 
with respect to the intrinsic axes.  
In this paper, we use the unit with $\hbar=1$. 
In the above equations, 
\begin{eqnarray}
\beta^2 W(\bg) &=& D_{\beta\beta}(\bg)D_{\gamma\gamma}(\bg) - D_{\beta\gamma}^2(\bg), 
\label{eq:BM_W}
\\
R(\bg) &=& D_1(\bg)D_2(\bg)D_3(\bg), 
\label{eq:BM_R}
\end{eqnarray} 
and $D_k(\bg)$~$(k=1,2,3)$ are the rotational inertial functions related to
the moments of inertia by 
\begin{equation}
\cJ_k(\bg)=4\beta^2D_k(\bg) \sin^2(\gamma-2\pi k/3). 
\label{eq:rotational-inertia} 
\end{equation}

If all inertial masses  
$(D_{\beta\beta}, D_{\gamma\gamma}\beta^{-2}, D_1, D_2, D_3)$ 
are replaced by a common constant $D$ and $D_{\beta\gamma}$ is ignored,  
the above ${\hat T}_{\rm vib}$ is reduced to 
\begin{equation}
{\hat T}_{\rm vib}=-\frac{1}{2D} \left( {\frac{1}{\beta^4}} 
\dbeta \beta^4 \dbeta 
+{\frac{1}{\beta^2 \sin 3\gamma}}\dgamma \sin 3\gamma \dgamma \right). 
\label{eq:Bohr-H-52}
\end{equation}
Such a drastic approximation may be valid only for small-amplitude vibrations 
about a spherical HFB equilibrium. 
The need to go beyond this simplest approximation for 
the inertia masses has been pointed out \cite{boh75}.
For recent experimental data and phenomenological analyses of this problem,
we refer \cite{jol09,bon15} and references therein.

The collective Schr\"odinger equation is  
\begin{equation}
({\hat T}_{\rm vib}+{\hat T}_{\rm rot}+V(\bg)) \Psi_{\alpha IM}(\beta,\gamma,\Omega)
= E_{\alpha I}\Psi_{\alpha IM}(\beta,\gamma,\Omega). 
\label{eq:collective-schrodinger}
\end{equation}
The collective wave function in the laboratory frame, 
$\Psi_{\alpha IM}(\bg,\Omega)$, 
is a function of $\beta$, $\gamma$ and a set of three Euler angles $\Omega$. 
It is specified by the total angular momentum $I$, its projection onto the
$z$-axis in the laboratory frame $M$, and $\alpha$ that distinguishes the eigenstates
possessing the same values of $I$ and $M$. 
With the rotational wave function ${\cal D}^I_{MK}(\Omega)$, they are written as
\begin{equation}
 \Psi_{\alpha IM}(\bg,\Omega) = 
 \sum_{K={\rm even}}\Phi_{\alpha IK}(\bg)\langle\Omega|IMK\rangle,
\label{eq:collective WF}
\end{equation}
where
\begin{equation}
\langle \Omega | IMK \rangle = 
\sqrt{\frac{2I+1}{16\pi^2 (1+\delta_{K0})}} [{\cal D}^{I}_{MK}(\Omega) + (-)^I {\cal D}^I_{M,-K}(\Omega)].
\end{equation}
The vibrational wave functions in the body-fixed frame, $\Phi_{\alpha IK}(\bg)$,  
are normalized as
\begin{equation}
 \int d\beta d\gamma \sqrt{G(\bg)} |\Phi_{\alpha I}(\bg)|^2 = 1, 
\end{equation}
where  
\begin{equation}
 |\Phi_{\alpha I}(\bg)|^2 \equiv \sum_{K={\rm even}} |\Phi_{\alpha IK}(\bg)|^2,
\end{equation}
and the volume element is given by $\sqrt{G(\bg)} d\beta d\gamma$ with 
\begin{equation} 
 G(\bg) = 4\beta^8 W(\bg)R(\bg) \sin^2 3\gamma. 
\label{eq:BM_G}
\end{equation}
Thorough discussions of symmetries of the collective wave functions 
and the boundary conditions for solving the collective Schr\"odinger equation are given in 
Refs.~\cite{boh75, pro09,kum67,bel65}. 

Inserting (\ref{eq:collective WF}) into the collective Schr\"odinger equation 
(\ref{eq:collective-schrodinger}),  
we obtain the eigenvalue equation for vibrational wave functions  
\begin{eqnarray}
 \left[ \hat{T}_{\rm vib} + V(\bg) \right] \Phi_{\alpha IK}(\bg)
 + \sum_{K'={\rm even}}
 \bra{IMK}\hat{T}_{\rm rot}\ket{IMK'} \Phi_{\alpha IK'}(\bg)  = E_{\alpha I}
 \Phi_{\alpha IK}(\bg). 
\label{eq:BMvibration}
\end{eqnarray} 
Solving this equation, we obtain quantum spectra and collective wave functions.  
It is then straightforward to calculate electromagnetic transition probabilities 
among collective excited states.  
We recapitulate some basic formulae in Appendix \ref{sec:E2transition}.
\\ 

\noindent
{\bf Historical note} 
\\

The simple expression (\ref{eq:Bohr-H-52}) with a constant mass parameter $D$
for the vibrational kinetic energy is valid for harmonic vibrations about 
a spherical equilibrium point of the mean field, as derived in the 1952 paper \cite{boh52} 
by transforming the collective Hamiltonian for harmonic shape vibrations to 
the body-fixed frame defined as the instantaneous principal axis frame of 
the vibrating density distribution.   
Combined with the irrotational mass parameter $D_{\rm irrot}$ resulting 
from modeling the vibrational flow by that of the irrotational fluid, 
it is sometimes referred to the Bohr liquid-drop Hamiltonian. 
It should be emphasized, however,  that 
the inadequacy of the irrotational fluid model for the low-frequency quadrupole excitations 
was recognized from early on.  

In the Preface to the second edition (March 1, 1957) of the 1953 paper 
\cite{boh53},   
Bohr and Mottelson wrote: 
``As a first orientation, one attempted to employ for these parameters 
obtained from a liquid drop model, 
but already the early analysis of various nuclear properties 
showed the limitation of this comparison. 
The inadequacy of the liquid drop estimates was especially clearly 
brought out by the comparison of the 
nuclear moment of inertia with the deformations deduced from the rate of 
the electric quadrupole rotational transitions."
``An improved understanding of the collective nuclear properties 
has come from the efforts to derive these directly from the motion 
of the nucleons; this analysis has revealed the important influence 
of the nuclear shell structure on the collective motion."
``The inadequacy of the liquid drop model with irrotational flow 
implies that the collective coordinates considered 
as functions of the nucleonic variables are of more general form 
than (II.2) , ..." 
[(II.2) is the famous definition of the collective parameters 
$\alpha_{\lambda, \mu}$ in terms of the polar coordinates of individual particles.] 

Indeed, if we assume that a collective coordinate corresponds to 
a {\it local} one-body operator in the coordinate space
(such as the mass-multipole operator), we obtain a collective mass parameter 
associated with an irrotational velocity field (see p. 510 of \cite{boh75} and \cite{rei90}).    
\\ 
\\

\noindent
{\bf Illustration of typical situations}\\

\begin{figure}[tbp]
\begin{center}
\includegraphics[width=0.6\textwidth,bb=0 0 896 599,clip]{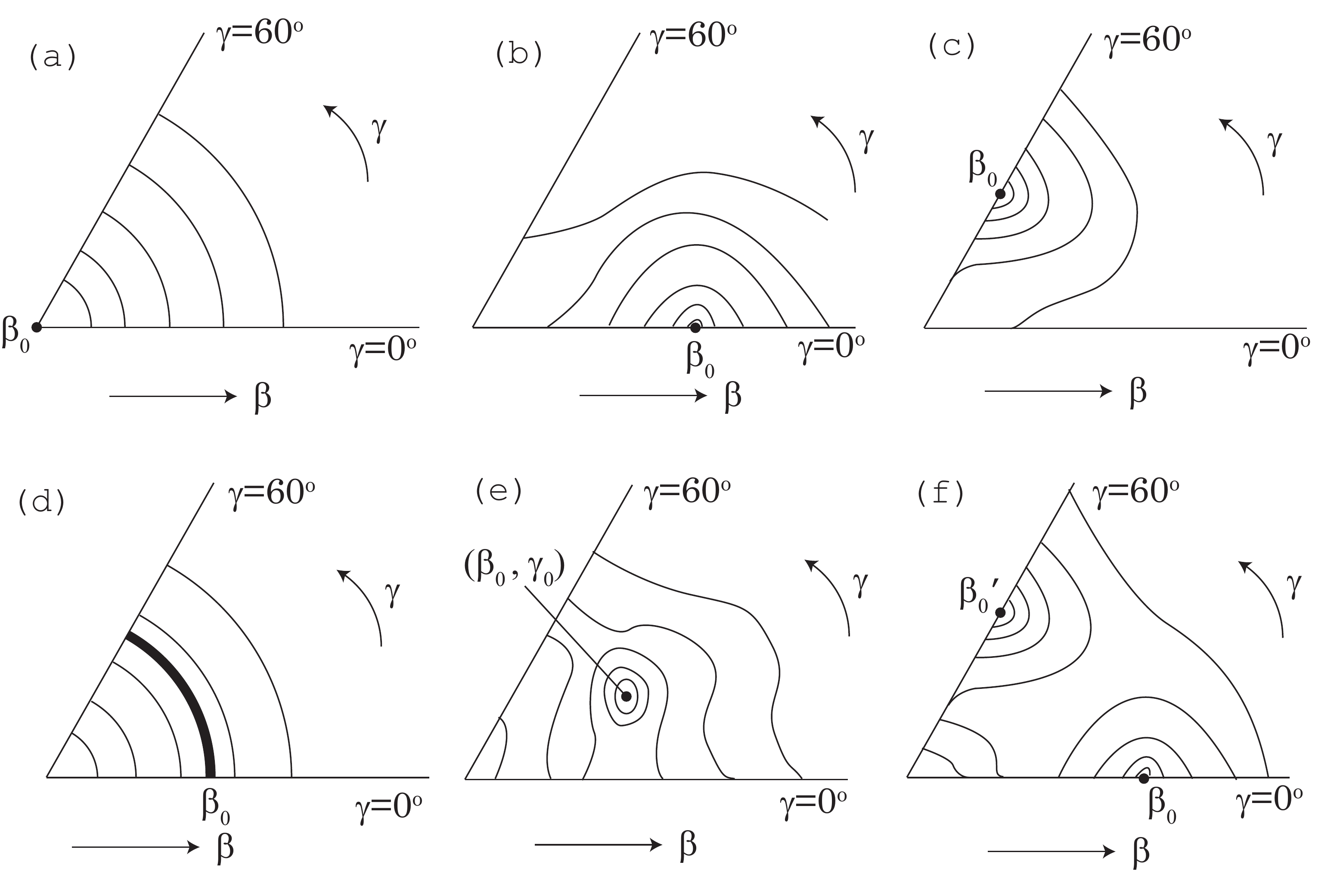}
\caption{
Illustration of typical patterns of the collective potential-energy surface $V(\bg)$,  
classified according to the location of the local minimum point(s) $(\beta_0, \gamma_0)$: 
(a) spherical, (b) prolate, (c) oblate, (d) $\beta_0 \ne 0$ in the $\beta$ direction
but the potential is flat with respect to $\gamma$
 (so-called $\gamma$-unstable situation \cite{wil56}), 
(e) triaxial, and (f) coexistence of the oblate and prolate minima.
}
\label{fig:PES}
\end{center}
\end{figure}

Figure \ref{fig:PES} illustrates typical patterns of the collective potential energy surface $V(\bg)$;  
these are classified according to the location of the local minimum.   
In the case that the potential energy $V(\bg)$ 
has a deep minimum at a finite value of $\beta$ and 
$\gamma=0^\circ$ (or $\gamma=60^\circ$), 
a regular rotational spectrum with the $I(I+1)$ pattern may appear. 
In addition to the ground band,  we can expect the $\beta$ and $\gamma$ bands to appear, 
where vibrational quanta with respect to the $\beta$ and $\gamma$ degrees of freedom 
are excited. 
Detailed investigations on the $\gamma$-vibrational bands over many nuclei 
have revealed, however, that they usually exhibit significant anharmonicities (non-linearlities) 
\cite{mat85b}.
Situations for the $\beta$-vibrational bands are quite mysterious.  
Recent experimental data indicate the need for a radical review of their characters 
\cite{hey11}. 
We shall discuss on this problem in Sec.~\ref{sec:OpenProblems}. 
The coexistence of two local minima at oblate and prolate shapes is a typical example 
of shape coexistence.  Experimental data indicate that the potential barrier between 
the two minima is, in many cases, low and the collective wave functions extend over 
the oblate and prolate regions through quantum tunneling (shape mixing). 
Also, there are many nuclei exhibiting intermediate features between 
the large-amplitude collective vibrations associated with the oblate-prolate shape coexistence 
and the rotational motions associated with the triaxial shape.  

We present in Appendix \ref{sec:TriaxialDynamics} a simple model that may be useful to understand  
several interesting limits of triaxial deformation dynamics in a unified perspective, 
including the axially symmetric rotor model, 
the $\gamma$-unstable model \cite{wil56}, 
the triaxial rigid rotor model \cite{dav58},  
and an ideal situation of the oblate-prolate shape coexistence. 
\\

\section{QRPA and its extensions} 
\label{sec:QRPAandBeyond}

In this section, we summarize the elementary concepts 
in microscopic theory of nuclear collective motion \cite{row70,rin80,bla86,ber94,row10}.  
We adopt the time-dependent mean-field picture. The main reason is that 
it provides a basis for a clear understanding of the correspondence between 
the quantum and classical aspects of the nuclear collective motions.   
Furthermore, this approach enables us to microscopically derive 
the collective coordinates and momenta on the basis of 
the time-dependent variational principle.   

We shall start from small-amplitude vibrations about the spherical 
equilibrium shape and then go to large-amplitude regime, 
where we need to consider full 5D quadrupole collective dynamics 
including 3D rotations restoring the broken symmetries 
as well as axially symmetric and asymmetric shape fluctuations.

\subsection{Collective motion as moving self-consistent mean field}

Let us consider even-even nuclei 
whose ground states consist of correlated nucleon pairs 
occupying time-reversal conjugate single-particle states.  
The Hartree-Fock-Bogoliubov (HFB) method is a generalized mean-field theory 
treating the formation of the HF mean field 
and superfluidity (nucleon pair condensate) in a self-consistent manner\cite{rin80,bla86,row10,ben03}, 
and yields the concept of quasiparticles as 
single-particle excitation modes in the presence of the pair condensate. 
Bohr and Mottelson opened the way to  
a unified understanding of single-particle and collective modes of motion 
of nuclei by introducing the concept of moving self-consistent mean field 
\cite{boh75,boh76,mot76}.    
The time-dependent extension of the HFB mean field, 
called the time-dependent HFB (TDHFB) theory,  
is suitable to formulate their ideas \cite{bel65,rin80,bla86,row10}. 

It is well known that the time evolution of the TDHF state vectors can be written    
as time-dependent unitary transformations (see {\it e.g.}, \cite{mar80,sak83}).   
It is called the generalized Thouless theorem. 
Adapting this theorem for nuclei with superfluidity,   
the TDHFB state vector $\ket{\phi(t)}$ 
may be written as \cite{mat86}: 
\begin{equation}
\ket{\phi(t)} = e^{i{\hat G}(t)}\ket{\phi(t=0)} = e^{i{\hat G}(t)}\ket{\phi_0}, 
\label{eq:TDHFBstatevector}
\end{equation}
\begin{equation}
i{\hat G}(t)= \sum_{(kl)} \left\{ g_{kl}(t) a_k^{\dag}a_l^\dag - g_{kl}^*(t) a_la_k \right\}, 
\label{eq:G_aa}
\end{equation}
where the HFB ground state $\ket{\phi_0}$ 
is a vacuum for quasiparticles $(a_k^{\dag}, a_l)$ , 
\begin{equation}
a_k|\phi_0\rangle = 0, 
\end{equation}
with the suffix $k$ distinguishing different quasiparticle states. 
(See Appendix \ref{sec:UnitaryTransformation} for more details.) 
The functions $g_{kl}(t)$ in the one-body operator ${\hat G}(t)$ 
is determined by the time-dependent variational principle
\begin{equation}
\delta \bra{\phi(t)}
\left( i\frac{\partial}{\partial t} - H \right) \ket{\phi(t)}=0.
\label{eq:TDVP}
\end{equation} 
The TDHFB states can be regarded as generalized coherent states,   
which are a kind of wave packets and cover the whole Hilbert space 
of a given Fermion many-body system \cite{suz83,yam87}. 
We call this space `the TDHFB phase space.' 
It may also be called `the TDHFB symplectic manifold' \cite{yam87,kur84}. 
This semiclassical concept is quite important because 
it provides a clear physical picture of collective dynamics. 
We shall see below that the unitary representation (\ref{eq:TDHFBstatevector}) 
is very convenient to develop a microscopic theory of nuclear collective motions.  

\subsection{Small-amplitude approximation (QRPA)} 

For small-amplitude vibrations around an HFB equilibrium point, 
one can make the linear approximation to the TDHFB equations 
and obtain the quasiparticle random phase approximation (QRPA).  
This is a starting point of microscopic theory of collective motion
\cite{bar60,mar60}. 
Expanding Eq.~(\ref{eq:TDHFBstatevector}) in a power series of ${\hat G}(t)$ 
and taking only the linear order, we obtain
\begin{equation}
\delta\langle\phi_0 \vert
[ H, i{\hat G} ] + \frac{\partial {\hat G}}{\partial t} 
\vert \phi_0 \rangle=0. 
\label{eq:eqG}
\end{equation} 
In place of the functions $g_{kl}(t)$ and $g_{kl}^*(t)$ in Eq.~(\ref{eq:G_aa}), 
let us introduce normal coordinates $q(t)=\{q^1(t),q^2(t),\cdots,q^f(t)\}$ and 
conjugate momenta $p(t)=\{p_1(t),p_2(t),\cdots,p_f(t)\}$, 
and represent ${\hat G}(t)$ in terms of the infinitesimal generators (${\hat Q}^i, {\hat P}_i$) 
of  $(p_i(t), q^i(t))$ as 
\begin{equation}
{\hat G}(t) = \sum_{i=1}^f \left( p_i(t) {\hat Q}^i - q^i(t) {\hat P}_i \right). 
\label{eq:G_QP}
\end{equation}
Here it is important to distinguish the {\it classical} dynamical variables $(q_i(t), p^i(t))$ 
from the {\it quantum} infinitesimal generators $({\hat Q}^i, {\hat P}_i)$.   
This representation is equivalent to Eq.~(\ref{eq:G_aa})  
if the number of normal coordinates, $f$,  is equal to 
the number of independent two-quasiparticle configurations $(kl)$. 
In reality, we shall be interested in only a few collective modes among the $f$ normal modes. 
For small-amplitude vibrations under consideration, the harmonic approximation 
holds; that is, time-dependence of $q^i(t)$ and $p_i(t)$ is given by
\begin{equation}
\dot{p}_i(t) = B_i \ddot{q}^i(t) = -C_i q^i(t), 
\label{eq:harmonic}
\end{equation}
where the $\dot{p}_i(t)$, $C_i$ and $B_i$ denote time-derivative of $p_i(t)$,   
the stiffness (restoring force) parameter, 
and the inertial mass for the normal mode 
(specified by the suffix $i$), respectively. 
Inserting (\ref{eq:G_QP}) into (\ref{eq:eqG}) and using (\ref{eq:harmonic}), 
we obtain the QRPA equation 
\begin{eqnarray}
~[~{\hat H}, ~{\hat Q}^i~ ] &=& -i B^i {\hat P}_i, ~~
\label{eq:QRPA_QP1} \\
~[~{\hat H}, ~{\hat P}_i~ ] &=&  i C_i {\hat Q}^i.  
\label{eq:QRPA_QP2}
\end{eqnarray}
where $B^i$ denotes the reciprocal of $B_i$, i.e., $B^i=1/B_i$. 
These equations determine the microscopic structure of ${\hat Q}^i$ and ${\hat P}_i$ 
as coherent superpositions of many two-quasiparticle excitations: 
Expressing them as sums over independent two-quasiparticles states $(kl)$,
\begin{eqnarray}
{\hat Q}^i &=& \sum_{(kl)} q_{kl}^i (a_k^{\dag}a_l^\dag + a_la_k) ,\\
{\hat P}_i  &=& i\sum_{(kl)} p_{kl}^i(a_k^{\dag}a_l^\dag - a_la_k) ,
\end{eqnarray}
and inserting these into (\ref{eq:QRPA_QP1}) and (\ref{eq:QRPA_QP2}), 
we obtain linear eigenvalue equations determining the frequency squared, 
$\omega_i^2=B^iC_i$, and the amplitudes $(q_{kl}^i, p_{kl}^i)$.  
Actually, we have to choose appropriate solutions among large number of solutions 
(the number of independent two-quasiparticle configurations). 
It is not difficult to identify them, however, because the solutions 
corresponding to low-frequency quadrupole vibrations 
appear much lower than twice the pairing gap, 2$\Delta$,  
(or the lowest two-quasiparticle excitation energy) and they are formed by coherent 
superpositions of many two-quasiparticle excitations. 
Because the time-evolution of the TDHFB state $\ket{\phi(t)}$ is determined by 
the normal coordinates and momenta ($q^i(t), p_i(t)$), we can write it as $\ket{\phi(q,p)}$.  
Using Eqs.~(\ref{eq:QRPA_QP1}), (\ref{eq:QRPA_QP2}), and (\ref{eq:QRPA_PQ}) below, 
we can easily calculate the expectation value of 
the microscopic Hamiltonian with respect to $\ket{\phi(q,p)}$: 
\begin{eqnarray}
\bra{\phi(q,p)} \hat{H} \ket{\phi(q,p)} 
= \bra{\phi_0} \hat{H} \ket{\phi_0} + \frac{1}{2}\sum_{i=1}^f \left (B^i p_i^2 +C_i {q^i}^2 \right ). 
\end{eqnarray} 
The increase of the total energy due to the vibrational motion,  
\begin{equation}
\Hc(q,p) \equiv \bra{\phi(q,p)} \hat{H} \ket{\phi(q,p)} -  \bra{\phi_0} \hat{H} \ket{\phi_0},  
\label{eq:HclassicalQRPA}
\end{equation}
may be identified as the classical vibrational Hamiltonian. 
Below we shall not care the ground-state energy (the second term in the r.h.s.), 
because it does not affect the equations of motion for ($q^i(t), p_i(t)$). 

For vibrational modes whose frequencies, $\omega_i = \sqrt{B^iC_i}$,  
are positive,  we can define the creation and annihilation operators  
$(\Gamma_i^\dag,\Gamma_i)$ of the excitation mode as 
\begin{equation}
\Gamma_i^\dag=\frac{1}{\sqrt{2}} 
\left( \sqrt{\frac{\omega_i}{B^i}} {\hat Q}^i - i \sqrt{\frac{B^i}{\omega_i}} {\hat P}_i \right) 
\end{equation}
and their Hermitian conjugates $\Gamma_i$. 
As is well known, they are written in terms of the quasiparticle operators as   
\begin{equation}
\Gamma_i^\dag  = \sum_{(kl)} (x_{kl}^i a_k^{\dag}a_l^\dag - y_{kl}^i a_la_k) ,
\end{equation}
and obey the QRPA equation of motion, 
\begin{equation}
~[~{\hat H}, ~\Gamma_i^\dag~ ] =  \omega_i \Gamma_i^\dag. 
\end{equation} 

It is worthy of notice that the $({\hat Q}^i, {\hat P}_i)$ representation possesses  
a wider applicability than the $(\Gamma_i^\dag,\Gamma_i)$ representation. 
First, for the ANG modes  
with  $\omega_i =0$, the former is valid while the latter is undefined. 
Note that their inertial masses, $B_i$ (inverse of $B^i$), are positive, whereas 
their frequency $\omega_i$ become to zero because 
the restoring-force parameters $C^i$ vanish. 
Second, the $({\hat Q}^i, {\hat P}_i)$ representation is valid 
also for unstable HFB equilibria 
where $C_i$ is negative and $\omega_i$  is imaginary. 
Obviously, we cannot define the creation and annihilation operators 
$(\Gamma_i^\dag,\Gamma_i)$ for imaginary $\omega_i$. 
We shall see that this is one of the key points when we try to extend 
the QRPA approach to non-equilibrium points far from the 
HFB local minima.  
\\

\noindent
{\bf Merits of the QRPA}\\

One of the beauties of the QRPA is that
it is able to determine the microscopic structures of 
collective coordinates and momenta in terms of a large number of 
microscopic (particle-hole, particle-particle, hole-hole) degrees of freedom. 
We can thus learn how collective vibrations are generated as 
coherent superpositions of many two-quasiparticle excitations. 
It is well known that two kinds of isoscalar quadrupole vibration 
appear exhibiting quite different characteristics;  
the low- (usually first excited $2^+$) and high-frequency (giant resonance) modes.  
Examining the microscopic structure of the low-frequency quadrupole vibrations, 
we see that the weights of two-quasiparticle excitations near the Fermi surface 
are much larger than those in the mass quadrupole operators (see, {\it e.g.}, \cite{nak99}). 
This example clearly shows the importance of describing collective modes
in a microscopic way. 

Another merit of the QRPA is that it yields the ANG modes as self-consistent solutions 
and determines their collective inertial masses.  
With use of the QRPA, we can restore the symmetries broken by the mean-field approximation.  
Furthermore, the QRPA fulfills the energy-weighted sum rules \cite{boh79}.   
\\


\noindent
{\bf Quantization condition}\\

In the QRPA, the following condition is customarily imposed 
to ortho-normalize the amplitudes $(q_{kl}^i, p_{kl}^i)$ or $(x_{kl}^i, y_{kl}^i)$.    
\begin{equation}
\langle\phi_0 \vert [{\hat Q}^i, {\hat P}_j ] \vert \phi_0\rangle = i \delta_{ij}, 
\label{eq:QRPA_PQ}
\end{equation}
or
\begin{equation}
\langle\phi_0 \vert [ \Gamma_i, \Gamma_j^\dag ] \vert \phi_0\rangle = \delta_{ij}. 
\label{eq:QRPA_Gamma}
\end{equation} 
We shall call this condition {\it canonical-variable condition}.  
It should be emphasized that, 
differently from the time-independent approaches, {\it e.g.} \cite{bar60}, 
these conditions cannot be derived within the standard framework of the TDHFB theory. 
For the derivation and justification of the canonical-variable conditions, 
we need to clarify the canonical structure of the TDHFB theory. 
We shall discuss on this point in Sec.~\ref{sec:SCC}.    

In this connection, we note that the inertial masses are not uniquely 
determined by (\ref{eq:QRPA_QP1}) and (\ref{eq:QRPA_QP2}), 
because the QRPA equations are invariant against 
the scale transformations ${\hat Q}^i \to  s^i{\hat Q}^i$ and  ${\hat P}_i  \to {\hat P}_i / s^i$ 
with arbitrary values of $s^i$. It is therefore possible to adopt the values of $s^i$ 
such that the collective inertial masses become unity. 
This arbitrariness is related to the freedom of scale transformations of 
the normal coordinates and momenta $(q^i, p_i)$. 

Because $(q^i, p_i)$ are canonical variables, we can make canonical quantization 
and obtain the quantum collective Hamiltonian, 
\begin{equation}
\hat{H}_{\rm QRPA} = \frac{1}{2}\sum_{i=1}^f \left (B^i (\hat{p}_i)^2 +C_i (\hat{q}^i)^2 \right ). 
\end{equation}
Here, the collective coordinates and momenta, $\hat{p}_i$ and $\hat{q}^i$, are quantum operators. 
It is important to note that the QRPA ground state after the quantization is different 
from the HFB ground state due to the quantum zero-point fluctuations. 
 
The necessity of canonical quantization in order to derive QRPA from the TDHFB theory,   
discussed above, is not necessarily emphasized  
in standard textbooks on theoretical nuclear physics. 
We shall see in Sec.~\ref{sec:SCC}, however, that the recognition of this point is essential to 
extend the basic ideas of the QRPA for small amplitude vibrations to LACM.    
\\

\noindent
{\bf Effective interactions for QRPA calculations}\\

In the investigation of low-energy excitation spectra, 
the pairing-plus-quadrupole (P+Q) model \cite{bes69,kis63,bar65}
and its extension \cite{sak89} have been playing the major roles. 
This phenomenological effective interaction 
represents the competition between the pairing correlations 
favoring the spherical symmetry 
and the quadrupole (particle-hole) correlations 
leading to the quadrupole deformation of the mean field \cite{boh75, mot98}. 

In recent years, QRPA calculations using density-dependent 
effective interactions \cite{vau72,vau73,dec80,neg82,sto07} have become possible. 
Density-dependent contact interactions 
such as the Skyrme interactions \cite{vau72,vau73,sto07}  
may be founded on the density functional theory (DFT) \cite{ben03}.  
From this point of view, the Skyrme interactions may be better called 
the Skyrme {\it energy density functionals} (EDFs).
Accordingly, the self-consistent calculations 
that use the same density-dependent contact interactions 
in solving the HFB equations for the ground state 
and the QRPA for excited states may be regarded as 
small-amplitude approximations of the time-dependent DFT (TDDFT) 
\cite {nak12}.   
A number of good textbooks on DFT and TDDFT 
are available, {\it e.g.}, \cite{fio03, mar06}. 
Note, however, there are conceptual differences between 
those for condensed-matter and those for nuclei, 
since the nucleus is a self-bound system without an external potential 
\cite{nak12}. 

For spherical mean fields, the QRPA matrix is blockdiagonal with respect to 
the angular momentum ($J$) and the parity ($\pi$) of two-quasiparticle configurations. 
Usually, the $J^{\pi}=2^+$ solution with lowest positive $\omega_i$ corresponds to 
the first excited quadrupole vibrational state.  
In this case, many calculations were performed \cite{ben03,vre05}. 
For axially symmetric deformed mean fields, 
the QRPA matrix is block-diagonal with respect to 
the $K$ quantum number (projection of angular momentum on the symmetry axis) 
and the parity of two-quasiparticle configurations. 
The $K^{\pi}=2^+ ~(0^+)$ solution with lowest positive $\omega_i$ may 
correspond to the first excited $\gamma~(\beta)$ vibrational state. 
It is well known, however, that the lowest $K^{\pi}=0^+$ solution contains 
an appreciable mixture of the pairing vibrational modes of protons and/or neutrons 
(sensitively depending on the deformed shell structure around the Fermi surface) \cite{boh75}. 
Moreover, as we shall discuss in Sec.~\ref{sec:OpenProblems}, 
recent experiments reveal mysterious characters of the lowest $K^{\pi}=0^+$ excitations. 
Although the dimension of the QRPA matrix is much larger than that in the spherical case,  
large scale QRPA calculations with modern EDFs
have been carried out also for deformed nuclei in recent years  
\cite{ter08,yos08,per08,art09,los10,ter10,ter11,yos11a,yos13}. 
In this way, it becomes one of the modern subjects in nuclear structure physics 
to carry out fully self-consistent QRPA calculations on the basis of DFT  
for superfluid (spherical and deformed) nuclei 
and treat low- and high-frequency vibrations (giant resonances)  
as well as the ground states in a unified way 
for all nuclei from the proton-drip line to the neutron-drip line. 

For triaxial mean fields breaking the axial symmetry, 
the dimension of the QRPA matrix further increases and 
it becomes computationally too heavy to diagonalize it at the present time.   
To overcome this problem, a new method of solving the QRPA  equations 
without recourse to diagonalization of the QRPA matrix 
has been developed in recent years
\cite{nak07,avo11,avo13}. 
It is called the {\it finite amplitude method},  and applied mainly to calculate 
strength functions for giant resonances \cite{sto11,ina13,mus14,lia13,nik13}. 
We shall suggest in Sec.~\ref{sec:microBM} that this method may be useful also for solving 
the {\it local} QRPA equations.  
\\

\noindent
{\bf Relations to spherical shell-model calculations}\\

The lowest $2^+$ vibrational states are obtained in the spherical shell model 
as coherent superpositions of many configurations. The {\it coherence} is indirectly confirmed 
by, {\it e.g.} the enhancements of the electric-quadrupole (E2) transition probabilities
$B(E2; 2_1^+ \to 0_1^+)$. 
In the QRPA, we can directly see the coherence in the QRPA amplitudes,
$x_{kl}^i$ and  $y_{kl}^i$ (or $q_{kl}^i$ and $p_{kl}^i$). 
In the time-dependent mean-field picture, {\it this coherence represents 
the correlations generating the self-consistent deformed mean field.}  
In this way, the TDHFB theory provides a transparent physical interpretation 
on the microscopic mechanism of emergence of nuclear collective motions.    


\subsection{Beyond the QRPA} 
\vspace{5mm}
\noindent
{\bf Boson expansion method}\\

Boson expansion method is well known as a useful microscopic method 
of describing anharmonic (non-linear) vibrations going 
beyond the harmonic approximation of the QRPA.   
In this approach, we first construct a collective subspace 
spanned by many-phonon states of vibrational quanta (determined by the QRPA)  
in the huge-dimensional shell-model space, and then map these many-phonon states 
one-to-one to many-boson states in an ideal boson space. 
Anharmonic effects neglected in the QRPA are treated as higher-order terms 
in the power-series expansion with respect to the boson creation and annihilation operators. 
Starting from the QRPA about a spherical shape, one can thus derive 
the 5D quadrupole collective Hamiltonian in a fully quantum mechanical manner.  
The boson expansion method has been successfully applied 
to low-energy quadrupole excitation spectra 
in a wide range of nuclei including those lying in regions of quantum phase transitions 
from spherical to deformed 
\cite{kle91,sak88}.  
\\

\noindent
{\bf Non-perturbative approaches to LACM}\\

The boson expansion about a single HFB local minimum is not suitable for treating a situation 
where a few local minima in the potential-energy surface $V(\bg)$ 
compete in energy.  In such situations the collective wave functions are not necessarily 
localized around a single minimum but tunnel through the potential barrier.    
We frequently encounter such situations, called `shape coexistence/mixing phenomena'  
in low-energy excitation spectra.  
The need to develop non-perturbative approaches capable of treating 
quantum many-body barrier penetrations is high also 
for treating large-amplitude collective motions in low-energy regions, 
such as spontaneous fissions and sub-barrier fusion reactions. 
It has been one of the longstanding fundamental subjects in nuclear structure physics 
to construct a microscopic theory of LACM  
by extending the QRPA concepts to arbitrary points in the $V(\bg)$ plane 
far from the HFB minima \cite{nak12,mat10,mat13}. 

State vectors of time-dependent mean field are kinds of generalized coherent states, 
and we can rigorously formulate the TDHFB as 
a theory of classical Hamiltonian dynamical system of large dimension 
\cite{yam87,kur01}.  
Because time-evolution of the mean field is determined by the classical Hamilton equations, 
we cannot describe, within the framework of the TDHFB, quantum spectra of low-lying states 
and macroscopic quantum tunneling phenomena 
such as spontaneous fissions and sub-barrier fusions. 
To describe these genuine quantum phenomena, 
we need to introduce a few collective variables determining the time-evolution of the 
mean field and quantize them. 
Succeeding and developing the ideas in microscopic theories of LACM acquired 
during 1970's-1990's, we have developed a new method, called the ASCC method 
\cite{mat00}, 
and shown its usefulness by applying it to shape coexistence/mixing phenomena 
\cite{hin08,hin09}.     
\\

\noindent
{\bf Introduction to the ASCC method}\\

Here we very briefly describe the basic ideas of the ASCC method \cite{mat00}. 
It will be presented in Sec.~\ref{sec:SCC} in a more systematic way. 
In this approach, assuming that the time evolution of the TDHFB state is determined by 
a few collective coordinates $q=(q^1,q^2,\cdots,q^f)$ and 
collective momenta $p=(p_1,p_2,\cdots,p_f)$, 
we write the TDHFB state as $\ket{\phi(t)}=\ket{\phi(q(t),p(t))}$. 
The TDHFB states $\ket{\phi(q,p)}$ constitute the $2f$-dimensional submanifold 
in the TDHFB phase space, which is called {\it collective submanifold}.
In the ASCC method, we further assume that $\ket{\phi(q,p)}$ 
can be written in a form
\begin{equation}
 \ket{\phi(q,p)}  =  \exp\left\{ i \sum_{i=1} ^f p_i \Qhat^i(q) \right\}
 \ket{\phi(q)} ,
\label{eq:ASCCstate1}
\end{equation}
where $\Qhat^i(q)$ are one-body operators corresponding to
infinitesimal generators of $p_i$ locally defined at the state $\ket{\phi(q)}$ 
which represents a TDHFB state $\ket{\phi(q,p)}$ at $p\rightarrow 0$. 
This state $\ket{\phi(q)}$ is called a {\it moving-frame HFB state.}  
Inserting (\ref{eq:ASCCstate1}) into the time-dependent variational principle, 
Eq.~(\ref{eq:TDVP}),  and considering that the time dependence is determined by 
the collective coordinates and momenta $(q,p)$, we obtain 
\begin{equation}
\delta\bra{\phi(q,p)} \left\{ 
i \sum_{i=1}^f \left( \dot{q^i}\frac{\partial}{\partial q^i} + \dot{p_i}\frac{\partial}{\partial p_i} \right)
 - \hat{H} \right\} \ket{\phi(q,p)}=0.  
\label{invariance principle1}
\end{equation}
We shall give a rigorous formulation to determine the microscopic structures 
of the infinitesimal generator $\Qhat^i(q)$ of $p_i$ on the basis of the time-dependent 
variational principle (\ref{invariance principle1}). 
We shall also introduce infinitesimal generators $\Phat_i(q)$ of $q^i$ 
and determine their microscopic structures.  
Furthermore, we shall formulate the theory such that  the 
collective variables $(q,p)$ can be treated as canonical variables.  

Quite recently, we have proposed a practical approximation scheme to the ASCC method. 
It is called the {\it local} QRPA (LQRPA) method \cite{hin10,sat11,hin11a,yos11b,sat12,mat14}. 
Here, the adjective ``{\it local}" means that it is locally defined around a point 
in the $(\bg)$ deformation space. More rigorously speaking, it is defined around a point   
on the collective submanifold embedded in the TDHFB phase space,  
and this point is mapped onto the $(\bg)$ space.  
The infinitesimal generators appearing in this method are nonlocal in the coordinate space.  
It may be regarded as an extension of the ordinary QRPA 
to non-equilibrium states, where the moving frame HFB states $\ket{\phi(q)}$
play a role analogous to the static HFB ground state $\ket{\phi_0}$. 
Because of this analogy it may be easy to understand the LQRPA method. 
In the next section, we show how this method is used  
for a microscopic derivation of the Bohr-Mottelson collective Hamiltonian. 
Fundamentals and validity 
of the LQRPA method will be discussed later in Sec.~\ref{sec:SCC}.  


\section{Microscopic derivation of the Bohr-Mottelson collective Hamiltonian}
\label{sec:microBM}

In this section, we derive the quadrupole collective Hamiltonian making use of the LQRPA method. 
We also discuss fundamental problems related to the microscopic derivation of the collective Hamiltonian. 

\subsection{Procedure for the microscopic derivation} 

Instead of treating the 5D collective coordinates simultaneously, 
we first calculate the collective inertial masses for two-dimensional (2D) vibrational motions 
corresponding to the $(\bg)$ deformation degree of freedom, and  
subsequently calculate the moments of inertia for 3D rotational motions 
at each point of $(\bg)$. 
We then derive the collective Hamiltonian for the 5D quadrupole collective dynamics 
and quantize it.  
\\

\noindent 
{\bf Microscopic calculation of the vibrational inertial masses}\\

We first derive two canonical coordinates $(q^1, q^2)$ that  
correspond to the $(\bg)$ vibrational degrees of freedom in the Bohr-Mottelson collective model. 
In this section we use the notion $q$ to represent $(q^1, q^2)$ 
and write the moving-frame HFB state as $\ket{\phi(q)}$.  

First, we solve the moving-frame HFB equations, 
\begin{eqnarray}
 \delta \bra{\phi(q)} \Hhat_{\rm M}(q) \ket{\phi(q)} = 0,  
\label{eq:CHFB} 
\end{eqnarray}
\begin{eqnarray}
 \Hhat_{\rm M}(q) = \Hhat -
 \sum_{\tau}\lambda^{(\tau)}(q)\Nt^{(\tau)} 
 - \sum_{m = 0, 2} \mu_{m}(q) \Dhatp_{2m}, 
\label{eq:H_CHFB} 
\end{eqnarray}
where $\Dhatp_{2m}$ and $\Nt^{(\tau)} \equiv \hat{N} ^{(\tau)}-N_0^{(\tau)}$ 
are the mass quadrupole operators and the number operators 
(measured from the expectation values at the ground state) 
for protons and neutrons ($\tau=$p,n), respectively.   
The quadrupole-deformation variables $(\bg)$ are defined through 
the expectation values of $\Dhatp_{2m}$ with respect to $\ket{\phi(q)}$:   
\begin{eqnarray}
 \beta\cos\gamma &=  \eta D^{(+)}_{20} (q)
= \eta \bra{\phi(q)} \Dhatp_{20} \ket{\phi(q)},  \label{eq:definition1}  \\
 \frac{1}{\sqrt{2}} \beta\sin\gamma &= \eta D^{(+)}_{22} (q)  
= \eta \bra{\phi(q)} \Dhatp_{22} \ket{\phi(q)}, 
\label{eq:definition2} 
\end{eqnarray}
where $\eta$ is a scaling factor with the dimension of $L^{-2}$.

\begin{figure}[tbp]
\begin{center}
\includegraphics[width=0.8\textwidth]{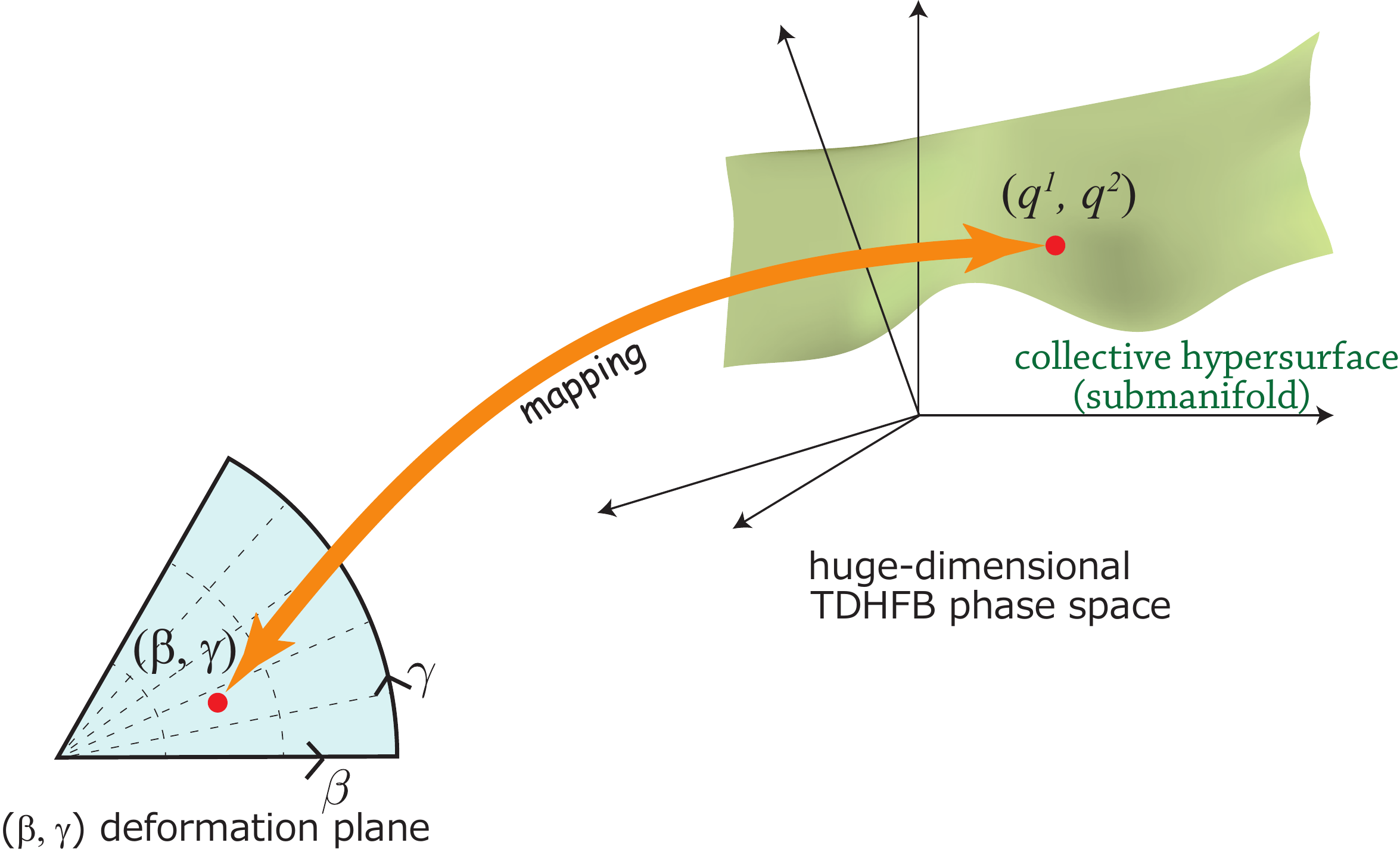}
\caption{
Illustration of the mapping of $(q^1, q^2)$ defined on the collective submanifold 
onto the $(\bg)$ deformation plane of the Bohr-Mottelson collective model. 
The collective submanifold is illustrated as a hypersurface 
in the huge-dimensional TDHFB phase space. 
}
\label{fig:mapping}
\end{center}
\end{figure}

Through the above definitions of $(\bg)$ we can make a one-to-one correspondence 
between $(q^1, q^2)$ and $(\bg)$. 
As illustrated in Fig.~\ref{fig:mapping}, this correspondence may be viewed as 
a mapping of the collective coordinates $(q^1, q^2)$ 
onto the $(\bg)$ plane of the Bohr-Mottelson collective model. 
For our purpose, it is sufficient to assume that this correspondence is one-to-one 
in the neighborhood of an arbitrary point $(q^1, q^2)$, 
because the collective inertial masses represent the inertia of the LACM 
for infinitesimal variation in time of the collective coordinates.    
Thus, the moving-frame HFB state $\ket{\phi(q)}$ may also be written as $\ket{\phi(\bg)}$. 
The solutions of Eq. (\ref{eq:CHFB}) for every point on the $(q^1,q^2)$ plane 
provide the moving-frame HFB states $\ket{\phi(\bg)}$ off 
the HFB ground state $\ket{\phi(\beta_0, \gamma_0)}$ 
at the local minimum $(\beta_0, \gamma_0)$ on the potential energy surface $V(\bg)$.   
\\

Next, we consider the TDHFB states of the form, Eq. (\ref{eq:ASCCstate1}), with $f=2$. 
Assuming that the collective motion is slow, 
we expand it in powers of $p$ and consider up to the second-order in $p$. 
Then, under certain approximations explained in Sec.~\ref{sec:SCC}, we obtain the 
following set of equations of motion for $\Qhat^i(q)$ and $\Phat_i(q)$. 
\begin{eqnarray}
 \delta \bra{\phi(q)} [ \Hhat_{\rm M}(q), \Qhat^i(q) ] 
 + i B^i(q) \Phat_i(q) \ket{\phi(q)} &=& 0,
\label{eq:LQRPA1} \\
 \delta \bra{\phi(q)} [ \Hhat_{\rm M}(q), \Phat_i(q)]
 - i C_i(q) \Qhat^i(q) \ket{\phi(q)} &=& 0, 
\label{eq:LQRPA2}
\end{eqnarray}
with the `weakly' canonical commutation relations, 
\begin{equation}
\bra{\phi(q)} \left[ \Qhat^i(q), \Phat_j(q) \right] \ket{\phi(q)} = i \delta_{ij}, 
\label{eq: QPcommutator1} 
\end{equation}
meaning that the canonical commutation relations hold only 
for their expectation values with respect to $\ket{\phi(q)}$.  
The equations (\ref{eq:LQRPA1}) and (\ref{eq:LQRPA2}) are called the LQRPA equations and 
may be regarded as generalizations of the QRPA eqs.    
(\ref{eq:QRPA_QP1}) and (\ref{eq:QRPA_QP2}) about the HFB ground state 
to those for a moving-frame HFB state $\ket{\phi(q)}$.  

Analogously to the $(\Qhat^i, \Phat_i)$ operators in the ordinary QRPA, 
the one-body operators $\Qhat^i(q)$ and $\Phat_i(q)$, called infinitesimal generators 
of collective motion, can be written as linear combinations of bilinear products 
of the local quasiparticle operators $(a_k^\dag, a_l)$ 
that are defined with respect to 
the moving-frame HFB state $\ket{\phi(q)}$ by  $a_k\ket{\phi(q)}=0$: 
\begin{eqnarray}
{\hat Q}^i(q) &=& \sum_{(kl)} q_{kl}^i(q) (a_k^{\dag}a_l^\dag + a_la_k) ,
\label{eq: LQRPA_Q}
\\
{\hat P}_i(q)  &=& i\sum_{(kl)} p_{kl}^i(q)(a_k^{\dag}a_l^\dag - a_la_k). 
\label{eq: LQRPA_P}
\end{eqnarray}
Because the collective coordinates $(q^1,q^2)$ 
corresponding to $(\bg)$ and their conjugate momenta $(p_1,p_2)$
are canonical variables, it is possible to make a scale transformation 
such that the collective masses relating $(p_1,p_2)$ 
to the time derivatives $(\dot{q}^1, {\dot{q}^2})$ of $(q^1,q^2)$ 
become unity.   
Thus, we can write the kinetic energy of vibrational motions as   
\begin{eqnarray}
 T_{\rm vib} = \frac{1}{2} \sum_{i=1,2} (p_i)^2 
 = \frac{1}{2} \sum_{i=1,2} (\dot{q}^i)^2 
\label{eq:Tvib}
\end{eqnarray}  
without loss of generality. 
\\

\noindent
{\bf Microscopic calculation of the rotational moments of inertia}\\

In a manner similar to the calculation of the vibrational inertial masses described above, 
we calculate, at every point on the $(\bg)$ plane, 
the rotational moments of inertia $\cJ_k$  for 3D rotational motions $(k=1,2,3)$. 
To treat the 3D rotational motions, we write rotating TDHFB states in the following form:
\begin{equation}
\ket{\phi(q,\varphi, \dot{\varphi})}  =  
       \exp \left[ i \sum_{k=1} ^3 \left\{  \cJ_k(q) \dot{\varphi}_k \hat{\Psi}^k(q) 
       - \varphi_k\hat{I}_k \right\} \right] \ket{\phi(q)} ,
\label{eq:RotatingTDHFBstate}
\end{equation} 
Here $\hat{\Psi}^k(q)$ are local angle operators conjugate to   
the angular-momentum operators $\hat{I}_k$ 
and satisfy the `weak' canonical commutation relations, 
\begin{equation}
\bra{\phi(q)} \left[ \hat{\Psi}^k(q), \hat{I}_l \right] \ket{\phi(q)} = i \delta_{kl}. 
\label{eq: QPcommutator2} 
\end{equation}
The set ($\hat{\Psi}^k(q)$, $\hat{I}_k$) corresponds to 
the infinitesimal generators ($\Qhat^i(q)$, $\Phat_i(q)$) for vibrational motions 
considered above.  The variables $\varphi_k$ and $\dot{\varphi}_k$ denote 
the rotational angles and their time derivatives.  
The set ($\varphi_k$, $\cJ_k(q) \dot{\varphi}_k$) corresponds to the set 
of collective coordinates and momenta $(q^i, p_i)$. 
The inverse of $B^i(q)$ corresponds to $\cJ_k(q)$. 
Needless to say, in contrast to $\Phat_i(q)$ for vibrational motions, 
the infinitesimal generators for rotational motions are 
the angular-momentum operators $\hat{I}_k$ {\it independent of} $q$, and 
the restoring-force parameters $C_k(q)$ are zero for rotational motions. 

Inserting (\ref{eq:RotatingTDHFBstate}) for $\ket{\phi(t)}$ 
in the time-dependent variational principle (\ref{eq:TDVP})
and considering only the linear-order terms with respect to 
$\hat{\Psi}^k(q)$ and $\hat{I}_k$, we obtain  
the LQRPA equations for 3D rotational motions:  
\begin{eqnarray}
 &\delta\bra{\phi(q)} [ \Hhat_{\rm M}(q), \hat{\Psi}^k(q) ] 
                               + i \frac{\hat{I}_k}{\cJ_k(q) } \ket{\phi(q)} = 0. 
\label{eq:LQRPA3} 
\end{eqnarray}
These equations are the same as the Thouless-Valatin equations \cite{tho62},  
except that we solve these equations 
not only at the equilibrium deformation $(\beta_0,\gamma_0)$ 
but also at every points on the $(\bg)$ plane off the equilibrium. 

Solving Eqs.~(\ref{eq:LQRPA3}) at every point on the $(q^1,q^2)$ plane 
and make a one-to-one mapping to the $(\bg)$ plane,  
we obtain the three moments of inertia  $\cJ_k(\bg)$,  
which determine the rotational masses $D_k(\bg)$  through Eq.~(\ref{eq:rotational-inertia}), 
and the rotational energy, 
\begin{equation}
T_{\rm rot}=\frac{1}{2}\sum_{k=1,2,3}\cJ_k(\bg)\dot{\varphi}_k^2, 
\end{equation} 
in the collective Hamiltonian (\ref{eq: BMcoll}).
If $D_k(\bg)$ are replaced with a constant, $D_k(\bg)=D$,  
then $\cJ_k(\bg)$ reduce to the moments of inertia for irrotational fluid.  
As mentioned in Sec.~2, this approximation may be valid only for harmonic vibrations  
about the spherical shape.  
\\

\noindent 
{\bf Derivation of the quadrupole collective Hamiltonian and its quantization}\\

Displacements of $(q^1,q^2)$ are related to variations of the expectation values 
$\Dp_{2m}$ of the mass quadrupole operators by
\begin{equation}
d\Dp_{2m} = \sum_{i=1,2} \frac{\del \Dp_{2m}}{\del q^i} dq^i,
  \quad\quad m=0,2 .
\end{equation}
This relation leads to the kinetic energy of vibrational motions given in terms of 
time derivatives of the quadrupole deformation,  
\begin{eqnarray}
 T_{\rm vib} = 
   \frac{1}{2} \sum_{m,m'=0,2} M_{mm'} \Ddotp_{2m}\Ddotp_{2m'}, 
\label{eq:TvibD}
\end{eqnarray}
where 
\begin{eqnarray}
 M_{mm'}(\bg) = \sum_{i=1,2} \frac{\del q^i}{\del
 \Dp_{2m}} \frac{\del q^i}{\del \Dp_{2m'}}. 
\label{eq:M_mm}
\end{eqnarray}
Taking time derivatives of Eqs.~(\ref{eq:definition1}) and (\ref{eq:definition2}), 
we can straightforwardly transform the expression (\ref{eq:TvibD}) to 
the form in terms of  ($\dot{\beta}$, $\dot{\gamma}$).  
The vibrational masses ($D_{\beta\beta}$, $D_{\beta\gamma}$,
$D_{\gamma\gamma}$) are then obtained from ($M_{00}$, $M_{02}$, $M_{22}$)  
through the following relations:  
\begin{align}
 D_{\beta\beta}=& \eta^{-2}
\left(
M_{00} \cos^2\gamma + \sqrt{2}M_{02}\sin\gamma\cos\gamma
 \right. \nonumber \\
&+ \left. \frac{1}{2}M_{22}\sin^2\gamma \right), \\ 
D_{\beta\gamma}=& 
 \beta \eta^{-2}
\left[
-M_{00}\sin\gamma\cos\gamma + \frac{1}{\sqrt{2}}M_{02}(\cos^2\gamma- \sin^2\gamma) 
 \right. \nonumber \\ 
&+ \left. \frac{1}{2} M_{22} \sin\gamma\cos\gamma \right], \\
D_{\gamma\gamma} =& \beta^2 \eta^{-2} 
\left( 
M_{00} \sin^2\gamma
- \sqrt{2} M_{02} \sin\gamma\cos\gamma 
 \right. \nonumber \\
&+  \left. \frac{1}{2} M_{22} \cos^2\gamma \right).
\end{align}

In this way, we can calculate, in a microscopic way, all the collective inertial masses 
appearing in the Bohr-Mottelson collective Hamiltonian (\ref{eq: BMcoll}). 
For quantization, we can apply the quantization scheme for the 5D curvilinear coordinates  
(so-called Pauli prescription, see Appendix A). 
After a somewhat lengthy but straightforward calculation,   
we obtain the quantized collective Hamiltonian (\ref{eq: quantized H_BMcoll}). 

\subsection{Discussions}
\label{sec:discussions}

Let us discuss some fundamental problems related to the microscopic derivation of 
the quadrupole collective Hamiltonian. 
\\ 
 
\noindent
{\bf Applicability of the Pauli prescription for quantization}\\

In the pioneering work of Baranger and Kumar toward microscopic derivation of 
the Bohr-Motttelson collective Hamiltonian, they wrote \cite{kum67} : 

``The next problem is that of quantizing Hamiltonian $H$. There is no unique way of 
doing this. Bohr uses the Pauli prescription, which is designed to give the right
answer when the variables can be transformed to Cartesian coordinates. But this is 
not the case here and therefore the Pauli prescription loses its only justification." 

In the 50 years since their work, we have now good prospects of justifying the use of 
the Pauli prescription. Because it is just the transformation of the Laplacian 
in Cartesian coordinates to that in the curvilinear coordinates, as Baranger and Kumar 
pointed out, the crucial question is whether or not we can derive the 5D collective coordinates 
which are Cartesian. As we have shown above, 
we have derived a local 5D canonical coordinate system on the collective submanifold 
embedded in the large-dimensional TDHFB phase space.  
(This concept will be further discussed in Sec.~\ref{sec:SCC}.)
In our view, to derive the kinetic energy term and the inertial masses, 
it is enough to define {\it a local coordinate system} at each point of the collective submanifold;  
that is,  it is unnecessary to define a {\it global} canonical coordinate system. 
It remains, however, as an interesting subject to develop a firm theoretical formulation 
to clarify the validity and limitation of the use of the Pauli prescription 
for quantization of collective coordinates.  
\\

\noindent
{\bf Treatment of 3D rotational motions}\\

It should be emphasized that we can define the local angle operators $\hat{\Psi}^k(q)$, 
although the global angle operators canonically conjugate to $\hat{I}_k$ do not exist.  
For the microscopic calculation of the moments of inertia $\cJ_k$, 
it is sufficient  to determine the microscopic structure of 
the local angle operators $\hat{\Psi}^k(q)$. 
This is because, similarly to the vibrational inertial masses, 
$\cJ_k(q)$ represents the inertia for an infinitesimal change of the rotational angles of 
the moving-frame HFB state $\ket{\phi(q)}$.   
It should be kept in mind that we use the expression (\ref{eq:RotatingTDHFBstate}) 
for rotating TDHFB states only for infinitesimal rotations, i.e., 
for very small rotational angles $\varphi_k$. 
For large $\varphi_k$, we have to consider higher-order effects 
associated with the non-Abelian nature of the angular momentum operators \cite{kan94}. 
Fortunately, it is unnecessary to consider such higher-order effects 
for our aim of evaluating the inertial masses for rotational motions.         
\\

\noindent
{\bf Effective interaction in the microscopic Hamiltonian}\\

The LQRPA method is quite general and it can be used for any microscopic 
Hamiltonian $\hat{H}$.  
Inserting Eqs. (\ref{eq: LQRPA_Q}) and (\ref{eq: LQRPA_P})  
into Eqs. (\ref{eq:LQRPA1}) and (\ref{eq:LQRPA2}), 
we obtain linear eigenvalue equations for the amplitudes $q_{kl}^i$ and $p_{kl}^i$. 
For effective interactions of separable type 
such as the P+Q force model, 
we can rewrite these equations into a form of dispersion equation 
determining the frequencies squared $\omega_i^2=B^iC_i$ and 
the amplitudes, $q_{kl}^i$ and $p_{kl}^i$ (see {\it e.g.}, \cite{sak97}).   
It is then easy to find the solutions satisfying the dispersion equation.  
For effective interactions of the Skyrme type or modern density functionals, 
we have to diagonalize the QRPA matrix of very large dimension.  
This is the case for deformed HFB states, especially for triaxial deformations, 
and the computation becomes heavy.  
Although a large-scale calculation is required,  
such an application of the LQRPA method with realistic interactions/functionals 
is a challenging future subject.
A step toward this goal has recently been carried out 
for axially symmetric cases \cite{yos11b}. 
To overcome this computational problem, 
the finite-amplitude method \cite{nak07,avo11,avo13} 
may be utilized. 
In particular, the recently developed technique \cite{hin13,hin15a,hin15b,hin15c}  
may be useful to find a few low-frequency solutions possessing strong collectivities.        
It is a great challenge to develop the LQRPA approach 
on the basis of the TDDFT and nuclear EDFs.    
\\

\noindent
{\bf Physical meaning of the collective inertial masses}\\

The pairing correlation plays a crucial role in determining the inertial masses of collective motion.
The reason may be understood microscopically as follows. 

The single-particle energies and wave functions are determined by the nuclear mean field.
The time-evolution of the mean field changes them and causes a number of single-particle level crossings.
The level crossing near the Fermi surface induces the change of the lowest-energy configuration.
Without the pairing, it is difficult for the system to rearrange 
to more energetically favorable configurations at the level crossing. 
In the presence of the pairing correlation, however, the nucleon pairs can make a hopping 
from up-sloping levels to down-sloping levels at the level crossing \cite{bar90}.
Such easiness/hardness of the configuration rearrangements at level crossings determines 
the adiabaticity/diabaticity of the collective motion.
The collective inertia represents a property of the system trying to keep a definite 
configuration during the collective motion.
Thus, the inertia becomes smaller for stronger pairing. 

In spherical mean fields, the pairing correlation acts for monopole nucleon pairs 
that couples to an angular momentum $J=0$.
In deformed mean fields, the nucleon pair becomes a superposition of 
multiple angular momenta $J$ because of the rotational symmetry breaking. 
In particular, the quadrupole $J=2$ pairing correlation plays an important role.  
The reason is understood as follows.  
When a mean field develops toward a larger prolate deformation, 
single-particle levels favoring the prolate deformation are pushed down, 
while those that favor the oblate deformation are pushed up. 
At the level crossing, the easiness/hardness of the rearrangement depends on 
the magnitude of the pairing matrix elements between the crossing single-particle levels.
The spacial overlaps between the single-particle wave functions of the
up-sloping and down-sloping levels are smaller than those at the spherical limit. 
Such reductions of the pairing matrix elements between the prolate-favoring and 
the oblate favoring levels are well described by taking into account 
the quadrupole pairing (in addition to the monopole pairing) \cite{bri05}.       
The Galilean invariance provides a link between the monopole and 
quadrupole pairing strengths \cite{sak90}. 
It is shown with the use of the ASCC and LQRPA methods~\cite{hin10, hin06} 
that the quadrupole pairing induces time-odd components 
(that change sign under time reversal) 
in the moving mean field and enhances the inertial masses. 
This indicates that the the collective dynamics associated with the pairing correlations 
is well described by these microscopic methods.    
More detailed investigation on the roles of the pairing in level crossing dynamics 
will prove fruitful for a deeper understanding of the microscopic mechanism 
determining the inertial masses.   
\\

\subsection
{Remarks on microscopic derivation of the particle-collective coupling Hamiltonian} 

In this review, we concentrate on the collective Hamiltonian $H_{\rm coll}$ in the 
unified model Hamiltonian (\ref{eq:H_unified}) of Bohr and Mottelson. 
Needless to say, it is a great challenge to develop a microscopic theory 
capable of treating the single-particle and collective motions in a unified manner. 
The particle-collective coupling Hamiltonian $H_{\rm coupl}$ in the unified model Hamiltonian 
may be derived by using the same concept of time-dependent self-consistent 
mean field which has been used in the microscopic derivation of 
the collective Hamiltonian $H_{\rm coll}$.  
As is well known, properties of single-particle motions are determined 
by the mean fields which are collectively generated by all nucleons constituting the nucleus. 
This implies that the dynamical time evolution of the mean field affects 
the single-particle motion and generates the particle-collective couplings.   

For small-amplitude vibrations about an equilibrium point of the HFB mean field, 
we can expand the single-particle Hamiltonian associated with mean field of the moving 
HFB state $\ket{\phi(\bg)}$ in terms of the vibrational amplitudes. 
We then obtain the particle-vibration coupling Hamiltonian in the linear order \cite{boh75, mot77}. 
To overcome the problem of over-completeness and non-orthogonality 
that arises from the use of the basis states 
consisting of both the single-particle modes (defined at the HFB minimum point)  
and the elementary modes of vibrations, 
the `Nuclear Field Theory (NFT)' has been developed since 1970's \cite{bor77}. 
The NFT has been used for microscopic analyses of anharmonicities of vibrational motions 
as well as the ``dressing" of single-particle motions due to the particle-vibration couplings. 
For these applications and recent achievements of the NFT,     
we refer the contribution by Broglia {\it et al.} to this Special Edition \cite{bro15}. 

A promising approach to derive the particle-vibration coupling Hamiltonian 
beyond the linear order is to derive the single-particle Hamiltonian 
in the moving self-consistent mean field and expand it in powers of collective variables. 
An interesting attempt in this direction was done by Yamada \cite{yam91} 
using 
the self-consistent collective coordinate (SCC) method 
with the $(\eta, \eta^*)$ expansion 
(described in Sec.~\ref{sec:SCCexpansion}).  
It is interesting to further develop this approach.  
Looking for future, it will certainly become an important fundamental subject 
in nuclear structure theory  
to derive the particle-vibration coupling Hamiltonian starting from the TDDFT.  

We should also remark the longstanding problem of deriving the particle-rotation 
coupling Hamiltonian starting from a microscopic many-body theory. 
In Ref.~\cite{shi01}, the single-particle motions in rapidly rotating mean field are 
described by means of the SCC method with a power-series expansion in the rotational frequency, 
and the alignments of single-quasiparticle and the rotational angular momenta are studied.  
Developing this line of approach, the SCC method may be used also 
for deriving the particle-rotation coupling Hamiltonian, 
but this subject remains for future.        
In our view, construction of a microscopic theory capable of treating the single-particle 
and collective motions in a unified manner, initiated by Bohr and Mottelson,  
still remains as the most fundamental and principal subject in nuclear structure dynamics.  
\\

\noindent
{\bf  Historical note}
\\

The construction of a self-consistent microscopic theory of collective motion 
capable of deriving the unified-model Hamiltonian of Bohr and Mottelson  
is a longstanding and difficult subject which 
always inspires the development of fundamental new concepts.    
Let us quote some remarks by Villars, which may be worthwhile to keep in mind: \\
 ``Although such a synthesis of the collective and the particle 
aspect of nuclear dynamics is rather easily achieved in words, 
by simply combining results borrowed from various models, 
a decent mathematical formulation of the same programme is 
far from easy.h in 1967 \cite{vil67}. \\
``It always appeared to this author that the proper formulation of 
a microscopic theory of nuclear collective motion is a strangely difficult 
subject.h~``Much is to be learned yet in the problem of formulating 
a consistent quantum theory of collective motion.h in 1982 \cite{vil82}. 

\newpage
\section{Illustrative examples}
\label{sec:Illustration} 
     
We here present some applications of the LQRPA method 
for deriving the 5D collective Hamiltonian. 
In the numerical examples below,  
the P+Q model Hamiltonian \cite{bar65} 
(including the quadrupole-pairing interaction) 
is employed in solving the LQRPA equations. 
The single-particle energies and the P+Q interaction strengths 
are determined such that the results of the Skyrme-HFB calculation  
for the ground states are best reproduced within the P+Q model 
(see Refs.~\cite{sat11, hin11a}  for details). 
More examples can be found 
for $^{68-72}$Se \cite{hin10,hin09}, 
 $^{72,74,76}$Kr \cite{sat11},  
 the $^{26}$Mg region \cite{hin11b}, 
 $^{30-34}$Mg \cite{hin11a}, 
  $^{58-68}$Cr \cite{yos11b}, 
 $^{58-66}$Cr \cite{sat12}, 
and 
 $^{128-132}$Xe, $^{130-134}$Ba \cite{hin12}.
\\

\noindent
{\it Oblate-prolate shape coexistence and fluctuations in $^{74}$Kr} 
\\

\begin{figure}[htbp]
\begin{center}
\includegraphics[width=\textwidth]{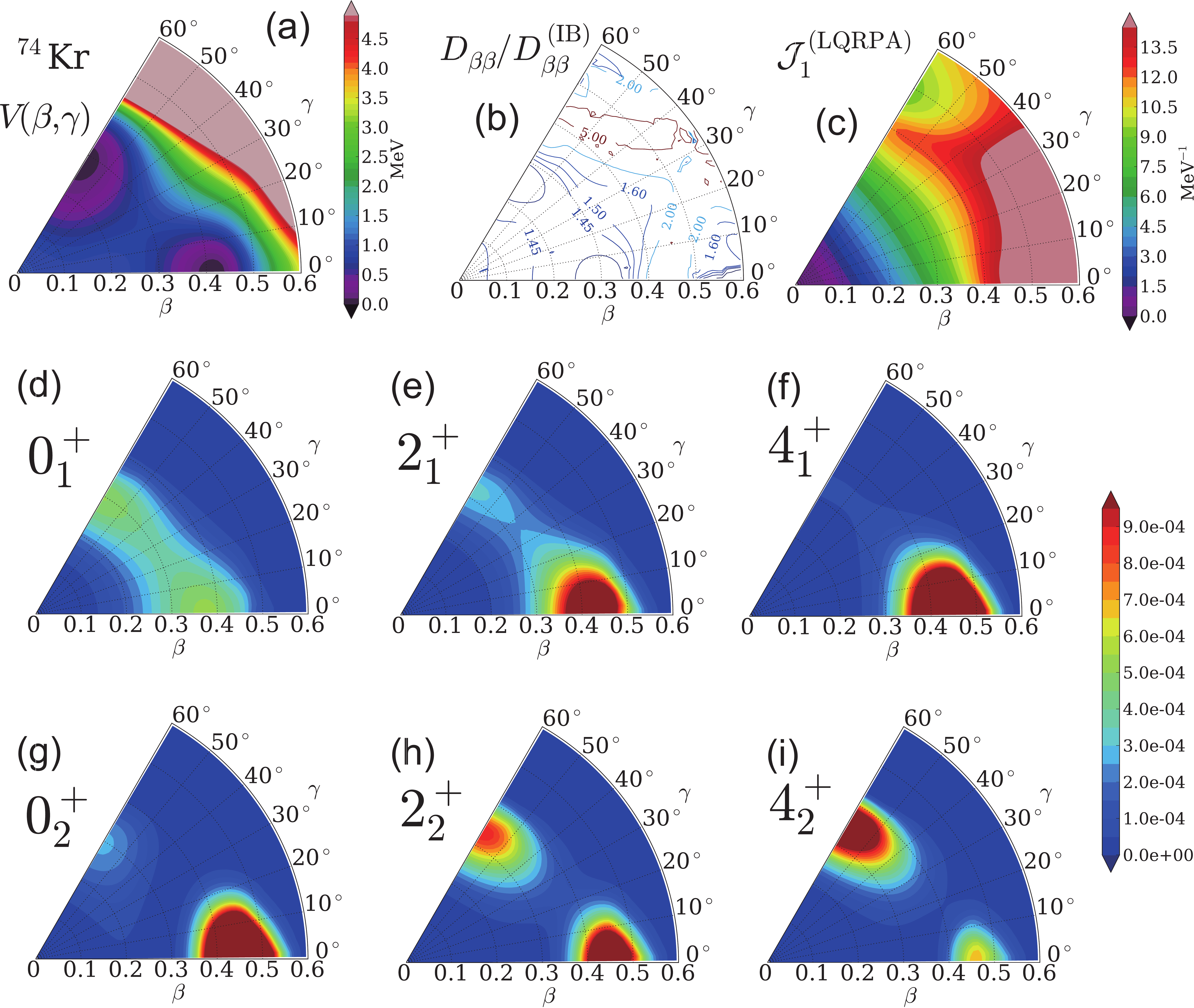}
\caption{ 
Application of the LQRPA method to the oblate-prolate shape coexistence/fluctuation 
phenomenon 
in $^{74}$Kr 
(from Ref.~\cite{sat11}). 
(a) Collective potential $V(\beta,\gamma)$, 
(b) Ratio of the collective inertial mass $D_{\beta\beta}(\bg)$ 
to the Inglis-Belyaev cranking mass. 
(c) The LQRPA moment of inertia $\mathcal{J}_1$ for rotation about the $x-$axis.
Vibrational wave functions squared, $\sum_K \beta^4|\Phi_{\alpha IK}(\beta,\gamma)|^2$, 
for (d) the $0_1^+$ state,  
(e) the $2_1^+$ state, 
(f) the $4_1^+$ state, 
(g) the $0_2^+$ state, 
(h) the $2_2^+$ state, 
and (i) the $4_2^+$ state. 
For the $\beta^4$ factor, see the text. 
}
\label{fig:74Kr_wave}
\end{center}
\end{figure}

The collective potential $V(\bg)$ depicted in Fig.~\ref{fig:74Kr_wave} exhibits two local minima. 
The prolate minimum is lower than the oblate minimum, and the spherical
shape is a local maximum.  
This figure also shows that the valley runs in the triaxially deformed region and the barrier 
connecting the oblate and prolate minima is low. 
Accordingly, one may expect large-amplitude quantum shape fluctuations 
to occur along the triaxial valley. 
In fact, the vibrational wave function of the ground  $0^+_1$ state has
bumps around the two potential minima, but
the wave function spreads over the entire $\gamma$ region along the potential valley. 
It is interesting to notice that, as the angular momentum increases,  
the localization of the vibrational wave functions in the $(\beta, \gamma)$ deformation plane 
develops; namely, the rotational effect plays an important role 
for the emergence of the shape-coexistence character.  
This development of localization results from the $\beta-\gamma$ dependence of the rotational moments
of inertia. One can clearly see the oblate-prolate asymmerty of the
moment of inertia $\mathcal{J}_1$ shown in Fig.~\ref{fig:74Kr_wave}(c). 
Due to this asymmetry, the localization on the prolate side develops in
the ground band. In the yrare band, although the vibrational wave
functions have a two-peak
structure, the localization on the oblate side
develops due to the orthogonality to the yrast states.

\begin{figure}[tbp]
\begin{center}
\includegraphics[width=0.9\textwidth]{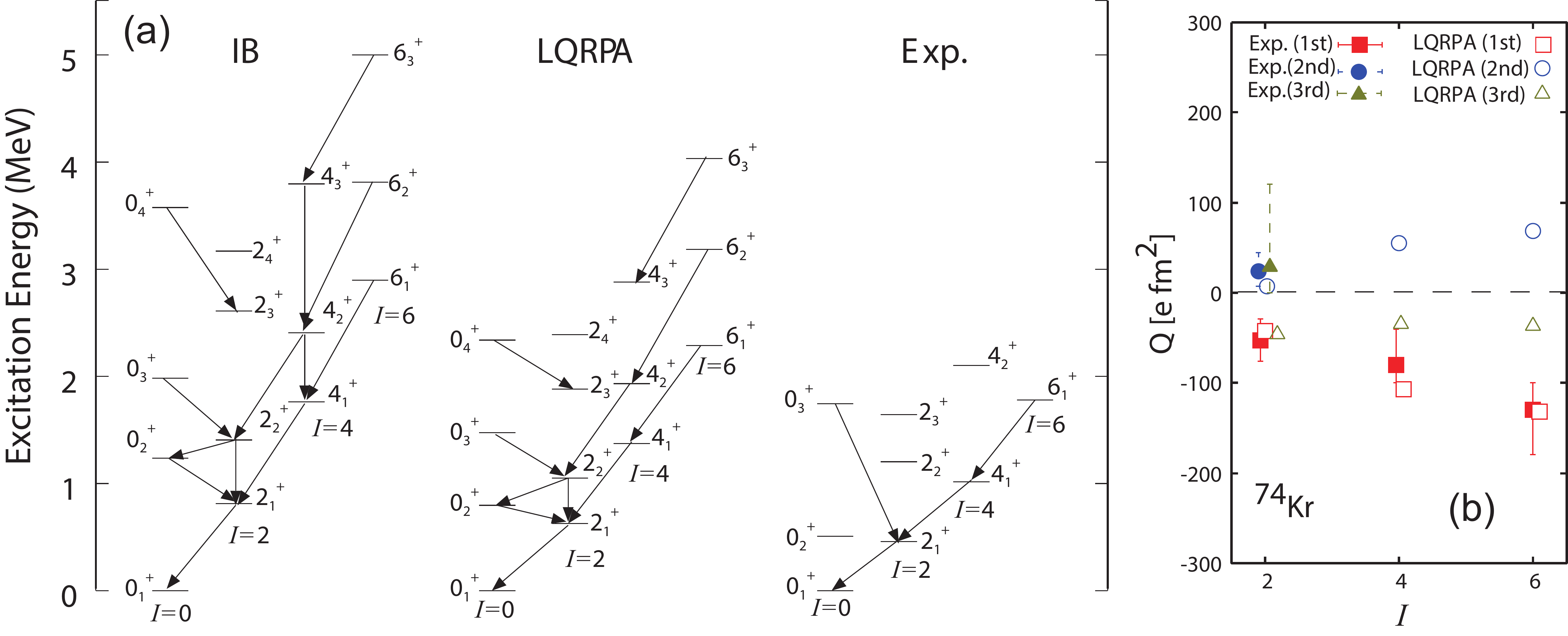}
\caption{
(a) Partial excitation spectrum calculated for $^{74}$Kr 
by means of the LQRPA method \cite{sat11} 
and experimental data \cite{cle07}.  
For comparison, the results calculated using the Inglis-Belyaev cranking masses (denoted
by IB) are also shown.
Only the levels with even angular momentum are shown (see Ref.  \cite{sat11} for the whole spectrum.)
The $E2$ transitions with $B(E2)$ larger than 50 Weisskopf units 
are indicated by arrows. 
(b) 
Spectroscopic quadrupole moments in unit of $e$fm$^2$ of the first
 (square), second (circle) and third (triangle) states for each angular
 momentum in $^{74}$Kr (from \cite{sat11}). 
Calculated values are shown by open symbols, while experimental data \cite{cle07} are indicated by filled symbols.
}
\label{fig:74Kr_spectra}
\end{center}
\end{figure}

We note that 
the rotational inertial functions $(D_1, D_2, D_3)$ and the pairing gaps  
significantly change as functions of $(\bg)$, as well as the vibrational
inertial masses $(D_{\beta\beta}, D_{\gamma\gamma},D_{\beta\gamma})$.
Due to the time-odd contributions of the moving HFB self-consistent field, 
the collective inertial masses calculated with the LQRPA method are $20-50\%$ larger than 
those evaluated with the Inglis-Belyaev cranking formula. 
Their ratios also change as functions of $(\bg)$ \cite{sat11}. 
As a consequence, as shown in Fig.~\ref{fig:74Kr_spectra}(a), 
the excitation spectrum calculated with the LQRPA masses is in much better 
agreement with experimental data than that with the Inglis-Belyaev cranking masses.
Figure \ref{fig:74Kr_spectra}(b) shows the spectroscopic quadrupole moments calculated with
the LQRPA masses for $^{74}$Kr. One sees that, aside from a minor
deviation for the $2_3$ state,
the calculated spectroscopic quadrupole moments are in excellent agreement with the experimental data. 
In particular, the signs and the increasing tendency of the
magnitudes with angular momentum in the ground band are well reproduced. 

\section{Some remarks on other approaches}
\label{sec:OtherApproaches}

In this section, we give short remarks on other methods 
widely used for microscopic calculation of collective inertial masses.   


\subsection{Constrained HFB + adiabatic perturbation} 
 
This method is convenient and widely used in the microscopic description of LACM 
\cite{lib99,yul99,pro04,del10,bar11}.  
It is based on the adiabatic assumption that the 
collective motion is much slower than the single-particle motion.
In this approach, we first postulate a few one-body operators $\hat{F_i}$ 
corresponding to collective coordinates $\alpha^i$, and solve 
the constrained HFB (or constrained HF + BCS) equation, 
\begin{equation}
     \delta \bra{\phi_0(\alpha)} \Hhat -
     \sum_i \mu^i(\alpha) \hat{F_i}\ket{\phi_0(\alpha)} = 0,
\end{equation}
to find the constrained HFB states $\ket{\phi_0(\alpha)}$. 
Here, $\mu^i(\alpha)$ are the Lagrange multipliers 
whose values are determined to fulfill the constraining conditions,  
\begin{equation}
\alpha^i = \bra{\phi_0(\alpha)}\hat{F_i}\ket{\phi_0(\alpha)}. 
\end{equation}
Assuming that the frequencies of the collective motion are much
smaller than those of non-collective two-quasiparticle excitation,   
we then calculate the collective kinetic energy
$T_{\rm{coll}}$ using the adiabatic perturbation theory: 
\begin{equation}
 T_{\rm{coll}}=\frac{1}{2} \sum_{ij} D_{ij}(\alpha)\dot{\alpha}^{i*}\dot{\alpha}^j, 
\end{equation}
where 
\begin{equation}
D_{ij}(\alpha)=2\sum_n
\frac{\bra{\phi_0(\alpha)}{\frac{\partial}{\partial \alpha^{i*}}}\ket{\phi_n(\alpha)}
\bra{\phi_n(\alpha)}{\frac{\partial}{\partial \alpha^{j}}} \ket{\phi_0(\alpha)}}
{E_n(\alpha)-E_0(\alpha)}
\label{IBcrankingMass1}
\end{equation}
are called Inglis-Belyaev cranking masses \cite{rin80}.
Here $\ket{\phi_0(\alpha)}$ and $\ket{\phi_n(\alpha)}$ represent the ground and 
two-quasiparticle excited states
for a given set of values $\alpha=\{\alpha^i\}$. 
In most of applications it is simplified furthermore by introducing an assumption that
the derivatives of the constrained HFB Hamiltonian with respect to $\alpha^i$ is
proportional to $\hat{F}_i$. 
Equation (\ref{IBcrankingMass1}) then reduces to 
\begin{equation}
D_{ij}(\alpha)=\frac{1}{2} \left[ {\cal M}_1^{-1}(\alpha)
 {\cal M}_3(\alpha) {\cal M}_1^{-1}(\alpha) \right]_{ij} 
 \label{IBcrankingMass2}
\end{equation}
with 
\begin{equation}
{\cal M}_k(\alpha)_{ij}=
\sum_n \frac{\bra{\phi_0(\alpha)}\hat{F_i}^\dag\ket{\phi_n(\alpha)}
\bra{\phi_n(\alpha)}\hat{F_j} \ket{\phi_0(\alpha)}}
{(E_n(\alpha)-E_0(\alpha))^k}. 
\end{equation}
In recent years, a systematic investigation on low-lying quadrupole spectra 
has been carried out in terms of the quadrupole collective Hamiltonian 
by using the Inglis-Belyaev cranking formula and the collective potential energies 
derived from the relativistic (covariant) density functionals  
\cite{nik09,li09,li10a,li10b,li11,nik11,fu13}.  

A problem of the Inglis-Belyaev cranking formula is that 
the collective inertial masses are underestimated 
\cite{dob95}. 
Moving mean fields induce time-odd components
that change sign under time reversal.
However, the Inglis-Belyaev cranking formula ignores their effects 
on the collective inertial masses.
By taking into account such time-odd corrections to the cranking masses,    
one can better reproduce low-lying spectra 
\cite{hin12}. 
For rotational moments of inertia, we may estimate the time-odd
corrections taking the limit of $\omega_\textrm{rot}\rightarrow 0$ for
the solution of the HFB equation in the rotating frame, that is defined by  
adding the cranking term $-\omega_\textrm{rot}\hat{J}_x$ to the 
constrained HFB Hamiltonian.  
Since this provides about $20-40$ \%\ enhancement from the Inglis-Belyaev
formula, the similar enhancement factors of $1.2-1.4$ have been often
utilized for vibrational inertial masses without solid justification.

\subsection{Adiabatic TDHF theory}

In 1960's, Belyaev, Baranger and Kumar started 
efforts to self-consistently derive the collective Hamiltonian  
using adiabatic approximation to time evolution of mean fields \cite{bel65, bar65}. 
In these pioneer works, they derived 
the quadrupole collective Hamiltonian using the P+Q force model 
\cite{bes69}.  
During 1970's, the time-dependent mean-field approach 
with the use of the P+Q force model 
was generalized to be applicable to any effective interaction. 
This advanced approach is called adiabatic TDHF (ATDHF) 
\cite{bar78,bri76,goe78}. 

In the ATDHF theory of Baranger and V\'en\'eroni \cite{bar78}, 
the density matrix $\rho(t)$ is written in the following form and 
expanded as a power series with respect to $\chi(t)$. 
\begin{eqnarray}
\rho(t)&=&e^{i\chi(t)}\rho_0(t)e^{-i\chi(t)}\\
          &=&\rho_0(t)+i[\chi(t),\rho_0(t)]-\frac{1}{2}[\chi(t), [\chi(t),\rho_0(t)]] + ... 
\end{eqnarray}
Here the matrix elements $\rho_{ij}(t)$ of $\rho(t)$ are defined by 
$\rho_{ij}(t)=\bra{\phi_{\rm HF}(t)}c_j^\dag c_i \ket{\phi_{\rm HF}(t)}$ 
with the time-dependent HF state $\ket{\phi_{\rm HF}(t)}$ 
and the nucleon creation and annihilation operators, $c_i^{\dag}$ and $c_j$,  
in the single-particle states $i$ and $j$. 
The above expansion is regarded as an adiabatic expansion with respect to $\chi(t)$ 
which plays the role of the collective momentum 
associated with the time-even density matrix $\rho_0(t)$.   
Baranger and V\'en\'eroni suggested a possibility of introducing collective coordinates 
as parameters that describe the time evolution of the density matrix $\rho_0(t)$.  
They discussed an iterative procedure to solve the ATDHF equations.  
This idea has not been realized until now, however.  
We note that the ATDHF does not reduce to the RPA in the small-amplitude limit 
if a few collective coordinates are introduced by hand. 
In fact it gives a collective mass different from the RPA \cite{gia80}. 

Villars developed another ATDHF theory with the aim of self-consistently determining  
the optimum collective coordinates on the basis of the time-dependent variational principle
 \cite{vil77}.  
In the same way as in the ASCC method described in Sec.~8,  
the TDHFB states are written in the form of Eq.~(\ref{eq:ASCCstate1}). 
Villars encountered a difficulty, however, that he could not get 
unique solutions of the basic equations determining the collective path. 
This problem was later solved by treating the second-order terms of the 
momentum expansion in a self-consistent manner 
(see Mukherjee and Pal \cite{muk81}, and Klein {\it et al.} \cite{kle91a,dan00}). 
It was shown that, when the number of collective coordinate is only one,  
a collective path maximally decoupled from non-collective degrees of freedom 
runs along a valley in the multi-dimensional potential-energy surface 
associated with the TDHF states.   

To describe low-frequency collective motions, 
it is necessary to take into account the pairing correlations. 
In other words, we need to develop the adiabatic TDHFB (ATDHFB) theory. 
This is one of the reasons why applications of the ATDHF have been restricted 
to collective phenomena where pairing correlations play minor roles such as 
low-energy collisions between spherical closed-shell nuclei 
\cite{goe83} 
and giant resonances 
\cite{gia80}. 
As discussed in Sec.~\ref{sec:discussions}, 
when large-amplitude shape fluctuations take place, 
single-particle level crossings often occur. 
To follow the adiabatic configuration across the level crossing points, 
the pairing correlation plays an essential role. 
Therefore, we need to develop the ATDHFB theory 
to describe low-frequency collective excitations.  

In the past, Dobaczewski and Skalski \cite{dob81} tried to develop the ATDHFB theory 
assuming the axially symmetric quadrupole deformation parameter $\beta$ 
as the collective coordinate. 
Quite recently, Li {\it et al.} \cite{li12a} 
tried to derive the 5D quadrupole collective Hamiltonian 
on the basis of the ATDHFB.  
The extension of ATDHF to ATDHFB is not straightforward, however. 
This is because, as will be discussed in Sec.~\ref{sec:SCC}, 
we need to decouple the number-fluctuation degrees of freedom from the LACM of interest,  
respecting the gauge invariance with respect the pairing rotational angles. 

\subsection{Generator coordinate method} 

The  generator coordinate method (GCM) has been used 
for a wide variety of nuclear collective phenomena 
\cite{rei87, egi04, ben08a}. 
Using the angular-momentum projector $\hat{P}_{IMK}$ 
and the neutron(proton)-number projector 
$\hat{P}_N$ ($\hat{P}_Z$),  we write the state vector as a superposition of the projected 
mean-field states with different deformation parameters $(\beta,\gamma)$,  
\begin{equation}
\ket{\Psi^i_{NZIM}} =  \int d\beta d\gamma \sum_K f^i_{NZIK} (\beta,\gamma) 
\hat{P}_N \hat{P}_Z \hat{P}_{IMK}  \ket{\phi(\beta,\gamma)}.  
\label{eq:GCM}
\end{equation}
Because the projection operators contain integrations, 
it has been a difficult task to carry out such high-dimensional numerical integrations 
in solving the Hill-Wheeler equation for the states $\ket{\phi(\beta,\gamma)}$
obtained by the constrained HFB method.    
In recent years, however, remarkable progress has been taking place, 
which makes it possible to carry out such large-scale numerical computations  
\cite{ben08b, rod10, yao10, yao11, yao14, rod14}. 
The HFB calculations with use of density-dependent effective interactions 
are better founded on density functional theory (DFT). 
Accordingly, the modern GCM calculation is referred to as `multi-reference DFT' 
\cite{ben08b}.  

We can derive a collective Schr\"odinger equation 
by making the gaussian overlap approximation (GOA) to the Hill-Wheeler equation 
\cite{gri57,oni75,roh12,roh13}. 
There is no guarantee, however, that dynamical effects associated 
with time-odd components of moving mean field are sufficiently taken into account 
in the collective inertial masses obtained through this procedure. 
It is well known for the case of center of mass motion that  
we need to use complex generator coordinates to obtain the correct mass.  
This fact indicates that collective momenta conjugate to collective coordinates
should also be treated as generator coordinates 
\cite{rin80, pei62}. 

A fundamental question is how to choose the optimal generator coordinates. 
With the variational principle,  
Holzwarth and Yukawa \cite{hol74} 
proved that the mean-field states parametrized by a single optimal generator coordinate 
run along a valley of the collective potential energy surface.  
This line of investigation stimulated 
the challenge toward constructing a microscopic theory of LACM \cite{rei79}. 
In this connection, 
we note that conventional GCM calculations parametrized by a few real generator coordinates 
do not reduce to the (Q)RPA in the small-amplitude limit. 
It should be distinguished from the case 
that all two-quasiparticle (particle-hole) degrees of freedom are treated 
as complex generator coordinates 
\cite{jan64}. 

It is very important to distinguish the 5D collective Hamiltonian obtained 
by making use of the GOA to the  GCM 
from that derived in the preceding section by using the LQRPA to the ASCC method.  
In the latter, the canonical conjugate pairs of collective coordinate and momentum 
are self-consistently derived on the basis of the time-dependent variational principle. 
The canonical formulation  enables us to adopt the standard canonical quantization procedure.  
Furthermore, effects of the time-odd components of the moving mean field 
are automatically taken into account in the collective inertial masses. 
It is therefore misleading to say as if the 5D collective Hamiltonian approach 
is an approximation to the full 5D (three Euler angles, $\beta$, and $\gamma$) GCM calculation.  
\\

\noindent
{\bf Additional remarks}\\

In view of the above points, it is desirable to carry out  
a systematic comparison of collective inertial masses evaluated by different approximations 
including the LQRPA (based on the ASCC method summarized in the next section), 
the adiabatic cranking methods, the ATDHFB, and the GCM+GOA
for a better understanding of their physical implications.   
In this connection, we notice that the results of the recent GCM calculation for $^{76}$Kr 
\cite{yao14}, 
using the particle-number and angular-momentum projected basis, Eq.~(\ref{eq:GCM}), 
are rather similar to those obtained by use of the Bohr-Mottelson collective Hamiltonian 
with the Inglis-Belyaev cranking masses,    
except for an overall overestimation of the excitation energies by about 20$\%$. 
This work casts an interesting question 
why the two different approaches yield rather similar results.   

\section{Fundamentals of microscopic theory of LACM} 
\label{sec:SCC}

In this section, we review the modern concept of LACM and 
the fundamental theory underling the LQRPA method 
used in Sec.~\ref{sec:microBM} to derive the Bohr-Mottelson collective Hamiltonian.

\subsection{Extraction of collective submanifold}

It is possible to formulate the TDHFB dynamics as the classical Hamilton equations
for canonical variables in the TDHFB phase space \cite{neg82,yam87,kur01}.
The dimension of this phase space is very large;  
twice of the number of all the two-quasiparticle pairs.
The TDHFB state vector $\ket{\phi(t)}$ can be regarded as a generalized
coherent state moving on a trajectory in the large-dimensional TDHFB phase space.  
For low-frequency collective motions, however,
we assume that the time evolution is governed by a few collective variables.

During the attempts to construct microscopic theory of LACM
since the latter half of the 1970's, significant progress has been achieved in the 
fundamental concepts of collective motion.
Especially important is the recognition that
microscopic derivation of the collective Hamiltonian is equivalent
to extraction of a collective submanifold embedded in the TDHFB phase space,
which is approximately decoupled
from other ``non-collective'' degrees of freedom.
From this point of view we can say that 
collective variables are nothing but {\it local canonical variables} 
which can be flexibly chosen on this submanifold.
Here, we recapitulate recent developments achieved on the basis of such
concepts. 

Attempts to formulate a LACM theory 
without assuming adiabaticity of large-amplitude collective motion 
were initiated by Rowe and Bassermann \cite{row76} and Marumori \cite{mar77} 
and led to the formulation of the SCC method 
by Marumori, Maskawa, Sakata, and Kuriyama \cite{mar80}.
In these approaches,
collective coordinates and collective momenta are treated on the same footing.  
In the SCC method, basic equations determining the collective submanifold are derived 
by requiring {\it maximal decoupling} of the collective motion of interest from 
other non-collective degrees of freedom. 
The collective submanifold 
is invariant with respect to the choice of 
the coordinate system, whereas the collective coordinates depend on it.  
The idea of coordinate-independent theory of collective motion
was developed also by Rowe \cite{row82}, and Yamamura and Kuriyama \cite{yam87}.
This idea had a significant impact on the fundamental question,
``{\it what are the collective variables?}''.
The SCC method was first formulated on the basis of the TDHF theory without pairing.
Later, it is extended to treat pairing correlations in superfluid nuclei 
on the basis of the TDHFB theory \cite{mat86}.
 
In the SCC method, the TDHFB state $\ket{\phi(t)}$
is written as $\ket{\phi(q,p)}$ under the assumption that the time evolution
is governed by a few collective coordinates $q=(q^1,q^2,\cdots,q^f)$  and
collective momenta $p=(p_1,p_2,\cdots,p_f)$.
The parametrization of the TDHFB state with the $2f$-degrees of freedom
$(q,p)$ means that we define a submanifold inside the TDHFB phase space, 
which is called ``{\it collective submanifold}.''  
Below, we summarize the basic equations that determine 
the collective submanifold on which 
the TDHFB state $\ket{\phi(q,p)}$ evolves in time. 
(For simplicity, we here omit the terms arising from the pairing-rotational 
degrees of freedom, which will be discussed in Sec.~\ref{sec:gauge_invariance}.)
\\
\newline\noindent 1.
{\it Invariance principle of the TDHFB equation} \\

We require that the TDHFB equation of motion is invariant 
in the collective submanifold.   
In a variational form, this requirement can be written as  
\begin{equation}
\delta\bra{\phi(q,p)} \left( i\frac{\partial}{\partial t} - \hat{H} \right) \ket{\phi(q,p)}=0.  
\label{invariance principle}
\end{equation}
Here, the variation $\delta$ is given by 
$\delta \ket{\phi(q,p)} =a_i^\dagger a_j^\dagger \ket{\phi(q,p)}$ 
in terms of the quasiparticle operators $(a_i^\dagger, a_j)$, 
which satisfy the vacuum condition, $a_i\ket{\phi(q,p)}=0$.  
Under the basic assumption,  we can replace the time derivative with 
\begin{equation}
\frac{\partial}{\partial t} = \sum_{i=1}^f
\left(\dot{q^i}\frac{\partial}{\partial q^i} + \dot{p_i}\frac{\partial}{\partial p_i} \right)
= \dot{q^i}\frac{\partial}{\partial q^i} + \dot{p_i}\frac{\partial}{\partial p_i},  
\end{equation}
Hereafter, to simplify the notation,
we adopt the Einstein summation convention
and remove $\sum_{i=1}^f$. 
Accordingly, we can rewrite Eq.~(\ref{invariance principle}) as
\begin{equation}
\delta\bra{\phi(q,p)} \left\{ 
\dot{q}^i \mathring{P}_i(q,p) 
- \dot{p}_i\mathring{Q}^i(q,p)  
 - \hat{H} \right\} \ket{\phi(q,p)}=0, 
\label{invariance principle2}
\end{equation}
where the local infinitesimal generators are defined by 
\begin{eqnarray}
\mathring{P}_i(q,p) \ket{\phi(q,p)} &=&   i \frac{\del}{\del q^i} \ket{\phi(q,p)}, 
\label{generator_P}
\\
\mathring{Q}^i(q,p) \ket{\phi(q,p)} &=& -i \frac{\del}{\del p_i} \ket{\phi(q,p)}. 
\label{generator_Q}
\end{eqnarray}
These are one-body operators which can be written as linear combinations of 
bilinear products $\{a_i^\dagger a_j^\dagger, a_j a_i \}$ 
of the quasiparticle operators defined with respect to $\ket{\phi(q,p)}$.  
\\
\newline\noindent 2.
{\it Canonicity conditions} \\

We require $q$ and $p$ to be canonical variables. 
According to 
the Frobenius-Darboux theorem \cite{eis33}, 
pairs of canonical variables $(q, p)$ exist 
for the TDHFB states $\ket{\phi(q,p)}$ satisfying the following 
{\it canonicity conditions}, 
\begin{eqnarray}
\bra{\phi(q,p)} \mathring{P}_i(q,p) \ket{\phi(q,p)} &=& p_i + \frac{\del S}{\del q^i}, 
\label{canonicity conditions1}
\\
\bra{\phi(q,p)} \mathring{Q}^i(q,p) \ket{\phi(q,p)} &=& -\frac{\del S}{\del p_i}, 
\label{canonicity conditions2}
\end{eqnarray} 
where $S$ is an arbitrary differentiable function of $q$ and $p$
\cite{mar80,yam87,kur84}.
By specifying the functional form of $S(q,p)$ and $S'(q',p')$
and demanding that the form of these equations be preserved,  
we can fix the type of allowed canonical transformations, $(q,p) \to (q',p')$ 
among the collective variables.  
We shall discuss typical examples 
in Sects. \ref{sec:SCCexpansion}  and \ref{sec:ASCC}, 
and call the canonicity conditions with a specified function $S(q,p)$ 
``{\it canonical-variable conditions.}"
Taking derivatives of Eqs.~(\ref{canonicity conditions1}) 
and (\ref{canonicity conditions2})
with respect to $p_i$ and $q^i$, respectively, 
we can easily confirm that the local infinitesimal generators satisfy 
the `weakly' canonical commutation relations, 
\begin{equation}
\bra{\phi(q,p)} \left[ \mathring{Q}^i(q,p),  \mathring{P}_j(q,p) \right] \ket{\phi(q,p)} 
= i \delta_{ij}. 
\label{weak canonical}
\end{equation}
 
Taking variations of Eq. (\ref{invariance principle2})
in the direction of the collective variables, $q$ and $p$,  
generated by  $\mathring{P}_i$ and $\mathring{Q}^i$,  
we obtain the Hamilton equations of motion, 
\begin{equation}
\frac{dq^i}{dt}=\frac{\partial \Hc}{\partial p_i}
,\quad
\frac{dp_i}{dt}=-\frac{\partial \Hc}{\partial q^i}. 
\label{Hamilton_eq}
\end{equation}
Here, the total energy $\Hc(q,p)\equiv \bra{\phi(q,p)} \hat{H} \ket{\phi(q,p)}$
plays the role of the classical collective Hamiltonian. \\
\newline\noindent 3.
{\it Equation of collective submanifold} \\

The variational principle (\ref{invariance principle2}) and
Eq.~(\ref{Hamilton_eq})
lead to the equation of collective submanifold:
\begin{equation}
\delta\bra{\phi(q,p)} \left\{
\hat{H} -  \frac{\partial \Hc}{\partial p_i}  \mathring{P}_i(q,p)
- \frac{\partial \Hc}{\partial q^i}\mathring{Q}^i(q,p)  \right\} \ket{\phi(q,p)}=0. 
\label{eq of submanifold1}
\end{equation}
Taking variations $\delta_{\perp}$ in the directions orthogonal to $q$ and $p$, 
we see that 
\begin{equation}
\delta_{\perp}\bra{\phi(q,p)} \hat{H} \ket{\phi(q,p)}=0.  
\end{equation}
This implies that the energy expectation value is stationary with respect to all variations 
except for those along directions tangent to the collective submanifold. 
In other words, the large-amplitude collective motion is decoupled from other modes of excitation. 

\subsection{Solution with $(\eta, \eta^*)$ expansion}
\label{sec:SCCexpansion}

In the original paper of the SCC method \cite{mar80}, 
the TDHFB state $\ket{\phi(q,p)}$ is written as 
\begin{equation}
\ket{\phi(q,p)} = U(q,p) \ket{\phi_0} = e^{i{\hat G}(q, p)} \ket{\phi_0}. 
\label{TDHFB_state_SCC}
\end{equation}
Here, $U(q,p)$ is a time-dependent unitary transformation 
written in terms of an Hermitian one-body operator $\hat{G}(q,p)$.
The HFB ground state $\ket{\phi_0}$ is taken as an initial state;  
$U(q,p)  =1$ at $(q,p)=(0,0)$.  

Using complex variables $\eta=(\eta_1,\eta_2,\cdots,\eta_f)$
defined by    
\begin{equation}
\eta_i = \frac{1}{\sqrt 2}(q^i+ip_i)
,\quad
\eta_i^* = \frac{1}{\sqrt 2}(q^i-ip_i), 
\end{equation}
we can rewrite the TDHFB state as
\begin{equation}
\ket{\phi(\eta,\eta^*)} = U(\eta,\eta^*) \ket{\phi_0} = e^{i{\hat G}(\eta, \eta^*)} \ket{\phi_0}. 
\end{equation}
Correspondingly, we define 
local infinitesimal generators, $\mathring{O}_i^\dag(\eta,\eta^*)$ and $\mathring{O}_i(\eta,\eta^*)$,  by
\begin{eqnarray}
\mathring{O}_i^\dag (\eta,\eta^*) \ket{\phi(\eta,\eta^*)} &=& \frac{\del}{\del \eta_i} \ket{\phi(\eta,\eta^*)} , 
\label{generator3}
\\
\mathring{O}_i (\eta,\eta^*)  \ket{\phi(\eta,\eta^*)} &=& -\frac{\del}{\del \eta_i^*}\ket{\phi(\eta,\eta^*)}  . 
\label{generator4}
\end{eqnarray}
Replacing $(q,p)$ by $(\eta,\eta^*)$,
the equation of collective submanifold (\ref{eq of submanifold1})
is rewritten as 
\begin{eqnarray}
\delta\bra{\phi_0}U^\dag(\eta,\eta^*)
\left\{
\hat{H} -
\frac{\partial \Hc}{\partial \eta_i^*} \mathring{O}_i^\dag(\eta,\eta^*)
-\frac{\partial \Hc}{\partial \eta_i}\mathring{O}_i(\eta,\eta^*)
  \right\}&& \nonumber\\
\times U(\eta,\eta^*)\ket{\phi_0}=0 . \quad\quad\quad &&
\label{eq of submanifold2}
\end{eqnarray}
Here, the variation is to be performed only for the HFB ground state $\ket{\phi_0}$. 

Let us consider the following canonical-variable conditions, 
\begin{eqnarray}
\bra{\phi(\eta,\eta^*)} \mathring{O}_i^\dag(\eta,\eta^*)\ket{\phi(\eta,\eta^*)} &=& \frac{1}{2}\eta_i^*, 
\label{canonical-variable conditions1}
\\
\bra{\phi(\eta,\eta^*)} \mathring{O}_i(\eta,\eta^*) \ket{\phi(\eta,\eta^*)} &=& \frac{1}{2}\eta_i, 
\label{canonical-variable conditions2} 
\end{eqnarray}
which are obtained by a specific choice of  $S=-\frac{1}{2}\sum_i q^i p_i$ in 
the canonicity conditions, (\ref{canonicity conditions1}) and (\ref{canonicity conditions2}). 
From Eqs.~(\ref{canonical-variable conditions1}) and (\ref{canonical-variable conditions2}), 
we can easily obtain the ``weak'' boson commutation relations, 
\begin{equation}
\bra{\phi(\eta,\eta^*)} \left[ \mathring{O}_i(\eta,\eta^*),   
\mathring{O}_j^\dag(\eta,\eta^*)\right] \ket{\phi(\eta,\eta^*)} = \delta_{ij}. 
\label{weak canonical2}
\end{equation}
We note that only linear canonical transformations among $\eta$ and $\eta^*$, 
which do not change the power of  $(\eta,\eta^*)$,   
are allowed under the conditions, 
(\ref{canonical-variable conditions1}) and (\ref{canonical-variable conditions2}).    
Therefore, these canonical-variable conditions    
are suitable for solving the variational equation (\ref{eq of submanifold2})  
by means of a power series expansion of ${\hat G}$ with respect to $(\eta,\eta^*)$:  
\begin{eqnarray}
{\hat G}(\eta,\eta^*) = {\hat G}_i^{(10)}\eta_i^* + {\hat G}_i^{(01)}\eta_i 
+ {\hat G}_{ij}^{(20)}\eta_i^*\eta_j^* +  {\hat G}_{ij}^{(11)}\eta_i^*\eta_j 
+ {\hat G}_{ij}^{(02)}\eta_i\eta_j + \cdots \quad\quad .
\end{eqnarray}
Requiring that the variational principle~(\ref{eq of submanifold2})
holds for every power, 
we can successively determine the one-body operator ${\hat G}^{(m, n)}$ 
with $m+n=1, 2, 3, \cdots$.  
This method of solution is called the ``$(\eta,\eta^*)$-expansion method."
Because $(\eta,\eta^*)$ are complex canonical variables, 
they are replaced by boson operators after the canonical quantization. 
The lowest linear order corresponds to the QRPA.  
Accordingly, the collective variables
$(\eta_i,\eta_i^*)$ correspond to a specific QRPA mode 
in the small-amplitude limit.
In the higher orders, however, 
the microscopic structure of ${\hat G}$ changes
as a function of $(\eta,\eta^*)$ 
due to the mode-mode coupling effects
among different QRPA modes.
In this sense, the $(\eta,\eta^*)$-expansion method may be regarded as
a dynamical extension of the boson expansion method \cite{mat85a}. 
Thus, it is a powerful method of treating anharmonic effects
originating from mode-mode couplings, 
as shown in its application to the two-phonon states 
of anharmonic $\gamma$ vibration \cite{mat85b,mat85c}. 
The SCC method was also used for derivation of
the 5D collective Hamiltonian and analysis of 
the quantum phase transition from spherical to deformed shapes \cite{yam93}
and for constructing diabatic representation
in the rotating shell model \cite{shi01}. 
The validity of the canonical quantization procedure, including a treatment of  
the ordering ambiguity problem, was examined in \cite{mat85a}. 
Description of the 3D rotational motions by means of the SCC method 
was discussed in \cite{kan94} from a viewpoint of constrained dynamical system.
\subsection{Solution with adiabatic expansion}
\label{sec:ASCC}

The $(\eta,\eta^*)$ expansion about a single HFB equilibrium point is
not suitable for treating situations where a few local minima having different shapes 
energetically compete in the HFB potential-energy surface 
and large-amplitude shape-mixing vibrations occur.   
It is also difficult to apply the expansion method
to a collective motion which goes far away from the equilibrium,
such as nuclear fission.
The time evolution of these low-energy LACM's in nuclei are usually slow (adiabatic) 
in comparison with the time scale of the single-particle motions.  
For describing adiabatic LACM 
extending over very far from the HFB equilibrium,
a new method of solution has been proposed \cite{mat00}.  
In this method, the basic equations of the SCC method are
solved by an expansion with respect to 
the collective momenta, keeping full orders in the collective coordinates.
It is called ``{\it adiabatic SCC (ASCC) method}.''
Similar methods have been proposed also by Klein, Walet, and Do Dang \cite{kle91a},
and Almehed and Walet \cite{alm04a}, 
but the gauge invariance in the TDHFB theory (discussed in Sec.~8.4 below) 
were not considered in these papers.   

A microscopic theory for adiabatic LACM is constructed by the ASCC method
in the following way. 
We assume that the TDHFB state  $\ket{\phi(q,p)}$
can be written in a form
\begin{equation}
 \ket{\phi(q,p)}  =  \exp\left\{ i p_i \Qhat^i(q) \right\}
 \ket{\phi(q)} ,
\label{eq:ASCCstate}
\end{equation}
where $\Qhat^i(q)$ are infinitesimal generators of $p_i$ 
locally defined at the state $\ket{\phi(q)}$ 
that represents a TDHFB state $\ket{\phi(q,p)}$ at $p\rightarrow 0$. 
This state $\ket{\phi(q)}$ is called a ``moving-frame HFB state." 

We use the following canonical-variable conditions different from 
(\ref{canonical-variable conditions1}) and (\ref{canonical-variable conditions2}), 
\begin{eqnarray}
\bra{\phi(q,p)} \mathring{P}_i(q,p) \ket{\phi(q,p)} &=& p_i, 
\label{canonical-variable conditions3}
\\
\bra{\phi(q,p)} \mathring{Q}^i(q,p) \ket{\phi(q,p)} &=& 0, 
\label{canonical-variable conditions4}
\end{eqnarray}
which are obtained by putting $S=$const. in 
the canonicity conditions (\ref{canonicity conditions1}) and (\ref{canonicity conditions2}). 
These canonical-variable conditions are suitable for the adiabatic expansion  
with respect to the collective momenta $p$, 
because only point transformations, $q\rightarrow q'(q)$
(more generally, similarity transformations) which do not mix $p$ and $q$, 
are allowed under the conditions, 
(\ref{canonical-variable conditions3}) and (\ref{canonical-variable conditions4}).   
We insert the above form of the TDHFB state (\ref{eq:ASCCstate})
into  the equation of collective submanifold (\ref{eq of submanifold2})
and the canonical variable conditions, 
(\ref{canonical-variable conditions3}) and (\ref{canonical-variable conditions4}), 
and make a power-series expansion in $p$. 
We can determine the microscopic structures of $\Qhat^i(q)$ and $\ket{\phi(q)}$  
by requiring that these equations hold for every power of $p$.   
We take into account up to the second order.  
The canonical variable conditions, 
(\ref{canonical-variable conditions3}) and (\ref{canonical-variable conditions4}), 
then yield the `weakly' canonical commutation relations, 
\begin{equation}
\bra{\phi(q)} \left[ \Qhat^i(q), \Phat_j(q) \right] \ket{\phi(q)} = i \delta_{ij}. 
\label{weak canonical3}
\end{equation}
Here, $\Phat_i(q)$ are infinitesimal generators of $q^i$, 
locally defined at the state $\ket{\phi(q)}$ by 
\begin{equation}
 \Phat_i(q) \ket{\phi(q)} = i \frac{\del}{\del q^i} \ket{\phi(q)}. 
\end{equation}
We also obtain $\bra{\phi(q)} \Qhat^i(q) \ket{\phi(q)} = 0$ and  
$\bra{\phi(q)} \Phat_i(q) \ket{\phi(q)} =0$, which are trivially satisfied. 
Note that  $\Qhat^i(q)$ and $\Phat_i(q)$ operate on $\ket{\phi(q)}$, 
while $\mathring{Q}^i(q,p)$ and $\mathring{P}_i(q,p)$ on $\ket{\phi(q,p)}$. 

The time derivatives, $\dot{q}^i$ and $\dot{p}_i$, are determined by 
the Hamilton equations of motion (\ref{Hamilton_eq}) 
with the classical collective Hamiltonian
$\Hc(q,p)$ expanded with respect to $p$ up to the second order,
\begin{equation}
\Hc(q,p)= V(q) + \frac{1}{2} B^{ij}(q) p_i p_j ,
\label{ASCC_collective_Hamiltonian}
\end{equation}
where
\begin{equation}
V(q) = \Hc(q,p=0)
,\quad
B^{ij}(q)=\left.\frac{\partial^2 \Hc}{\partial p_i \partial p_j}\right|_{p=0} .
\end{equation}
The collective inertial tensors $B_{ij}(q)$ are defined as the inverse matrix
of $B^{ij}(q)$, ~$B^{ij} B_{jk}=\delta^i_k$.
Under these preparations,
the following equations, which constitute the core of the ASCC method,
can be derived \cite{mat00}.
Here, to further simplify the expression, we show the case for normal systems
with TDHF (see the next subsection about the extension to TDHFB).

\medskip
\noindent
{1.~{\it Moving-frame HF(B) equation}
\begin{equation}
 \delta\bra{\phi(q)}\Hhat_{\rm M}(q)\ket{\phi(q)} = 0, 
\label{eq:mfHFB}
\end{equation} 
where $\Hhat_{\rm M}(q)$ represents the Hamiltonian in the frame attached to the moving mean field, 
\begin{equation}
  \Hhat_{\rm M}(q) =  \Hhat 
 -  \frac{\del V}{\del q^i}\Qhat^i(q), 
\end{equation}
and is called ``moving-frame Hamiltonian."
\\

\noindent
{2.~{\it Moving-frame (Q)RPA equations}\\
(or ``{\it Local harmonic equations}'')
\begin{eqnarray}
&&\delta\bra{\phi(q)} \left[\Hhat_{\rm M}(q), \Qhat^i(q)\right]
- \frac{1}{i} B^{ij}(q) \Phat_j(q)   \quad\quad
\nonumber \\ 
&& \quad\quad\quad\quad\quad
+\frac{1}{2}\left[\frac{\del V}{\del q^j}\Qhat^j(q),
 \Qhat^i(q)\right]
\ket{\phi(q)} = 0, \quad
\label{eq:ASCC1}
\\
&& \delta\bra{\phi(q)} [\Hhat_{\rm M}(q),
 \frac{1}{i}\Phat_i(q)] - C_{ij}(q) \Qhat^j(q)
\nonumber \\
&&
- \frac{1}{2}\left[\left[\Hhat_{\rm M}(q), \frac{\del V}{\del
 q^k}\Qhat^k(q)\right], B_{ij}(q) \Qhat^j(q)\right] 
\ket{\phi(q)} = 0, \quad\quad
\label{eq:ASCC2}
\end{eqnarray}
where  
\begin{eqnarray}
 C_{ij}(q) &=& \frac{\del^2 V}{\del q^i \del q^j} - \Gamma_{ij}^k\frac{\del V}{\del q^k}, \\
 \Gamma_{ij}^k(q) &=& \frac{1}{2} B^{kl}\left( \frac{\del B_{li}}{\del q^j}
 + \frac{\del B_{lj}}{\del q^i} - \frac{\del B_{ij}}{\del q^l} \right). 
\end{eqnarray} 
The double-commutator term in Eq.~(\ref{eq:ASCC2}) arises from the $q$-derivative 
of the infinitesimal generators $\Qhat^i(q)$ and represents the curvatures of the collective submanifold.  
Diagonalizing the matrix, $B^{ik} C_{kj}$, at each point of $q$,
we may identify the local normal modes and eigen-frequencies 
$\omega_i(q)$ of the moving-frame QRPA equations.

Extension from TDHF to TDHFB for superfluid nuclei
can be achieved by introducing the number fluctuation $n \equiv N-N_0$
and their conjugate angle $\varphi$ as additional collective variables \cite{mat00} 
(see Sec.~\ref{sec:gauge_invariance}). 

Solving Eqs.~(\ref{eq:mfHFB}), (\ref{eq:ASCC1}), and (\ref{eq:ASCC2}) self-consistently, 
we can determine the microscopic expressions of 
the infinitesimal generators, $\Qhat^i(q)$ and $\Phat_i(q)$,  
in bilinear forms of the quasiparticle creation and annihilation operators 
defined locally with respect to $\ket{\phi(q)}$. 
These equations reduce to the HF(B) and (Q)RPA equations 
at the equilibrium point where $\del V/\del q^i=0$. 
Therefore, they are regarded as natural extensions of the well-known HFB-QRPA equations 
to non-equilibrium states. 

Some key points of the ASCC method are noted below: 
\\
\noindent
(i) {\it Meaning of adiabatic approximation}\\
The term ``adiabatic approximation'' is frequently used for different meanings.
In the present context, we use this term for 
the approximate solution of the variational equation
(\ref{invariance principle2})
by taking into account up to the second order 
in an expansion with respect to the collective momenta $p$.  
It is important to note that the effects of finite frequency of the LACM 
are taken into account through the moving-frame QRPA equation.
No assumption is made, such as that the kinetic energy of LACM is much smaller 
than the lowest two-quasiparticle excitation energy at every point of $q$.
\\
\noindent
(ii) {\it Difference from the constrained HFB equations}\\
The moving-frame HFB equation (\ref{eq:mfHFB}) resembles 
the constrained HFB equation. 
An essential difference is that the infinitesimal generators $\Qhat^i(q)$ 
are here self-consistently determined together with $\Phat_i(q)$
as solutions of the 
moving-frame QRPA equations, (\ref{eq:ASCC1}) and (\ref{eq:ASCC2}),  
at every point of the collective coordinate $q$. 
Thus, contrary to constrained operators in the constrained HFB theory, 
their microscopic structure changes as functions of $q$. 
The optimal ``constraining'' operators are locally determined at each $q$. 
The collective submanifold embedded in the TDHFB phase space is 
extracted in this way.
\\ 
\noindent
(iii) {\it Canonical quantization}\\
The collective inertia tensors $B_{ij}(q)$ take a diagonal form  
when the classical collective Hamiltonian is represented in terms of 
the local normal modes of the moving-frame QRPA equations.  
We can then make a scale transformation of the collective coordinates $q$ 
such that  they become unity. 
The kinetic energy term in the resulting collective Hamiltonian 
depends only on $p$. 
Thus, there is no ordering ambiguity between $q$ and $p$ 
in the canonical quantization procedure.   
\\
\noindent
(iv) {\it Collective inertial mass}\\
Although the collective submanifold is invariant against coordinate
transformations, $q \rightarrow q'(q)$,
the collective inertial tensors $B_{ij}(q)$ depends on the adopted
coordinate system. 
The scale of the coordinates can be arbitrarily chosen as far as the
canonical-variable conditions are satisfied. 
To obtain physical insights
and to examine the effects of time-odd components in the mean field, 
however, it is convenient to adopt a conventional coordinate system,
such as the quadrupole $(\beta,\gamma)$ variables. 


\subsection{Inclusion of the pairing rotation and gauge invariance}
\label{sec:gauge_invariance}

In the QRPA at the HFB equilibrium,
the ANG modes such as the number fluctuation (pairing rotational)
modes are decoupled from other normal modes.
Thereby, the QRPA restores the gauge invariance (number conservation) 
broken in the HFB mean field \cite{bri05}.
It is desirable to retain this merit of the QRPA beyond the small-amplitude regime.   
Otherwise, spurious number-fluctuation modes would heavily mix in the LACM of interest. 
It is possible to achieve this aim by using the SCC method for superfluid nuclei \cite{mat86}.

Introducing the number-fluctuation $n=N-N_0$ and the gauge angle $\varphi$ 
(conjugate to $n$) as additional collective variables,
we generalize the TDHFB state (\ref{eq:ASCCstate}) to 
\begin{eqnarray}
\ket{\phi(q,p,\varphi,n)}  &=&  e^{-i\varphi\Nhat} \ket{\phi(q,p,n)} ,
\label{eq:ASCCstate2}  
\\
\ket{\phi(q,p,n)} &=& e^{i \left[ p_i \Qhat^i(q) 
+ 
 n \That(q) \right]} \ket{\phi(q)}. 
\label{eq:ASCCstate3}  
\end{eqnarray}
Here $\hat{N}$ and $\That(q)$ denote the nucleon-number operator and  
the infinitesimal generator of $n$, respectively, 
and $N_0$ is a reference value of the nucleon number $N$. 
In the generalized TDHFB state, (\ref{eq:ASCCstate2}),   
the number operator $\hat{N}$ and the state vector $\ket{\phi(q,p,n)}$
may be regarded as an infinitesimal generator of the gauge angle $\varphi$  
and an intrinsic state with respect to the pairing rotational motion, respectively.   
It is straightforward to extend the equation for the collective submanifold  
(\ref{invariance principle2}) as  
\begin{eqnarray}
\delta \bra{\phi(q,p,\varphi,n)}
\left\{
i \dot{q}^i \frac{\partial}{\partial q^i}
+ i\dot{p}_i \frac{\partial}{\partial p_i} \right.
\quad\quad\quad\quad\quad &&\nonumber \\
\left.
+ i\dot{\varphi} \frac{\partial}{\partial\varphi}
- \Hhat
\right\}
\ket{\phi(q,p,\varphi,n)}  = 0. &&
\label{invariance principle4}
\end{eqnarray} 
Note that $\dot{n}=0$, 
because the Hamilton equations for the canonical conjugate pair $(n, \varphi)$ are   
\begin{equation}
\dot{\varphi}=\frac{\partial \Hc}{\partial n}, 
\quad 
\dot{n}=-\frac{\partial \Hc}{\partial \varphi}, 
\end{equation}
and the classical collective Hamiltonian 
$\Hc(q,p,\varphi,n)\equiv \bra{\phi(q,p,\varphi,n)} \hat{H} \ket{\phi(q,p,\varphi,n)}$ 
does not depend on $\varphi$. 
Making a power-series expansion with respect to $n$ as well as $p$ 
and considering up to the second order, 
we can determine $\That(q)$ simultaneously with $\Qhat^i(q)$ and $\Phat_i(q)$ 
such that the moving-frame equations become invariant 
against the rotation of the gauge angle $\varphi$. 
In fact, we introduce two sets of $(\Nhat, \That(q))$ to describe the pairing rotations 
of neutrons and protons, separately. 

Writing the time derivative  $\dot{\varphi}$ of the gauge angle as $\lambda$, 
we can easily confirm that the term proportional to $\dot{\varphi}$ in (\ref{invariance principle4}) 
leads to an operator $\lambda\hat{N}$ on the intrinsic state $\ket{\phi(q,p,n)}$. 
In this form, $\dot{\varphi}$ corresponds to the chemical potential in the BCS theory 
of superconductivity. The term, $\lambda\hat{N}$, in the BCS theory is usually interpreted 
as a constraining term to impose the condition that $\bra{\phi(q,p,n)}\hat{N}\ket{\phi(q,p,n)}=N$. 
It should be emphasized, however, that this term is naturally derived 
by introducing the concept that the moving-frame TDHFB state, $\ket{\phi(q,p,n)}$, 
is an intrinsic state with respect to the pairing-rotational motion of the gauge angle $\varphi$. 
In the microscopic approach under discussion, the `chemical potential' $\lambda$
plays a role analogous to the rotational velocities $\dot{\varphi}_k$ 
in Eq.~(\ref{eq:RotatingTDHFBstate}) for the rotational motions in the 3D coordinate space; 
that is, they are not introduced as Lagrange multipliers but dynamical variables. 

Hinohara {\it et al.} investigated the gauge-invariance properties of the ASCC equations 
and extended the infinitesimal generators $\Qhat^i(q)$ to include
quasiparticle creation-annihilation $(a_i^\dag a_j)$ parts 
in addition to two-quasiparticle creation $(a_i^\dag a_j^\dag)$ and annihilation $(a_j a_i)$ parts 
\cite{hin07}. 
This is the reason why Eqs. (\ref{eq:ASCC1}) and (\ref{eq:ASCC2}) 
are written in a more general form than those originally given in \cite{mat00}. 
The gauge invariance of the ASCC method implies that 
we need to fix the gauge in numerical applications.
A convenient  procedure for the gauge fixing is given in \cite{hin07}.
A more general consideration on the gauge symmetry of the ASCC method 
is given from a viewpoint of constrained dynamical systems in a recent paper \cite{sat15}. 

\subsection{Solution with the LQRPA method} 

The LQRPA method used in Sec.~\ref{sec:microBM} for the microscopic derivation of the 
Bohr-Mottelson collective Hamiltonian   
may be regarded as a non-iterative solution of 
(\ref{eq:mfHFB}) - (\ref{eq:ASCC2}) in the ASCC method, 
without the consistency in the generator $\Qhat^i(q)$ between
the moving-frame HFB equation and the moving-frame QRPA equations. 
It may also be regarded as a first-step of the iterative procedure 
for solving the self-consistent equations.  
Equation (\ref{eq:CHFB}) corresponds to the moving-frame HFB equation
(\ref{eq:mfHFB}) with $\Qhat^i(q)$ replaced by global one-body
operators $\Dhatp_{2m}$.
It is worth noting that the moving frame HFB Hamiltonian
$\Hhat_\textrm{M}$
contains terms, $-\lambda^{(\tau)} \Nhat^{(\tau)} - \mu_m \Dhatp_{2m}$,
which naturally appear from the ASCC equations
with the approximation to replace
$\Qhat^i(q)$ in $\Hhat_M(q)$ by $\Dhatp_{2m}$.
In fact, the origin of these terms are not constraints, but the time-derivative terms 
in Eq. (\ref{invariance principle4}). 
The LQRPA equations, (\ref{eq:LQRPA1}) and (\ref{eq:LQRPA2}),  are obtained 
by ignoring the curvature term in the moving-frame QRPA equations, 
(\ref{eq:ASCC1}) and (\ref{eq:ASCC2}).  

The validity of the LQRPA method was examined  
for the cases where a well-defined valley (collective path)  
exists in the collective potential $V(\bg)$ \cite{hin10}. 
The rotational and vibrational inertial masses calculated 
by using the LQRPA method were compared with those 
obtained by the fully self-consistent ASCC calculations.   
It was confirmed that they agree very well, indicating that 
the LQRPA is a good approximation to the ASCC calculation  
along the collective path on the $(\bg)$ plane. 
The accuracy of the LQRPA method on the full $(\bg)$ plane 
may be checked by making an iterative calculation; that is, 
by solving Eq.~(\ref{eq:CHFB}) replacing $\Dhatp_{2m}$ 
with the solutions $\Qhat^i(q)$ 
of the LQRPA eqs.~(\ref{eq:LQRPA1}) and (\ref{eq:LQRPA2}),
and evaluate the deviations from the result of the lowest-order LQRPA calculation. 
This task remains for future, however.   
\section{Open problems in quadrupole collective dynamics}
\label{sec:OpenProblems}

Nowadays, the domain of quadrupole collective phenomena 
awaiting applications of the Bohr-Mottelson collective model is enormously increasing 
covering wide regions from low to highly excited states, 
from small to large angular momenta, 
and from proton-drip line to neutron-drip line. 
Among many interesting subjects, we here remark only a few.

\subsection{Shape coexistence, pairing fluctuation and mysterious $0^+$ states } 

As mentioned in Sec.~\ref{sec:QuadrupoleCollectivePhenomena}, 
when two different HFB equilibrium shapes coexist in the same energy region, 
large-amplitude shape mixings through the potential barriers take place.   
These phenomena may be regarded as a kind of  macroscopic quantum tunneling 
where the potential barrier itself is generated as a 
consequence of the dynamics of the self-bound quantum system. 
For instance, two strongly distorted rotational bands built on 
the oblate and prolate shapes, 
which seem to coexist and interact with each other,
have been found in $^{68}$Se \cite{hey11,hin10}.   
Such phenomena are widely seen in low-energy spectra from light to heavy nuclei 
\cite{hey11}. 
We have applied the Bohr-Mottelson collective Hamiltonian
to some of these shape coexistence/mixing phenomena 
with the use of the collective inertial masses   
microscopically calculated by means of the LQRPA method.  
An illustrative example is presented for $^{74}$Kr in Sec.~\ref{sec:Illustration}.  

One of the issues related to the shape coexistence/fluctuation 
is to clarify the nature of deformation in neutron-rich nuclei 
around $^{32}$Mg having the magic number  $N=20$ of the spherical shell model \cite{hey11}.  
In the P+Q model,  the major properties of low-lying states 
in open-shell nuclei are determined by the competition 
between the pairing (particle-particle, hole-hole) and 
quadrupole (particle-hole) correlations acting among nucleons in partially filled major shells. 
On the other hand, in situations 
where the pairing and quadrupole correlations {\it across} the spherical major shells 
play the major role, such as in neutron-rich Mg isotopes around $^{32}$Mg,   
the two different correlations seem to act coherently 
and generate interesting collective phenomena 
where large-amplitude fluctuations in the monopole and 
quadrupole pairing gaps as well as the quadrupole shape 
take place simultaneously \cite{hin11a}.    

In some nuclei, the first excited  $0^+$ state appears 
below the first excited $2^+$ state.   
An example is the first excited $0^+$ state of $^{72}$Ge 
which is known from old days but still poorly understood. 
This anomaly occurs in the vicinity of $N=40$ 
where the $g_{9/2}$ shell starts to be partially filled (due to the pairing). 
It has been pointed out 
\cite{iwa76,wee81,taz81,tak86} 
that the mode-mode coupling between the $0^+$ member of the 
two quadrupole-phonon triplet and the neutron pairing vibration 
becomes especially strong near $N=40$ and generates  
such anomalous $0^+$ states with extremely low excitation energy. 

As reviewed by Hyde and Wood \cite{hey11} 
and by Garrett \cite{gar01}, 
the nature of the low-lying excited $0^+$ states systematically found in recent experiments,  
in addition to those known from old days, is not well understood. 
It is thus quite challenging to apply, in a systematic ways,  
the Bohr-Mottelson collective Hamiltonian approach to all of these data, 
from light to heavy and from stable to unstable nuclei, 
and explore the limit of the applicability. 
Considering the suggestion \cite{iwa76,wee81,taz81,tak86} about 
the coupling effects with pairing vibrations,  
one of the basic questions is ``{\it under what situations 
we need to extend the 5D collective Hamiltonian to 7D 
by explicitly treating the proton and neutron pairing gaps as dynamical variables.}"    

We should mention about a few fundamental subjects 
that are closely related to the shape coexistence phenomena in low-lying states:
In the decays of superdeformed rotational bands \cite{yos01}, 
macroscopic quantum tunnelings through self-consistently generated barriers 
are very clearly seen.   
Needless to say, microscopic description of spontaneous fissions is 
a longstanding yet modern fundamental subject of nuclear structure physics 
{\cite{bra72,neg82}. 
Recent experimental progress in deep sub-barrier fusion reactions \cite{bac14} 
provides another modern problem of macroscopic tunnelings in finite quantum systems.       

\subsection{Vibrational and rotational modes at high angular momentum} 

As a nucleus rotates rapidly, excitations of aligned quasiparticles take place 
\cite{fra01,sat05}.
Rapid rotation changes the deformation and shell structure of the mean field.  
The pair field also disappears eventually at high spin \cite{shi89}. 
These structural changes in the high-spin yrast states   
significantly affect the properties of vibrational motions built on them  
(the yrast state is the 'ground' states for given angular momenta).
Unfortunately experimental data for low-frequency shape vibrations 
in the vicinity of the high-spin yrast states have not been accumulated enough.
Considering the role of the BCS pairing in forming the collective low-frequency 
quadrupole vibrations built on the ground state,
existence of low-frequency collective vibrations built on the high-spin yrast states 
is actually not evident, 
since we expect that the role of the pairing is much less in high-spin states.
On the other hand, we expect that the vibrations could compete with rotations 
in high-spin states because the rotational frequency increases with the angular momentum, 
and becomes comparable to the vibrational frequencies \cite{boh81}.

Discovery of superdeformed bands \cite{nol88,jan91} shed a new light 
on the above situation.
In superdeformed states, a new shell structure 
called {\it superdeformed shell structure} emerges 
and it creates new low-frequency octupole vibrations 
on superdeformed states at high angular momentum \cite{nak96,ros01}. 
These vibrational modes simultaneously break 
the axial symmetry and the reflection symmetry. 
Moreover, some experimental data for $\gamma$-vibrations (quadrupole shape vibration 
that breaks the axial symmetry) at high spin have
been reported \cite{pat03, oll11}.  
It has been discussed for a long time that 
the triaxial deformation may be realized at high spin states 
due to the weakening of the pairing correlations. 
When the mean field breaks the axial symmetry, a new rotational mode 
called {\it wobbling motion} is expected to emerge. 
Observation of the wobbling rotational band is therefore 
a clear signature of the occurrence of the triaxial deformation in the mean field.  
About 15 years ago, the first experimental data on the wobbling band were obtained 
\cite{ode01} (see also \cite{jen02, gor04}). 
Their properties have been theoretically analyzed from various points of view 
\cite{ham03, mat04, sho09, fra14}.
These investigations show that 
the aligned quasiparticle plays a crucial role in the emergence of the wobbling motion. 
There is another new phenomenon expected to emerge in the axial-symmetry-broken 
nuclei under certain conditions:
{\it chiral rotation} and its experimental signature, chiral doublet bands \cite{fra01}.
Experimental search for chiral doublet bands and its precursor phenomena 
called {\it chiral vibrations} \cite{alm11} is currently in progress.

The Bohr-Mottelson collective Hamiltonian as reviewed in this paper is not 
applicable to quadrupole collective phenomena at high angular momenta. 
This is because the collective inertial masses and the collective potential $V(\bg)$ 
are calculated at low angular momenta.  
It seems, however, possible to extend it to describe such high-spin phenomena. 
We have learned through the success of the cranked shell model
\cite{fra01, sat05} 
that the concept of single-particle motion in a rotating mean field holds very well. 
This means that major effects of rapid rotation (Coriolis- and centrifugal-force effects) 
can be captured in the self-consistent mean-field 
by defining the single-particle motion 
in a rotating frame of reference attached to the rapidly rotating nucleus.  
In the extension of the self-consistent mean field to a rotating frame, 
the time-reversal symmetry is broken, 
but it opens a new dimension in nuclear structure physics. 
{\it 
In the history of nuclear structure physics, we have been successfully extending  
the concept of the single-particle motion to a more general mean-field. 
Such extensions have been achieved by breaking some symmetries of the self-consistent mean field. 
}
Let us recall that 
extension of the concept of single-particle excitation (with spontaneous breaking of the symmetry)  
and appearance of new collective excitation (restoring the broken symmetry)  
are {\it dual concepts} that underline the quantum many-body theory of nuclear structure.  

\subsection{Low-frequency collective excitations in nuclei near the neutron drip line} 

The mean field in unstable nuclei near the neutron drip line possesses  
new features associated with the large neutron to proton ratio, 
the formation of neutron skin, the weak binding of single-particles states 
near the Fermi surface, the excitation of neutron pairs into the continuum, etc.
The collectivity of surface vibrations may change reflecting 
the modification of shell structure \cite{ham12} 
and the variation of pairing properties \cite{mat13c}. 
Thus, the QRPA method has been extended to properly treat 
the excitations into the continuum \cite{mat01}.   
The extended version is called {\it continuum QRPA}, and 
it has been applied to weakly bound unstable nuclei \cite{miz09,ser09,mat10b}.  
The particle-vibration coupling theory has also been extended to include 
the continuum effects by means of the continuum QRPA method \cite{miz12}.     

In stable nuclei, overlaps of different single-particle wave functions 
become maximum at the surface and generate a strong coherence 
among many quasiparticle excitations \cite{boh69}.    
In the weak binding situation, single-particle wave functions significantly 
extend from the surface (half-density radius) to the low-density region 
and acquire strong individualities.   
It is an open problem how the pairing correlation in such a situation  
acts to generate the collectivity of vibrational modes.  
Nowadays, it is one of the central subject in nuclear structure-reaction theory 
to carry out fully self-consistent HFB+QRPA calculations 
using the same energy density functional and 
simultaneously taking into account 
the deformation, pairing and excitations into the continuum
\cite{mat10b}.     
From such microscopic calculations, for instance, it is suggested 
\cite{yos08} that a strong coherence among the quadrupole shape fluctuation and 
the fluctuations of the monopole and quadrupole pairing gaps 
may generate a collective vibration unique to the weakly bound neutron-rich nuclei.    

At the present time, the major efforts are devoted to clarifying the properties of 
the ground states and a few excited states of nuclei near the drip line. 
In the coming future, more experimental data on excitation spectra will be obtained 
with the progress of ambitious experimental projects now ongoing in the world. 
We shall then encounter a variety of phenomena that cannot be understood within the 
small-amplitude approximation for collective motions. 
It will become necessary to explore the nature of collective motions in nuclei 
near the drip line by an extension of the collective Hamiltonian approach 
reviewed in this paper.  

\section{Concluding remarks}
\label{sec:Conclusion}

We have reviewed recent approaches to microscopically derive 
the Bohr-Mottelson collective Hamiltonian 
on the basis of the time-dependent self-consistent mean field.   
The {\it moving} self-consistent mean-field is the key concept 
to the unified understanding of the single-particle and collective motions in nuclei.  
We hope that this paper fits the aim of this special edition 
for the 40 year anniversary of Nobel Prize 1975. 

Although the progress achieved during these 40 years with the Bohr-Mottelson 
collective Hamiltonian is spectacular, 
many interesting subjects of fundamental significance are awaiting our challenge  
in our road toward understanding quantum collective dynamics in nuclei.  
As we briefly remarked in the preceding section, it will be very interesting to  
explore the limits of applicability of the Bohr-Mottelson collective Hamiltonian
by systematically applying it to shape coexistence/fluctuation/mixing phenomena.   
At the present time, the quadrupole collective Hamiltonian is used mainly 
for low-spin states. 
It seems possible, however, to extend the microscopic approach reviewed 
in this paper to collective phenomena at high-spin states  
by taking into account the effects of rapid rotation 
from the beginning in the self-consistent mean fields. 
In a similar manner, 
it will be interesting to extend the collective Hamiltonian approach to describe  
low-lying excited states in neutron-rich unstable nuclei, 
by taking into account the effects of weak binding and continuum coupling 
in constructing the self-consistent mean fields. 
These extensions will open new dimensions 
in quantum collective dynamics of nuclear structure. 
Finally, it should be emphasized that one of the great challenges 
is to calculate the collective inertial masses using the LQRPA method 
on the basis of the density functional theory.

\section*{Acknowledgments}

It is a great pleasure for us to contribute this paper to the Special Physica Scripta Edition 
for the 40 year anniversary of Nobel Prize '75.  
In this occasion, two of the authors (K. M. and M. M.) wish to express their deep appreciations 
to Aage Bohr and Ben R. Mottelson for giving us invaluable ideas and inspirations 
to explore nuclear structure dynamics during our stay (1976-1978 and 1990-1991, respectively) 
in the Niels Bohr Institute, Copenhagen. 
We thank T. Ichikawa for carefully checking the manuscript. 
This work was supported in part by JSPS KAKENHI Grants No. 24105006,
No. 25287065, No. 26400268, and No. 15H03674.
The numerical calculations were carried out on
RIKEN Cluster of Clusters (RICC) facility.


\appendix

\section{Quantization in curvilinear coordinates}
\label{sec:PauliPrescription}

For Cartesian coordinates $q=(q^1,q^2,...,q^f)$ in a $f$-dimensional space, 
the kinetic energy in classical mechanics is given by 
$T=\frac{1}{2}\sum_{i=1}^f (\dot{q}^i)^2$ in a unit with mass $m=1$, 
where $\dot{q}^i$ are time derivatives (velocities) of $q^i$. 
After the canonical quantization, we obtain the kinetic energy operator
\begin{equation}
\hat{T} = -\frac{1}{2}\sum_{i=1}^f \frac{\del^2}{\del {q^i}^2} = -\frac{1}{2}\Delta 
\end{equation}  
in the unit with $\hbar=1$, where $\Delta$ is the Laplacian 
in the Cartesian coordinates.  

For the curvilinear coordinates $x=(x^1,x^2,...,x^f)$ in a $f$-dimensional curved space,  
the line element squared may be written as 
\begin{eqnarray}
ds^2=\sum_{i,j}g_{ij}(x)dx_i dx_j   
\end{eqnarray}
with $g_{ij}(x)=g_{ji}(x)$, 
using the metric tensor $\{g_{ij}(x)\}$ characterizing the curved space. 
The kinetic energy in classical mechanics is then given by
\begin{eqnarray}
T&=&\frac{1}{2} \left(\frac{ds}{dt}\right)^2 \nonumber \\
&=&\frac{1}{2}  \sum_{i,j}g_{ij}(x) \frac{dx^i}{dt} \frac{dx^j}{dt}. 
\end{eqnarray} 
Note that the metric tensor $\{g_{ij}(x)\}$ depends on the coordinate $x$. 

According to the Pauli prescription for quantization in curvilinear coordinates, 
the corresponding kinetic energy operator in quantum mechanics is given by  
\begin{eqnarray}
\hat{T} &=& -\frac{1}{2}\Delta \nonumber \\
&=& -\frac{1}{2\sqrt{g(x)}}\sum_{i,j}\frac{\partial}{\partial x^i} 
\sqrt{g(x)} g^{ij}(x)\frac{\partial}{\partial x^j},
\end{eqnarray}
where $g(x)$ denotes the determinant of the metric tensor, $g(x)$=det $\{g_{ij}(x)\}$,  
and $g^{ij}(x)$ are the components of the inverse matrix $\{g_{ij}(x)\}^{-1}$.   
This expression is obtained in a straightforward way by rewriting  
the Laplacian $\Delta$ in the curvilinear coordinates. 
The Schr{\"o}dinger equation is written as 
\begin{equation}
\left( -\frac{1}{2\sqrt{g(x)}}\sum_{i,j}\frac{\partial}{\partial x^i} 
          \sqrt{g(x)} g^{ij}(x)\frac{\partial}{\partial x^j} +V(x) \right) \psi(x) = E\psi(x). 
\end{equation}
The normarization of the wave function is 
\begin{equation}
\int |\psi(x)|^2 d\tau =1 
\end{equation}
with the volume element 
$d\tau=\sqrt{g(x)}dx\equiv\sqrt{g(x^1,x^2,...,x^f)}dx^1dx^2 \cdot\cdot\cdot dx^f$. 

For the Bohr-Mottelson collective model, $f=5$ and the five collective variables 
consist of the $(\bg)$ deformation variables and the three Euler angles $\vartheta_k$; 
that is,  
 ($x^1=\beta, x^2=\gamma, x^3=\vartheta_1, x^4=\vartheta_2, x^5=\vartheta_3$).
The three components of the angular velocity 
(time-derivatives of the rotational angle) on the intrinsic axes, $\dot{\varphi}_k$ 
appearing in the classical expression of the rotational energy 
$T_{\rm rot}=\frac{1}{2}\sum_{k=1}^3\cJ_k(\bg)\dot{\varphi}_k^2$,   
are related with the time derivatives of the Euler angles $\dot{\vartheta}_k$ by   
\begin{equation}
\dot{\varphi}_k=\sum_{k'=1}^3 V_{kk'}\dot{\vartheta}_{k'} 
\end{equation}
with 
\begin{equation}
V_{kk'}=
\left(
\begin{array}{ccc}
-\sin \vartheta_2 \cos \vartheta_3 & \sin \vartheta_3 &    0 \\
  \sin \vartheta_2 \sin \vartheta_3 & \sin \vartheta_3  &    0 \\
  \cos \vartheta_2                        &          0              &    1             
\end{array}
\right). 
\label{eq:Euler_angles} 
\end{equation}
After the quantization, the classical expression $T_{\rm rot}$ for the rotational energy 
becomes to
$\hat{T}_{\rm rot}=\sum_k\frac{\hat{I}_k^2}{2\cJ_k(\bg)}$, 
where the components of the angular-momentum operator 
on the intrinsic axes, $(\hat{I}_1,\hat{I}_2,\hat{I}_3)$, are represented 
in terms of the Euler angles $(\vartheta_1,\vartheta_2,\vartheta_3)$ 
and the derivatives with respect to them.  
In the same way, 
we obtain, after carrying out somewhat lengthy but straightforward calculations,
the quantum operator $\hat{T}_{\rm vib}$ for the kinetic energy of the
vibrational motion, given in Eq.~(\ref{eq:quantumTvib}),  
and the determinant of the metric tensor,  
\begin{eqnarray}
g(\beta, \gamma, \vartheta_1, \vartheta_2, \vartheta_3) 
          = G(\bg) \sin^2 \vartheta_2, 
\end{eqnarray} 
with $G(\bg)$ given by Eq.~(\ref{eq:BM_G}). 
Note that the determinant does not depend on $\vartheta_1$ and $\vartheta_3$.  
For the definitions of the Euler angles and more details of the calculation, 
see, {\it e.g.}, Chap.~6 in the textbook of Eisenberg and Greiner \cite{eis87}. 

\section{Calculation of $E2$ transitions and moments} 
\label{sec:E2transition}

The electric quadrupole ($E2$) operators in the body-fixed frame are given 
as a sum of neutron and proton contributions
with effective charges $e_{\rm eff}^{(\tau)}$,
\beq
\hat D^{(E2)} _{m}=\sum_{\tau=n, p}e^{(\tau)}_{\rm eff}\hat D^{(\tau)}_{2m},
\eeq
where $\hat D^{(\tau)}_{2m}$ are the quadrupole operator of neutrons and protons,
and $\sum_{\tau}\hat D^{(\tau)}_{2m}=\hat D_{2m}$.
The $E2$ operator in the laboratory frame is related with that in the intrinsic frame as
\beq
\hat D^{\prime (E2)}_{m}=\sum_{m'} \mathcal{D}^2_{mm'}(\Omega)\hat D^{(E2)}_{m'}, 
\eeq
where $\mathcal{D}$ are Wigner's rotational matrices.
The experimental observables such as the $B(E2)$ 
and the spectroscopic quadrupole moment $Q$ are defined as
\begin{eqnarray}
  B(E2;\alpha I \rightarrow \alpha'I')
=\left(2I+1\right)^{-1}|\brared{\alpha I}\hat D'^{(E2)} \ketred{\alpha' I'}|^2 \label{eq:BE2}, \\
  Q(\alpha I)=\sqrt{\frac{16\pi}{5}}\bra{\alpha,I,M=I}\hat D'^{(E2)}_0\ket{\alpha, I, M=I}.
\end{eqnarray}
Here, the reduced matrix element in Eq.~(\ref{eq:BE2}) is defined 
with the Wigner-Eckart theorem,
\begin{equation}
\bra{\alpha,I,M=I}\hat D'^{(E2)}_0\ket{\alpha ', I', M'=I}
=\begin{pmatrix}
  I & 2 & I' \\
 -I & 0 & I
\end{pmatrix}
\brared{\alpha I}\hat D'^{(E2)} \ketred{\alpha' I'}.
\end{equation}
Substituting Eq. (\ref{eq:collective WF}) into  $\ket{\alpha, I, M}$, we obtain \cite{kum67}
\begin{eqnarray}
&\brared{\alpha I} \hat D'^{(E2)} \ketred{\alpha ' I'} \notag\\
=& \sqrt{(2I+1)(2I'+1)}(-)^I\sum_{K}
\left\{
\begin{pmatrix}
  I & 2 & I' \\
 -K & 0 & K
\end{pmatrix}\right.
\bra{\Phi_{\alpha,I,K}}\hat D^{(E2)}_{0+}\ket{\Phi_{\alpha', I',K}} \notag\\
&+\sqrt{(1+\delta_{K0})}
\left[
\begin{pmatrix}
  I & 2 & I' \\
 -K-2 & 2 & K
\end{pmatrix}\right.
\bra{\Phi_{\alpha,I,K+2}}\hat D^{(E2)}_{2+}\ket{\Phi_{\alpha', I',K}} \notag\\
&+\left.\left.(-)^{I+I'}
\begin{pmatrix}
  I & 2 & I' \\
 K & 2 & -K-2
\end{pmatrix}
\bra{\Phi_{\alpha,I,K}}\hat D^{(E2)}_{2+}\ket{\Phi_{\alpha', I',K+2}} \right]\right\},
\label{eq:redmat}
\end{eqnarray}
with $\hat D^{(E2)}_{m+}=(\hat D^{(E2)}_{m}+\hat D^{(E2)}_{-m})/2$.

The quadrupole matrix elements between the intrinsic states are evaluated using the
collective wave functions as
\begin{equation}
\bra{\Phi_{\alpha,I,K}}\hat D^{(E2)}_{m+}\ket{\Phi_{\alpha', I',K'}}
=\int d\beta d\gamma \sqrt{G(\beta,\gamma)} \Phi^*_{\alpha IK}(\beta,\gamma)
D^{(E2)}_{m+}(\beta,\gamma) \Phi_{\alpha' I'K'}(\beta,\gamma),
\end{equation}
where
\begin{equation}
D^{(E2)}_{m+}(\beta,\gamma)=\bra{\phi(\beta,\gamma)}\hat D^{(E2)}_{m+}
\ket{\phi(\beta,\gamma)}.
\end{equation}

\section{Illustration of triaxial deformation dynamics} 
\label{sec:TriaxialDynamics}

We consider a simple model that may be useful to understand several interesting limits
of triaxial deformation dynamics in a unified perspective.
The model discussed below includes several situations, such as
the axially symmetric rotor model,
the $\gamma$-unstable model \cite{wil56},
the triaxial rigid rotor model \cite{dav58},
and oblate-prolate shape coexistence in an ideal situation.
This model enables us to describe the smooth change between these extreme situations
by changing a few parameters.
Here we show only the simplest example, referring Ref.~\cite{sat10} 
for more general cases.
To focus our attention on the $\gamma$-degree of freedom, 
we fix the $\beta$-degree of freedom
at a constant value $\beta_0$ in the Bohr-Mottelson collective Hamiltonian 
(\ref{eq: quantized H_BMcoll})
and parametrize the collective potential $V(\bg)$ as
$V(\beta_0,\gamma)=V_{\beta}(\beta_0) + V_{\gamma}(\beta_0,\gamma)$
with
\begin{equation}
V_{\gamma}(\beta_0,\gamma)=V_0(\beta_0)\sin^2 3\gamma+V_1(\beta_0) \cos3\gamma.
\label{eq:1Dpotential}
\end{equation}
This form is readily obtained by expanding $V(\bg)$
in powers of the basic second- and third-order invariants,
$\beta^2$ and $\beta^3\cos 3\gamma$,
and keeping up to the second order in $\beta^3\cos 3\gamma$.

When $V_1=0$, the collective potential is symmetric with respect to 
the reflection about $\gamma=30^\circ$.
For positive $V_0$, two minima appear at the oblate ($\gamma=60^\circ$) 
and prolate ($\gamma=0^\circ$) shapes.
They are degenerated and separated by a barrier located at $\gamma=30^\circ$.
For negative $V_0$, on the other hand, the barrier top at $\gamma=30^\circ$ 
turns out to be the single minimum,
and it becomes deeper as $|V_0|$ increases.
The term $V_1$ breaks the oblate-prolate symmetry, and controls 
the magnitude of the symmetry breaking.
For positive (negative) $V_1$, the oblate (prolate) shape becomes 
the minimum (when $V_0$ is positive).

Let us discuss the simplest case where $V_1=0$ and 
the collective inertial masses 
$(D_{\beta \beta}, D_{\gamma \gamma}\beta_0^{-2}, D_1, D_2, D_3)$
are replaced by a common constant $D$,
and $D_{\beta \gamma}$ is ignored.
In this case, both the collective potential and the moments of inertia $\cJ(\bg)$
are symmetric with respect to the reflection about $\gamma=30^\circ$,
so that the collective Hamiltonian possesses the oblate-prolate symmetry.
Furthermore, $D$ and $\beta_0$ appear only in the form $(2D\beta_0^2)^{-1}$ 
in the kinetic energy.
Therefore the ratio $2D\beta_0^2V_0$ is a single quantity that enters in 
the collective Schr\"odinger equation
(\ref{eq:collective-schrodinger}) and determines the dynamics.
A particular case of $V_0=0$ is known to be 
the Wilets-Jean $\gamma$-unstable model \cite{wil56}.
In this case the excitation spectra just scale with the factor $(2D\beta_0^2)^{-1}$.

\begin{figure}[htbp]
\begin{center}
\includegraphics[width=0.8\textwidth]{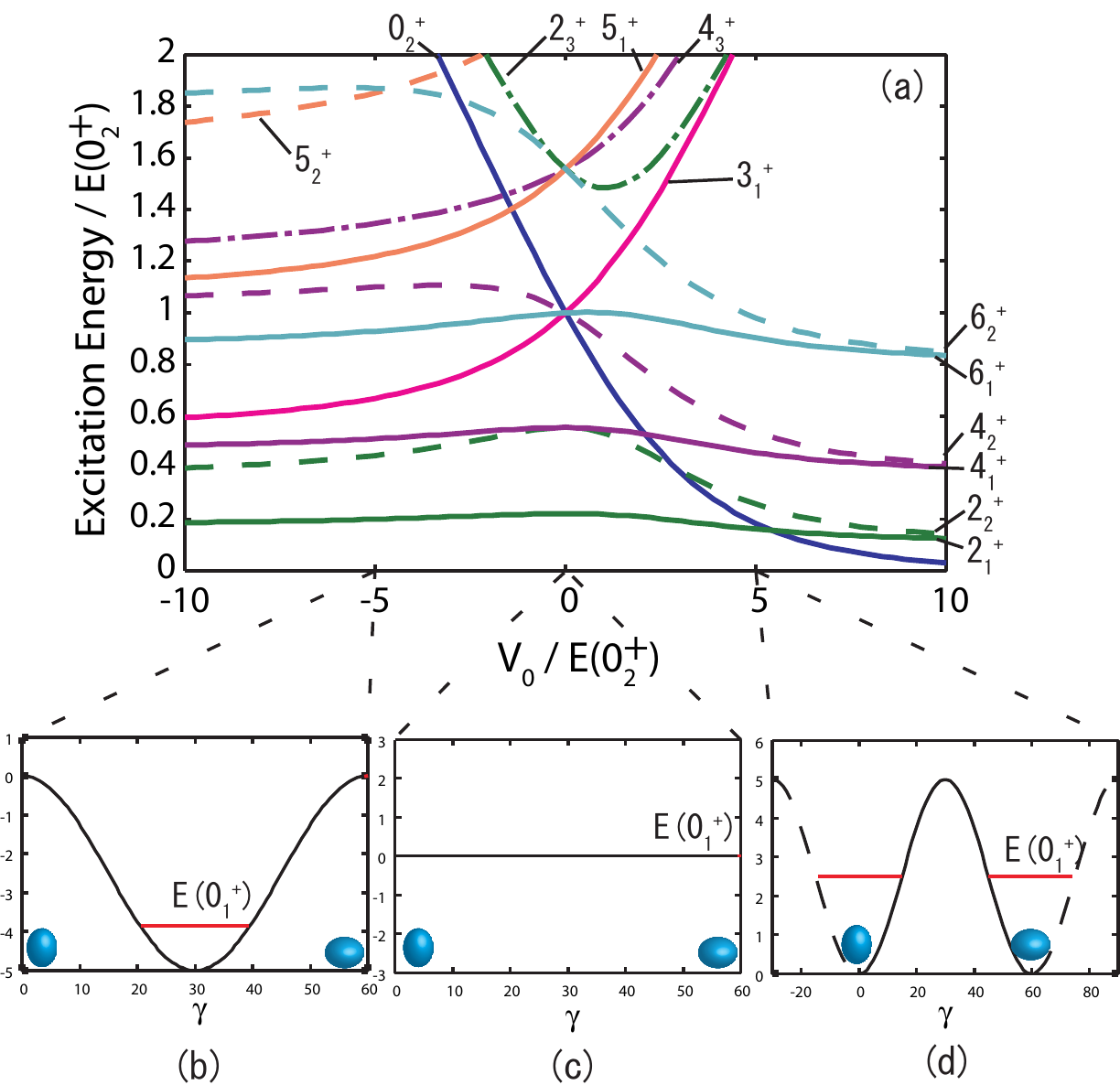}
\caption{
  (a): Excitation energies as functions of the parameter $V_0$. 
Both the excitation energies and the $V_0$ are normalized by the
excitation energy of the second $0^+$ state for $V_0=0$.
  (b),(c),(d): The collective potentials $V(\gamma)$ and the ground state energies 
  $E(0_1^+)$ at $V_0/E(0_2^+) =$ -5.0 (b), 0.0 (c), and 5.0 (d).
  Note that $E(0_1^+)=0$ for $V_0=0$.
  The collective potential is a periodic function of $60^\circ$ in $\gamma$, and
  only the region $0^\circ\le \gamma\le 60^\circ$ is drawn with a solid line in (d).
  This figure is taken from~\cite{sat10}.
}
\label{fig:triaxial-spectra}
\end{center}
\end{figure}

Figure~\ref{fig:triaxial-spectra} shows excitation spectra as functions of $V_0$.
The excitation energies are normalized with the excitation energy of 
the second $0^+$ state (first excited $0^+$ state)
$E(0_2^+)$ at $V_0=0$ (which is 1.8 MeV for $\beta_0^2=0.1$ and $D=50$ MeV$^{-1}$).
Because of the scaling property of the collective Schr\"odinger equation, 
this figure is valid for any value of $(2D\beta_0^2)^{-1}$.
In the lower panels, the collective potentials $V(\beta_0,\gamma)$ are shown 
for three extreme situations, namely
1) a triaxially deformed case with a deep minimum at $\gamma=30^\circ$,
2) a $\gamma$-unstable case, 
and 
3) an ideal case of oblate-prolate shape coexistence with two degenerated minima.
Note that the collective potential $V(\beta,\gamma)$ is a periodic function of 
$60^\circ$ in $\gamma$.
The solid line in Fig.~\ref{fig:triaxial-spectra}(d) shows the region 
$0^\circ\le \gamma\le 60^\circ$.

When $V_0$ is positive, a doublet structure appears with increasing 
the barrier-height parameter $V_0$.
In other words, an approximately degenerated pair of eigenstates emerges 
for every angular momentum when $V_0/E(0_2^+)\gg 1$.
This is a well-known doublet pattern in the double-well potential problem.
We can associate this doublet structure with the oblate-prolate symmetry 
as seen in Fig.~\ref{fig:triaxial-spectra}(d).
Furthermore, we notice a very unique behavior of the $0_2^+$ state; 
when $V_0$ decreases in the positive-$V_0$ side,
its energy rises more rapidly than those of the yrare $2_2^+, 4_2^+$, and $6_2^+$ states.
It crosses with $E(2_2^+)$ at $V_0/E(0_2^+)\simeq 3$, and finally at $V_0=0$, 
the $0_2^+$ state is degenerated
with $4_2^+$ and $6_1^+$ states, as expected in the Wilets-Jean model \cite{wil56}. 

In the negative-$V_0$ side, the excitation energies of $3_1^+$ and $5_1^+$ states 
rapidly decrease with decreasing $V_0$,
and when the potential minimum at $\gamma=30^\circ$becomes very deep,  
the spectrum becomes similar with that of the Davydov-Filippov
rigid triaxial rotor model \cite{dav58}.
\\

\section{Time-dependent unitary transformation of the HFB state vectors} 
\label{sec:UnitaryTransformation}

Let us first consider the TDHF case. 
It is convenient to define the particle-hole concept with respect to 
the HF ground state $|\phi_{\rm{HF}}\rangle$ for doubly even nuclei by 
\begin{eqnarray}
c^\dag_i &=& (1-n_i) c^\dag_i + n_i c^\dag_i =  a^\dag_i +  b_{\bar i}, \nonumber \\
c_{\bar i} &=& (1-n_i) c_{\bar i}  + n_i c_{\bar i} = a_{\bar i} - b_i^\dag.  
\label{Eq:particle-hole}
\end{eqnarray} 
Here $c^\dag_i$ and $c_{\bar i}$ are the nucleon creation and annihilation operators 
in the HF states $i$ and its time-conjugate states ${\bar i}$, respectively, 
and $n_i$ is 1 or 0 according to whether a pair of the HF states $(i,{\bar i})$ is occupied or unoccupied. 
The nucleon operators ($c^\dag_i$, $c_{\bar i}$) correspond to the particle operators 
($a^\dag_i, a_{\bar i}$) for unoccupied space and 
the hole operators ($b_{\bar i}, b^\dag_i$) for the occupied space.  
Obviously, the HF ground state is a vacuum for the particles and holes:  
\begin{eqnarray}
a_i|\phi_{\rm{HF}}\rangle=b_j|\phi_{\rm{HF}}\rangle=0. 
\end{eqnarray}

According to the Thouless theorem \cite{tho60}, another HF state $|\phi_{\rm{HF}}(t)\rangle$ 
non-orthogonal to  $|\phi_{\rm{HF}}\rangle$ can be written as 
\begin{equation}
|\phi_{\rm{HF}}(t)\rangle = N(t) \exp (\sum_{ij} z_{ij}(t) a_i^{\dag}b_j^\dag)|\phi_{\rm{HF}}\rangle 
\end{equation}
with the normalization constant  $N(t)$.  
It may be more convenient to describe the same HF state as a unitary transformation 
of $|\phi_{\rm{HF}}\rangle$ \cite{mar80,row80} :  
\begin{equation}
|\phi_{\rm{HF}}(t)\rangle = e^{i{\hat G_{\rm{HF}}}(t)}|\phi_{\rm{HF}}(t=0)\rangle 
\label{Eq:TDHF}
\end{equation}
with
\begin{equation}
i{\hat G_{\rm{HF}}}(t)= \sum_{ij} (f_{ij}(t) a_i^{\dag}b_j^\dag - f_{ij}^*(t) b_ja_i). 
\end{equation}
Here, $|\phi_{\rm HF}\rangle$ is denoted $|\phi_{\rm{HF}}(t=0)\rangle$ to emphasize 
that Eq.~(\ref{Eq:TDHF}) can be regarded as a time-dependent unitary transformation 
describing the time evolution of the TDHF state vectors. 
In this generalized form, in contrast to the original Thouless theorem, 
even the HF states orthogonal to $|\phi_{\rm{HF}}(t=0)\rangle$ can be described. 
\\

It is straightforward to generalize the above formulation to the TDHFB case 
including the pairing correlations.  
The particle-hole concept in the HF theory is replaced by the quasiparticle concept,  
which is introduced through the generalized Bogoliubov transformations \cite{rin80}, 
\begin{eqnarray}
c^\dag_i &=&  \sum_j (u_{ij}^*a_j^\dag +  v_{ij}a_j),    \nonumber \\
c_i &=&  \sum_j (u_{ij}a_j + v_{ij}^* a_j^\dag ),  
\label{Eq:quasiparticle}
\end{eqnarray} 
(separately for protons and neutrons) in the HFB theory. 
(The use of the same notation ($a^\dag_i, a_{\bar i}$) for the quasipartcles in the HFB theory
and the particles in the HF theory may not cause any confusion.)
The particle-hole pair creation and annihilation operators 
$(a_i^{\dag}b_j^\dag, b_ja_i)$ are then replaced by the two-quasipartcle creation and 
annihilation operators $(a_i^{\dag}a_j^\dag, a_ja_i)$.  
Similarly to Eq.~(\ref{Eq:TDHF}) in the TDHF case, 
the time-evolution of the TDHFB state $\ket{\phi(t)}$ can be described as  
a time-dependent unitary transformation \cite{suz83,rin77}: 
\begin{equation}
\ket{\phi(t)} = e^{i{\hat G}(t)} \ket{\phi(t=0)},  
\label{Eq:TDHFB}
\end{equation}
where $i{\hat G}(t)$ is a one-body anti-Hermitian operator given by 
\begin{equation}
i{\hat G}(t)= \sum_{(ij)} (g_{ij}(t) a_i^{\dag}a_j^\dag - g_{ij}^*(t) a_ja_i). 
\end{equation}
Here, the sum is taken over independent two quasiparticle configurations $(ij)$. 
For the HFB state at $t=0$, one may choose the HFB ground state $\ket{\phi_0}$   
which satisfies the vacuum condition for the quasiparticles:  
\begin{equation}
a_i\ket{\phi_0} = 0. 
\end{equation}
It is important to note that Eq.~(\ref{Eq:TDHFB}) is valid for any choice of the initial HFB state 
$\ket{\phi(t=0)}$, if the quasiparticle operators  in $i{\hat G}(t)$ are defined with respect to 
$\ket{\phi(t=0)}$.  

Because ${\hat G}(t)$ is a one-body operator, it is possible to define 
quasiparticle creation and annihilation operators $\{a_i^\dag(t), a_j(t)\}$
with respect to $\ket{\phi(t)}$ as follows: 
\begin{eqnarray}
a_i(t) &=& e^{i{\hat G}(t)} a_i e^{-i{\hat G}(t)}    \nonumber \\
       &=& a_i + [ i{\hat G}, a_i ] 
          + \frac{1}{2} [ i{\hat G}, [ i{\hat G}, a_i ] ] 
          + \frac{1}{6}[ i{\hat G}, [ i{\hat G}, [ i{\hat G}, a_i ]]] + \cdot\cdot\cdot   \nonumber \\ 
       &=& \sum_j ( U_{ji}(t) a_j + V_{ji}(t) a_j^\dag ). 
\end{eqnarray}
The matrices, $U(t)$ and $V(t)$, composed of the amplitudes $U_{ij}(t)$ and $V_{ij}(t)$,  
are given by \cite{sak83,rin77} 
\begin{eqnarray}
U^T(t) &=& \cos \sqrt{G^\dag G}, \\
V^T(t) &=& G^\dag \frac{\sin \sqrt{G G^\dag}} {\sqrt{G G^\dag}}, 
\end{eqnarray}
where $G$ is a matrix composed of the components $g_{ij}$. 
Obviously, the quasiparticle operators $\{a_i^\dag(t), a_j(t) \}$ satisfy the vacuum condition 
for  $\ket{\phi(t)}$: 
\begin{eqnarray}
a_i(t)\ket{\phi(t)} &=& e^{i{\hat G}(t)} a_i e^{-i{\hat G}(t)} e^{i{\hat G}(t)} \ket{\phi(t=0)} \nonumber \\
                        &=& e^{i{\hat G}(t)} a_i \ket{\phi(t=0)} \nonumber \\
                        &=& 0.
\end{eqnarray}


\end{document}